\pdfoutput=1

\documentclass[11pt,a4paper,twoside]{article}

\usepackage[utf8]{inputenc}
\usepackage[T1]{fontenc}
  
\usepackage[inner=2.5cm,outer=1.5cm,top=3cm,bottom=2cm]{geometry}
\usepackage[colorlinks]{hyperref}
\usepackage{
sectsty,
multicol,
alphalph,
cite,
tocloft,
bbold,
color,
colortbl,
caption,
subcaption,
listings,
multirow,
arydshln,
appendix,
titlesec,
bold-extra,
authblk,
setspace
}

\usepackage{amsmath,amssymb,amsfonts,slashed,mathtools,ulem,centernot,empheq,bm,array}
\usepackage[makeroom]{cancel}

\usepackage{feynmp-auto,graphics,graphicx,tikz,pgfplots,pdfpages}

\allowdisplaybreaks

\renewcommand{\arraystretch}{1} 

 \hypersetup{
 colorlinks=true
 ,citecolor=red
 ,urlcolor=blue
 ,anchorcolor=blue
 ,citecolor=red
 ,filecolor=blue
 ,linkcolor=blue
 ,menucolor=blue
 ,linktocpage=true
 ,pdfproducer=medialab
 }

\usetikzlibrary{patterns}
\usetikzlibrary{calc}
\usetikzlibrary{tikzmark}

\newcommand*\widefbox[1]{\fbox{\hspace{0.25cm}#1\hspace{0.25cm}}}

\allsectionsfont{\boldmath} 

\pgfplotsset{compat=1.18}

\newcommand\titlemath[1]{\texorpdfstring{\(#1\)}{xx}}

\title{Critical properties of three-dimensional many-flavour QEDs}

\author[$\dagger$]{S.\ Metayer}
\author[$\dagger$]{S.\ Teber}
\affil[$\dagger$]{Sorbonne Universit\'e, CNRS, Laboratoire de Physique Th\'eorique et Hautes Energies, LPTHE, F-75005 Paris, France.}
\date{}                     
\begin{document}

\newcommand{\com}[1]{\textcolor{red}{{#1}}} 
\newcommand{\coms}[1]{\textcolor{magenta}{{#1}}} 

\newcommand{\ol}[1]{\textcolor{green}{{#1}}} 


\newcommand{\eg}{{\it e.g.}}
\newcommand{\ie}{{\it i.e.}}
\newcommand{\etal}{{\it et al.\ }}

\newcommand{\eps}{\ensuremath{\varepsilon}}
\newcommand{\ep}{\ensuremath{\varepsilon}}
\newcommand{\ve}{\ensuremath{\varepsilon}}
\newcommand{\veps}{\ensuremath{\varepsilon}}
\newcommand{\al}{\alpha}
\newcommand{\bra}{\langle }
\newcommand{\ket}{\rangle }
\newcommand{\Tr}{\mbox{Tr}}
\newcommand{\Ord}{{\rm O}}
\newcommand{\K}[1]{\mathcal{K}\left(#1\right)}
\newcommand{\EP}{\mathcal{E}}
\newcommand{\I}{\ensuremath{{\mathrm{i}}}}
\newcommand{\D}{\ensuremath{\mathrm{d}}}
\newcommand{\fsl}[1]{{\centernot{#1}}}
\newcommand{\cE}{\ensuremath{\mathcal{E}}}
\newcommand{\cL}{\ensuremath{\mathcal{L}}}
\newcommand{\cC}{\ensuremath{\mathcal{C}}}
\newcommand{\cN}{\ensuremath{\mathcal{N}}}
\newcommand{\xib}{\bar{\xi}}
\newcommand{\Lpt}{\tilde{L}}
\newcommand{\KK}[1]{\mathcal{K}\!\left[#1\right]}
\def\Sp{{\slashed p}}
\def\Sq{{\slashed q}}
\def\Sk{{\slashed k}}
\def\ra{{\rightarrow}}
\newcommand{\C}{\mathbb{C}}
\newcommand{\N}{\mathbb{N}}
\newcommand{\Q}{\mathbb{Q}}
\newcommand{\R}{\mathbb{R}}
\newcommand{\Z}{\mathbb{Z}}
\newcommand{\pE}{p_{{}_{\!E}}}
\newcommand{\kE}{k_{{}_{\!E}}}
\newcommand{\abs}[1]{\left|#1\right|}
\newcommand{\Nf}{\ensuremath{N_{\hspace{-0.4mm}f}}}
\newcommand{\nf}{\ensuremath{n_{\hspace{-0.4mm}f}}}
\newcommand{\Nfr}{{\Nf}_{\!\!r}}
\newcommand{\nfr}{{\nf}_{\!\!r}}
\newcommand{\Nc}{\ensuremath{N_{\hspace{-0.2mm}c}}}
\newcommand{\nc}{\ensuremath{n_{\hspace{-0.2mm}c}}}

\makeatletter
\newcommand{\vast}{\bBigg@{4}}
\newcommand{\Vast}{\bBigg@{5}}
\makeatother

\def\be{\begin{equation}}
\def\ee{\end{equation}}
\def\ba#1\ea{\begin{align}#1\end{align}}
\def\bs{\begin{subequations}}
\def\es{\end{subequations}}
\def\nonum{\nonumber}

\definecolor{mygray}{gray}{0.8}

\def\bl{\begin{adjustwidth}{-\extralength}{0cm}}
\def\el{\end{adjustwidth}}

\newcommand{\MasterTwoA}{
    \parbox[c][10mm][c]{25mm}{
        \centering
        \begin{fmfgraph*}(25,20)
            \fmfleft{i}
            \fmfright{o}
            \fmf{plain}{i,v1}
            \fmf{plain}{o,v5}
            \fmfn{phantom}{v}{5}
            \fmffreeze
            \fmfforce{(0.1w,0.5h)}{v1}
            \fmfforce{(0.9w,0.5h)}{v5}
            \fmf{plain,left}{v1,v3}
            \fmf{plain,right}{v1,v3}
            \fmf{plain,left}{v3,v5}
            \fmf{plain,right}{v3,v5}
            \fmfdot{v1,v3,v5}
        \end{fmfgraph*}
    }
}

\newcommand{\MasterTwoB}{
    \parbox[c][10mm][c]{15mm}{
        \centering
        \begin{fmfgraph*}(15,20)
            \fmfleft{i}
            \fmfright{o}
            \fmf{plain}{i,v1}
            \fmf{plain}{o,v3}
            \fmfn{plain}{v}{3}
            \fmffreeze
            \fmfforce{(0.1w,0.5h)}{v1}
            \fmfforce{(0.9w,0.5h)}{v3}
            \fmf{plain,left}{v1,v3}
            \fmf{plain,right}{v1,v3}
            \fmfdot{v1,v3}
        \end{fmfgraph*}
    }
}

\newcommand{\generalprop}[1]{
\parbox[c][0mm][c]{10mm}{
\centering
\begin{fmfgraph*}(10,10)
\fmfleft{i}
\fmfright{o}
\fmf{plain,label=$#1$}{i,o}
\end{fmfgraph*}
}
}

\newcommand{\vdot}{
\parbox[c][0mm][c]{0mm}{
\centering
\begin{fmfgraph*}(0,0)
\fmfleft{v}
\fmfdot{v}
\end{fmfgraph*}
}}

\newcommand{\generaloneloop}[2]{
\parbox[c][15mm][c]{20mm}{
\centering
\begin{fmfgraph*}(10,10)
\fmfleft{i}
\fmfright{o}
\fmfleft{ve}
\fmfright{vo}
\fmffreeze
\fmfforce{(-0.3w,0.5h)}{i}
\fmfforce{(1.3w,0.5h)}{o}
\fmfforce{(0w,0.5h)}{ve}
\fmfforce{(1.0w,0.5h)}{vo}
\fmffreeze
\fmf{plain}{i,ve}
\fmf{plain,left,label=$#1$,l.d=0.1h}{ve,vo}
\fmf{plain,left,label=$#2$,label.side=left,l.d=0.1h}{vo,ve}
\fmf{plain}{vo,o}
\fmffreeze
\fmfdot{ve,vo}
\end{fmfgraph*}
}
}

\newcommand{\generaloneloopnolegs}[2]{
\parbox[c][20mm][c]{20mm}{
\centering
\begin{fmfgraph*}(10,10)
\fmfleft{i}
\fmfright{o}
\fmfleft{ve}
\fmfright{vo}
\fmffreeze
\fmfforce{(-0.3w,0.5h)}{i}
\fmfforce{(1.3w,0.5h)}{o}
\fmfforce{(0w,0.5h)}{ve}
\fmfforce{(1.0w,0.5h)}{vo}
\fmffreeze
\fmf{plain,left,label=$#1$,l.d=0.1h}{ve,vo}
\fmf{plain,left,label=$#2$,label.side=left,l.d=0.1h}{vo,ve}
\fmffreeze
\fmfdot{ve,vo}
\end{fmfgraph*}
}
}

\newcommand{\generaltwoloop}[5]{
\parbox[c][20mm][c]{20mm}{
\centering
\begin{fmfgraph*}(12,10)
\fmfleft{i}
\fmfright{o}
\fmfleft{ve}
\fmfright{vo}
\fmftop{vn}
\fmftop{vs}
\fmffreeze
\fmfforce{(-0.3w,0.5h)}{i}
\fmfforce{(1.3w,0.5h)}{o}
\fmfforce{(0w,0.5h)}{ve}
\fmfforce{(1.0w,0.5h)}{vo}
\fmfforce{(.5w,1.1h)}{vn}
\fmfforce{(.5w,-0.1h)}{vs}
\fmffreeze
\fmf{plain}{i,ve}
\fmf{plain,left}{ve,vo}
\fmf{phantom,left,label=$#1$,l.d=-0.1w}{ve,vn}
\fmf{phantom,right,label=$#2$,l.d=-0.1w}{vo,vn}
\fmf{plain,left}{vo,ve}
\fmf{phantom,left,label=$#3$,l.d=-0.1w}{vo,vs}
\fmf{phantom,right,label=$#4$,l.d=-0.1w}{ve,vs}
\fmf{plain,label=$#5$,l.d=0.05w}{vs,vn}
\fmf{plain}{vo,o}
\fmffreeze
\fmfdot{ve,vn,vo,vs}
\end{fmfgraph*}
}
}

\newcommand{\doublebubble}[4]{
\parbox[c][15mm][c]{25mm}{
\centering
\begin{fmfgraph*}(25,20)
\fmfleft{i}
\fmfright{o}
\fmf{plain}{i,v1}
\fmf{plain}{o,v5}
\fmfn{phantom}{v}{5}
\fmffreeze
\fmfforce{(0.1w,0.5h)}{v1}
\fmfforce{(0.9w,0.5h)}{v5}
\fmf{plain,left,label=$#1$,l.d=0.05w}{v1,v3}
\fmf{plain,right,label=$#2$,l.d=0.05w}{v1,v3}
\fmf{plain,left,label=$#3$,l.d=0.05w}{v3,v5}
\fmf{plain,right,label=$#4$,l.d=0.05w}{v3,v5}
\fmfdot{v1,v3,v5}
\end{fmfgraph*}
}
}

\newcommand{\sunset}[3]{
\parbox[c][15mm][c]{15mm}{
\centering
\begin{fmfgraph*}(15,20)
\fmfleft{i}
\fmfright{o}
\fmf{plain}{i,v1}
\fmf{plain}{o,v3}
\fmffreeze
\fmfforce{(0.1w,0.5h)}{v1}
\fmfforce{(0.9w,0.5h)}{v3}
\fmf{plain,left,label=$#1$,l.d=0.05w}{v1,v3}
\fmf{plain,label=$#2$,l.d=0.05w}{v3,v1}
\fmf{plain,left,label=$#3$,l.d=0.05w}{v3,v1}
\fmfdot{v1,v3}
\end{fmfgraph*}
}
}

\newcommand{\eye}[4]{
\parbox[c][15mm][c]{15mm}{
\centering
\begin{fmfgraph*}(15,20)
\fmfleft{i}
\fmfright{o}
\fmf{plain}{i,v1}
\fmf{plain}{o,v3}
\fmfn{phantom}{v}{3}
\fmffreeze
\fmfforce{(0.1w,0.5h)}{v1}
\fmfforce{(0.9w,0.5h)}{v3}
\fmf{plain,label=$#1$,l.d=0.05w,left}{v1,v3}
\fmf{plain,right}{v1,v3}
\fmf{phantom,label=$#2$,l.d=-0.001w,right=0.5}{v1,v4}
\fmf{phantom,label=$#3$,l.d=-0.001w,right=0.5}{v4,v3}
\fmffreeze
\fmfforce{(0.5w,0.2h)}{v4}
\fmf{plain,label=$#4$,l.d=-0.001w,right=0.5}{v4,v1}
\fmfdot{v1,v3,v4}
\end{fmfgraph*}
}
}

\newcommand{\GraphGeneraThreeLoopLA}{
\parbox[c][20mm][c]{34mm}{
\centering
\begin{fmfgraph*}(30,25)
\fmfleft{i}
\fmfright{o}
\fmf{plain}{i,v1}
\fmf{plain}{v2,o}
\fmf{phantom,right,tension=0.1,tag=1}{v1,v2}
\fmf{phantom,right,tension=0.1,tag=2}{v2,v1}
\fmf{phantom,tension=0.1,tag=3}{v1,v2}
\fmfdot{v1,v2}
\fmfposition
\fmfipath{p[]}
\fmfiset{p1}{vpath1(__v1,__v2)}
\fmfiset{p2}{vpath2(__v2,__v1)}
\fmfi{plain,label=$\alpha_6$,l.d=0.02w}{subpath (0,length(p1)/3) of p1}
\fmfi{plain,label=$\alpha_5$,l.d=0.04w}{subpath (length(p1)/3,2*length(p1)/3) of p1}
\fmfi{plain,label=$\alpha_4$,l.d=0.02w}{subpath (2*length(p1)/3,length(p1)) of p1}
\fmfi{plain,label=$\alpha_3$,l.d=0.02w}{subpath (0,length(p2)/3) of p2}
\fmfi{plain,label=$\alpha_2$,l.d=0.04w}{subpath (length(p2)/3,2*length(p2)/3) of p2}
\fmfi{plain,label=$\alpha_1$,l.d=0.02w}{subpath (2*length(p2)/3,length(p2)) of p2}
\fmfi{plain,label=$\alpha_7$,l.d=0.01w}{point length(p1)/3 of p1 -- point 2*length(p2)/3 of p2}
\fmfi{plain,label=$\alpha_8$,l.d=0.01w,label.side=left}{point 2*length(p1)/3 of p1 -- point length(p2)/3 of p2}
\myvert{point length(p1)/3 of p1}
\myvert{point 2*length(p1)/3 of p1}
\myvert{point length(p2)/3 of p2}
\myvert{point 2*length(p2)/3 of p2}
\end{fmfgraph*}
}
}

\newcommand{\GraphGeneraThreeLoopBE}{
\parbox[c][20mm][c]{34mm}{
\centering
\begin{fmfgraph*}(30,25)
\fmfleft{i}
\fmfright{o}
\fmf{plain}{i,v1}
\fmf{plain}{v2,o}
\fmf{phantom,right,tension=0.1,tag=1}{v1,v2}
\fmf{phantom,right,tension=0.1,tag=2}{v2,v1}
\fmf{phantom,tension=0.1,tag=3}{v1,v2}
\fmfdot{v1,v2}
\fmfposition
\fmfipath{p[]}
\fmfiset{p2}{vpath1(__v1,__v2)}
\fmfiset{p1}{vpath2(__v2,__v1)}
\fmfiset{p3}{vpath3(__v2,__v1)}
\fmfi{plain,label=$\alpha_3$,l.d=0.02w}{subpath (0,length(p1)/4) of p1}
\fmfi{plain,label=$\alpha_2$,l.d=0.04w}{subpath (length(p1)/4,3*length(p1)/4) of p1}
\fmfi{plain,label=$\alpha_1$,l.d=0.02w}{subpath (3*length(p1)/4,length(p1)) of p1}
\fmfi{plain,label=$\alpha_5$,l.d=0.02w}{subpath (0,length(p2)/2) of p2}
\fmfi{plain,label=$\alpha_4$,l.d=0.02w}{subpath (length(p2)/2,length(p2)) of p2}
\fmfi{plain,label=$\alpha_7$,label.side=left,l.d=0.001w}{point length(p1)/4 of p1 -- point length(p3)/2 of p3}
\fmfi{plain,label=$\alpha_6$,label.side=right,l.d=0.001w}{point 3*length(p1)/4 of p1 -- point length(p3)/2 of p3}
\fmfi{plain,label=$\alpha_8$,l.d=0.02w}{point length(p2)/2 of p2 -- point length(p3)/2 of p3}
\myvert{point length(p1)/4 of p1}
\myvert{point 3*length(p1)/4 of p1}
\myvert{point length(p3)/2 of p3}
\myvert{point length(p2)/2 of p2}
\end{fmfgraph*}
}
}

\newcommand{\GraphGeneraThreeLoopNP}{
\parbox[c][20mm][c]{34mm}{
\centering
\begin{fmfgraph*}(30,25)
\fmfleft{i}
\fmfright{o}
\fmf{plain}{i,v1}
\fmf{plain}{v2,o}
\fmf{phantom,right,tension=0.1,tag=1}{v1,v2}
\fmf{phantom,right,tension=0.1,tag=2}{v2,v1}
\fmf{phantom,tension=0.1,tag=3}{v1,v2}
\fmfdot{v1,v2}
\fmfposition
\fmfipath{p[]}
\fmfiset{p2}{vpath1(__v1,__v2)}
\fmfiset{p1}{vpath2(__v2,__v1)}
\fmfi{plain,label=$\alpha_3$,l.d=0.02w}{subpath (0,length(p1)/4) of p1}
\fmfi{plain,label=$\alpha_2$,l.d=0.04w}{subpath (length(p1)/4,3*length(p1)/4) of p1}
\fmfi{plain,label=$\alpha_1$,l.d=0.02w}{subpath (3*length(p1)/4,length(p1)) of p1}
\fmfi{plain,label=$\alpha_6$,l.d=0.02w}{subpath (0,length(p2)/4) of p2}
\fmfi{plain,label=$\alpha_5$,l.d=0.04w}{subpath (length(p2)/4,3*length(p2)/4) of p2}
\fmfi{plain,label=$\alpha_4$,l.d=0.02w}{subpath (3*length(p2)/4,length(p2)) of p2}
\fmfi{plain,label=~~$\alpha_7$,label.side=left,l.d=0.05w}{point length(p1)/4 of p1 -- point length(p2)/4 of p2}
\fmfi{plain,label=~~$\alpha_8$,label.side=left,l.d=+0.05w}{point 3*length(p1)/4 of p1 -- point 3*length(p2)/4 of p2}
\myvert{point length(p1)/4 of p1}
\myvert{point 3*length(p1)/4 of p1}
\myvert{point length(p2)/4 of p2}
\myvert{point 3*length(p2)/4 of p2}
\end{fmfgraph*}
}
}

\newcommand{\GRAPHLOAa}{
\parbox[c][20mm][c]{33mm}{
\centering
\begin{fmfgraph*}(23,23)
\fmfleft{i}
\fmfright{o}
\fmf{photon}{i,v1}
\fmf{fermion,left,label=$k$}{v1,v2}
\fmf{fermion,left,label=$k-p$}{v2,v1}
\fmf{photon}{v2,o}
\fmfforce{(0w,0.5h)}{i}
\fmfforce{(1w,0.5h)}{o}
\fmfforce{(0.25w,0.5h)}{v1}
\fmfforce{(0.75w,0.5h)}{v2}
\fmflabel{$\mu$}{i}
\fmflabel{$\nu$}{o}
\marrow{a}{down}{-90}{$p$}{i,v1}
\fmfdot{v1,v2}
\end{fmfgraph*}
}}

\newcommand{\GRAPHLOAb}{
\parbox[c][20mm][c]{33mm}{
\centering
\begin{fmfgraph*}(23,23)
\fmfleft{i}
\fmfright{o}
\fmf{photon}{i,v1}
\fmf{selectron,left,label=$k$}{v1,v2}
\fmf{selectron,left,label=$k-p$}{v2,v1}
\fmf{photon}{v2,o}
\fmfforce{(0w,0.5h)}{i}
\fmfforce{(1w,0.5h)}{o}
\fmfforce{(0.25w,0.5h)}{v1}
\fmfforce{(0.75w,0.5h)}{v2}
\fmflabel{$\mu$}{i}
\fmflabel{$\nu$}{o}
\marrow{a}{down}{-90}{$p$}{i,v1}
\fmfdot{v1,v2}
\end{fmfgraph*}
}}

\newcommand{\GRAPHLOeps}{
\parbox[c][23mm][c]{33mm}{
\centering
\begin{fmfgraph*}(23,23)
\fmfleft{i}
\fmfright{o}
\fmf{epsscalar}{i,v1}
\fmf{fermion,left,label=$k$}{v1,v2}
\fmf{fermion,left,label=$k-p$}{v2,v1}
\fmf{epsscalar}{v2,o}
\fmfforce{(0w,0.5h)}{i}
\fmfforce{(1w,0.5h)}{o}
\fmfforce{(0.25w,0.5h)}{v1}
\fmfforce{(0.75w,0.5h)}{v2}
\fmflabel{$\mu$}{i}
\fmflabel{$\nu$}{o}
\marrow{a}{down}{-90}{$p$}{i,v1}
\fmfdot{v1,v2}
\end{fmfgraph*}
}}

\newcommand{\GRAPHLOlambda}{
\parbox[c][23mm][c]{25mm}{
\centering
\begin{fmfgraph*}(23,23)
\fmfleft{i}
\fmfright{o}
\fmf{photino}{i,v1}
\fmf{fermion,left}{v1,v2}
\fmf{selectron,left}{v2,v1}
\fmf{photino}{v2,o}
\fmfforce{(0w,0.5h)}{i}
\fmfforce{(1w,0.5h)}{o}
\fmfforce{(0.25w,0.5h)}{v1}
\fmfforce{(0.75w,0.5h)}{v2}
\marrow{a}{down}{-90}{$p$}{i,v1}
\Marrow{b}{down}{bot}{$k-p$}{v2,v1}{20}
\Marrow{c}{up}{top}{$k$}{v1,v2}{20}
\fmfdot{v1,v2}
\end{fmfgraph*}
}}

\newcommand{\GRAPHLOlambdabis}{
\parbox[c][23mm][c]{25mm}{
\centering
\begin{fmfgraph*}(23,23)
\fmfleft{i}
\fmfright{o}
\fmf{photino}{i,v1}
\fmf{fermion,right}{v2,v1}
\fmf{selectron,right}{v1,v2}
\fmf{photino}{v2,o}
\fmfforce{(0w,0.5h)}{i}
\fmfforce{(1w,0.5h)}{o}
\fmfforce{(0.25w,0.5h)}{v1}
\fmfforce{(0.75w,0.5h)}{v2}
\marrow{a}{down}{-90}{$p$}{i,v1}
\Marrow{b}{down}{bot}{$k-p$}{v2,v1}{20}
\Marrow{c}{up}{top}{$k$}{v1,v2}{20}
\fmfdot{v1,v2}
\end{fmfgraph*}
}}

\newcommand{\GRAPHLOpsia}{
\parbox[c][18mm][c]{33mm}{
\centering
\begin{fmfgraph*}(30,30)
\fmfleft{i}
\fmfright{o}
\fmf{fermion}{i,v1}
\fmf{fermion,label=$k$}{v1,v2}
\fmf{fermion}{v2,o}
\fmf{photon,left,label=$p-k$}{v1,v2}
\fmf{phantom,left,label=\circled{1},label.dist=0}{v1,v2}
\fmfforce{(0w,0.5h)}{i}
\fmfforce{(1w,0.5h)}{o}
\fmfforce{(0.25w,0.5h)}{v1}
\fmfforce{(0.75w,0.5h)}{v2}
\marrow{a}{down}{-90}{$p$}{i,v1}
\fmfdot{v1,v2}
\end{fmfgraph*}
}}

\newcommand{\GRAPHLOpsib}{
\parbox[c][18mm][c]{33mm}{
\centering
\begin{fmfgraph*}(30,30)
\fmfleft{i}
\fmfright{o}
\fmf{fermion}{i,v1}
\fmf{fermion,label=$k$}{v1,v2}
\fmf{fermion}{v2,o}
\fmf{epsscalar,right,label=$p-k$}{v2,v1}
\fmf{phantom,right,label=\circled{1},label.dist=0}{v2,v1}
\fmfforce{(0w,0.5h)}{i}
\fmfforce{(1w,0.5h)}{o}
\fmfforce{(0.25w,0.5h)}{v1}
\fmfforce{(0.75w,0.5h)}{v2}
\marrow{a}{down}{-90}{$p$}{i,v1}
\fmfdot{v1,v2}
\end{fmfgraph*}
}}

\newcommand{\GRAPHLOpsic}{
\parbox[c][18mm][c]{33mm}{
\centering
\begin{fmfgraph*}(30,30)
\fmfleft{i}
\fmfright{o}
\fmf{fermion}{i,v1}
\fmf{selectron,label=$k$}{v1,v2}
\fmf{fermion}{v2,o}
\fmf{photino,right,label=$p-k$}{v2,v1}
\fmf{phantom,right,label=\circled{1},label.dist=0}{v2,v1}
\fmfforce{(0w,0.5h)}{i}
\fmfforce{(1w,0.5h)}{o}
\fmfforce{(0.25w,0.5h)}{v1}
\fmfforce{(0.75w,0.5h)}{v2}
\marrow{a}{down}{-90}{$p$}{i,v1}
\fmfdot{v1,v2}
\end{fmfgraph*}
}}

\newcommand{\GRAPHLOphia}{
\parbox[c][18mm][c]{33mm}{
\centering
\begin{fmfgraph*}(30,30)
\fmfleft{i}
\fmfright{o}
\fmf{selectron}{i,v1}
\fmf{selectron,label=$k$}{v1,v2}
\fmf{selectron}{v2,o}
\fmf{photon,right,label=$k-p$}{v2,v1}
\fmf{phantom,right,label=\circled{1},label.dist=0}{v2,v1}
\fmfforce{(0w,0.5h)}{i}
\fmfforce{(1w,0.5h)}{o}
\fmfforce{(0.25w,0.5h)}{v1}
\fmfforce{(0.75w,0.5h)}{v2}
\marrow{a}{down}{-90}{$p$}{i,v1}
\fmfdot{v1,v2}
\end{fmfgraph*}
}}

\newcommand{\GRAPHLOphib}{
\parbox[c][18mm][c]{33mm}{
\centering
\begin{fmfgraph*}(30,30)
\fmfleft{i}
\fmfright{o}
\fmf{selectron}{i,v1}
\fmf{fermion,label=$k$}{v1,v2}
\fmf{selectron}{v2,o}
\fmf{photino,right,label=$k-p$}{v2,v1}
\fmf{phantom,right,label=\circled{1},label.dist=0}{v2,v1}
\fmfforce{(0w,0.5h)}{i}
\fmfforce{(1w,0.5h)}{o}
\fmfforce{(0.25w,0.5h)}{v1}
\fmfforce{(0.75w,0.5h)}{v2}
\marrow{a}{down}{-90}{$p$}{i,v1}
\fmfdot{v1,v2}
\end{fmfgraph*}
}}


\newcommand{\GRAPHLOpsiaSD}{
\parbox[c][18mm][c]{33mm}{
\centering
\begin{fmfgraph*}(30,30)
\fmfleft{i}
\fmfright{o}
\fmf{fermion}{i,v1}
\fmf{fermionSD,label=$k$}{v1,v2}
\fmf{fermion}{v2,o}
\fmf{photonSD,left,label=$k-p$}{v1,v2}
\fmfforce{(0w,0.5h)}{i}
\fmfforce{(1w,0.5h)}{o}
\fmfforce{(0.25w,0.5h)}{v1}
\fmfforce{(0.75w,0.5h)}{v2}
\marrow{a}{down}{-90}{$p$}{i,v1}
\fmfblob{.1w}{v2}
\fmfdot{v1}
\end{fmfgraph*}
}}

\newcommand{\GRAPHLOpsibSD}{
\parbox[c][18mm][c]{33mm}{
\centering
\begin{fmfgraph*}(30,30)
\fmfleft{i}
\fmfright{o}
\fmf{fermion}{i,v1}
\fmf{fermionSD,label=$k$}{v1,v2}
\fmf{fermion}{v2,o}
\fmf{epsscalarSD,right,label=$k-p$}{v2,v1}
\fmfforce{(0w,0.5h)}{i}
\fmfforce{(1w,0.5h)}{o}
\fmfforce{(0.25w,0.5h)}{v1}
\fmfforce{(0.75w,0.5h)}{v2}
\marrow{a}{down}{-90}{$p$}{i,v1}
\fmfblob{.1w}{v2}
\fmfdot{v1}
\end{fmfgraph*}
}}

\newcommand{\GRAPHLOpsicSD}{
\parbox[c][18mm][c]{33mm}{
\centering
\begin{fmfgraph*}(30,30)
\fmfleft{i}
\fmfright{o}
\fmf{fermion}{i,v1}
\fmf{selectronSD,label=$k$}{v1,v2}
\fmf{fermion}{v2,o}
\fmf{photinoSD,right,label=$k-p$}{v2,v1}
\fmfforce{(0w,0.5h)}{i}
\fmfforce{(1w,0.5h)}{o}
\fmfforce{(0.25w,0.5h)}{v1}
\fmfforce{(0.75w,0.5h)}{v2}
\marrow{a}{down}{-90}{$p$}{i,v1}
\fmfblob{.1w}{v2}
\fmfdot{v1}
\end{fmfgraph*}
}}

\newcommand{\GRAPHLOpsiaSDbis}{
\parbox[c][18mm][c]{33mm}{
\centering
\begin{fmfgraph*}(30,30)
\fmfleft{i}
\fmfright{o}
\fmf{fermion}{i,v1}
\fmf{fermionSD,label=$k$}{v1,v2}
\fmf{fermion}{v2,o}
\fmf{photon,left,label=$k-p$}{v1,v2}
\fmf{phantom,right,label=\circled{1},label.dist=0}{v2,v1}
\fmfforce{(0w,0.5h)}{i}
\fmfforce{(1w,0.5h)}{o}
\fmfforce{(0.25w,0.5h)}{v1}
\fmfforce{(0.75w,0.5h)}{v2}
\marrow{a}{down}{-90}{$p$}{i,v1}
\fmfdot{v1,v2}
\end{fmfgraph*}
}}

\newcommand{\GRAPHLOpsibSDbis}{
\parbox[c][18mm][c]{33mm}{
\centering
\begin{fmfgraph*}(30,30)
\fmfleft{i}
\fmfright{o}
\fmf{fermion}{i,v1}
\fmf{fermionSD,label=$k$}{v1,v2}
\fmf{fermion}{v2,o}
\fmf{epsscalar,right,label=$k-p$}{v2,v1}
\fmf{phantom,right,label=\circled{1},label.dist=0}{v2,v1}
\fmfforce{(0w,0.5h)}{i}
\fmfforce{(1w,0.5h)}{o}
\fmfforce{(0.25w,0.5h)}{v1}
\fmfforce{(0.75w,0.5h)}{v2}
\marrow{a}{down}{-90}{$p$}{i,v1}
\fmfdot{v1,v2}
\end{fmfgraph*}
}}

\newcommand{\GRAPHLOpsicSDbis}{
\parbox[c][18mm][c]{33mm}{
\centering
\begin{fmfgraph*}(30,30)
\fmfleft{i}
\fmfright{o}
\fmf{fermion}{i,v1}
\fmf{selectronSD,label=$k$}{v1,v2}
\fmf{fermion}{v2,o}
\fmf{photino,right,label=$k-p$}{v2,v1}
\fmf{phantom,right,label=\circled{1},label.dist=0}{v2,v1}
\fmfforce{(0w,0.5h)}{i}
\fmfforce{(1w,0.5h)}{o}
\fmfforce{(0.25w,0.5h)}{v1}
\fmfforce{(0.75w,0.5h)}{v2}
\marrow{a}{down}{-90}{$p$}{i,v1}
\fmfdot{v1,v2}
\end{fmfgraph*}
}}

\newcommand{\GRAPHNLOAa}{
\parbox[c][20mm][c]{30mm}{
\centering
\begin{fmfgraph*}(30,30)
\fmfleft{i}
\fmfright{o}
\fmf{photon}{i,v1}
\fmf{photon}{v2,o}
\fmf{selectron,right}{v2,v1}
\fmf{selectron,right}{v1,v2}
\fmf{photon}{v1,v2}
\fmf{phantom,label=\circled{1},label.dist=0}{v1,v2}
\fmfforce{(0w,0.5h)}{i}
\fmfforce{(1w,0.5h)}{o}
\fmfforce{(0.25w,0.5h)}{v1}
\fmfforce{(0.75w,0.5h)}{v2}
\fmffreeze
\fmfdot{v1,v2}
\end{fmfgraph*}
}}

\newcommand{\GRAPHNLOAbcde}{
\parbox[c][20mm][c]{30mm}{
\centering
\begin{fmfgraph*}(30,30)
\fmf{photon}{i,v1}
\fmf{photon}{v2,o}
\fmf{phantom,right,tension=0.1,tag=1}{v1,v2}
\fmf{phantom,right,tension=0.1,tag=2}{v2,v1}
\fmf{phantom,tension=0.1,tag=3}{v1,v2}
\fmfdot{v1,v2}
\fmfforce{(0w,0.5h)}{i}
\fmfforce{(1w,0.5h)}{o}
\fmfforce{(0.25w,0.5h)}{v1}
\fmfforce{(0.75w,0.5h)}{v2}
\fmffreeze
\fmfposition
\fmfipath{p[]}
\fmfiset{p1}{vpath1(__v1,__v2)}
\fmfiset{p2}{vpath2(__v2,__v1)}
\fmfi{selectron}{subpath (0,length(p1)/2) of p1}
\fmfi{selectron}{subpath (length(p1)/2,length(p1)) of p1}
\fmfi{selectron}{subpath (0,length(p2)/2) of p2}
\fmfi{selectron}{subpath (length(p2)/2,length(p2)) of p2}
\fmfi{photon}{point length(p1) of p1 -- point length(p2)/2 of p2}
\fmfi{phantom,label=\circled{1},label.dist=0}{point length(p1) of p1 -- point length(p2)/2 of p2}
\myvert{point length(p2)/2 of p2}
\end{fmfgraph*}
}}

\newcommand{\GRAPHNLOAfg}{
 \parbox[c][20mm][c]{30mm}{
\centering
\begin{fmfgraph*}(30,30)
\fmfleft{i}
\fmfright{o}
\fmf{photon}{i,v1}
\fmf{photon}{v2,o}
\fmf{phantom,right,tag=1}{v1,v2}
\fmf{phantom,right,tag=2}{v2,v1}
\fmf{phantom,tag=3}{v1,v2}
\fmfforce{(0w,0.5h)}{i}
\fmfforce{(1w,0.5h)}{o}
\fmfforce{(0.25w,0.5h)}{v1}
\fmfforce{(0.75w,0.5h)}{v2}
\fmffreeze
\fmfposition
\fmfipath{p[]}
\fmfiset{p1}{vpath1(__v1,__v2)}
\fmfiset{p2}{vpath2(__v2,__v1)}
\fmfiset{p3}{vpath3(__v1,__v2)}
\fmfi{selectron}{subpath (0,length(p1)) of p1}
\fmfi{selectron}{subpath (0,length(p2)/4) of p2}
\fmfi{selectron}{subpath (length(p2)/4,3length(p2)/4) of p2}
\fmfi{selectron}{subpath (3length(p2)/4,length(p2)) of p2}
\fmfi{photon}{point length(p2)/4 of p2 .. point length(p3)/2 of p3 .. point 3length(p2)/4 of p2}
\fmfi{phantom,label=\circled{1},label.dist=0}{point length(p2)/4 of p2 .. point length(p3)/2 of p3 .. point 3length(p2)/4 of p2}
\fmfdot{v1,v2}
\myvert{point length(p2)/4 of p2}
\myvert{point 3length(p2)/4 of p2}
\end{fmfgraph*}
}}

\newcommand{\GRAPHNLOAh}{
 \parbox[c][20mm][c]{30mm}{
\centering
\begin{fmfgraph*}(30,30)
\fmfleft{i}
\fmfright{o}
\fmf{photon}{i,v1}
\fmf{photon}{v2,o}
\fmf{phantom,right,tension=0.1,tag=1}{v1,v2}
\fmf{phantom,right,tension=0.1,tag=2}{v2,v1}
\fmf{phantom,tension=0.1,tag=3}{v1,v2}
\fmfdot{v1,v2}
\fmfforce{(0w,0.5h)}{i}
\fmfforce{(1w,0.5h)}{o}
\fmfforce{(0.25w,0.5h)}{v1}
\fmfforce{(0.75w,0.5h)}{v2}
\fmffreeze
\fmfposition
\fmfipath{p[]}
\fmfiset{p1}{vpath1(__v1,__v2)}
\fmfiset{p2}{vpath2(__v2,__v1)}
\fmfi{selectron}{subpath (0,length(p1)/2) of p1}
\fmfi{selectron}{subpath (length(p1)/2,length(p1)) of p1}
\fmfi{selectron}{subpath (0,length(p2)/2) of p2}
\fmfi{selectron}{subpath (length(p2)/2,length(p2)) of p2}
\fmfi{photon}{point length(p1)/2 of p1 -- point length(p2)/2 of p2}
\fmfi{phantom,label=\circled{1},label.dist=0}{point length(p1)/2 of p1 -- point length(p2)/2 of p2}
\myvert{point length(p1)/2 of p1}
\myvert{point length(p2)/2 of p2}
\end{fmfgraph*}
}}

\newcommand{\GRAPHNLOAij}{
 \parbox[c][20mm][c]{30mm}{
\centering
\begin{fmfgraph*}(30,30)
\fmfleft{i}
\fmfright{o}
\fmf{photon}{i,v1}
\fmf{photon}{v2,o}
\fmf{phantom,right,tag=1}{v1,v2}
\fmf{phantom,right,tag=2}{v2,v1}
\fmf{phantom,tag=3}{v1,v2}
\fmfforce{(0w,0.5h)}{i}
\fmfforce{(1w,0.5h)}{o}
\fmfforce{(0.25w,0.5h)}{v1}
\fmfforce{(0.75w,0.5h)}{v2}
\fmffreeze
\fmfposition
\fmfipath{p[]}
\fmfiset{p1}{vpath1(__v1,__v2)}
\fmfiset{p2}{vpath2(__v2,__v1)}
\fmfiset{p3}{vpath3(__v1,__v2)}
\fmfi{fermion}{subpath (0,length(p1)) of p1}
\fmfi{fermion}{subpath (0,length(p2)/4) of p2}
\fmfi{fermion}{subpath (length(p2)/4,3length(p2)/4) of p2}
\fmfi{fermion}{subpath (3length(p2)/4,length(p2)) of p2}
\fmfi{photon}{point length(p2)/4 of p2 .. point length(p3)/2 of p3 .. point 3length(p2)/4 of p2}
\fmfi{phantom,label=\circled{1},label.dist=0}{point length(p2)/4 of p2 .. point length(p3)/2 of p3 .. point 3length(p2)/4 of p2}
\fmfdot{v1,v2}
\myvert{point length(p2)/4 of p2}
\myvert{point 3length(p2)/4 of p2}
\end{fmfgraph*}
}}

\newcommand{\GRAPHNLOAk}{
 \parbox[c][20mm][c]{30mm}{
\centering
\begin{fmfgraph*}(30,30)
\fmfleft{i}
\fmfright{o}
\fmf{photon}{i,v1}
\fmf{photon}{v2,o}
\fmf{phantom,right,tag=1}{v1,v2}
\fmf{phantom,right,tag=2}{v2,v1}
\fmf{phantom,tag=3}{v1,v2}
\fmfforce{(0w,0.5h)}{i}
\fmfforce{(1w,0.5h)}{o}
\fmfforce{(0.25w,0.5h)}{v1}
\fmfforce{(0.75w,0.5h)}{v2}
\fmffreeze
\fmfposition
\fmfipath{p[]}
\fmfiset{p1}{vpath1(__v1,__v2)}
\fmfiset{p2}{vpath2(__v2,__v1)}
\fmfiset{p3}{vpath3(__v1,__v2)}
\fmfi{fermion}{subpath (0,length(p1)/2) of p1}
\fmfi{fermion}{subpath (0,length(p2)/2) of p2}
\fmfi{fermion}{subpath (length(p2)/2,length(p2)) of p2}
\fmfi{fermion}{subpath (length(p1)/2,length(p1)) of p1}
\fmfi{photon}{point length(p2)/2 of p1 .. point length(p2)/2 of p2}
\fmfi{phantom,label=\circled{1},label.dist=0}{point length(p2)/2 of p1 .. point length(p2)/2 of p2}
\fmfdot{v1,v2}
\myvert{point length(p1)/2 of p1}
\myvert{point length(p2)/2 of p2}
\end{fmfgraph*}
}}

\newcommand{\GRAPHNLOAlm}{
 \parbox[c][20mm][c]{30mm}{
\centering
\begin{fmfgraph*}(30,30)
\fmfleft{i}
\fmfright{o}
\fmf{photon}{i,v1}
\fmf{photon}{v2,o}
\fmf{phantom,right,tag=1}{v1,v2}
\fmf{phantom,right,tag=2}{v2,v1}
\fmf{phantom,tag=3}{v1,v2}
\fmfforce{(0w,0.5h)}{i}
\fmfforce{(1w,0.5h)}{o}
\fmfforce{(0.25w,0.5h)}{v1}
\fmfforce{(0.75w,0.5h)}{v2}
\fmffreeze
\fmfposition
\fmfipath{p[]}
\fmfiset{p1}{vpath1(__v1,__v2)}
\fmfiset{p2}{vpath2(__v2,__v1)}
\fmfiset{p3}{vpath3(__v1,__v2)}
\fmfi{fermion}{subpath (0,length(p1)) of p1}
\fmfi{fermion}{subpath (0,length(p2)/4) of p2}
\fmfi{fermion}{subpath (length(p2)/4,3length(p2)/4) of p2}
\fmfi{fermion}{subpath (3length(p2)/4,length(p2)) of p2}
\fmfi{epsscalar}{point length(p2)/4 of p2 .. point length(p3)/2 of p3 .. point 3length(p2)/4 of p2}
\fmfi{phantom,label=\circled{1},label.dist=0}{point length(p2)/4 of p2 .. point length(p3)/2 of p3 .. point 3length(p2)/4 of p2}
\fmfdot{v1,v2}
\myvert{point length(p2)/4 of p2}
\myvert{point 3length(p2)/4 of p2}
\end{fmfgraph*}
}}

\newcommand{\GRAPHNLOAn}{
 \parbox[c][20mm][c]{30mm}{
\centering
\begin{fmfgraph*}(30,30)
\fmfleft{i}
\fmfright{o}
\fmf{photon}{i,v1}
\fmf{photon}{v2,o}
\fmf{phantom,right,tag=1}{v1,v2}
\fmf{phantom,right,tag=2}{v2,v1}
\fmf{phantom,tag=3}{v1,v2}
\fmfforce{(0w,0.5h)}{i}
\fmfforce{(1w,0.5h)}{o}
\fmfforce{(0.25w,0.5h)}{v1}
\fmfforce{(0.75w,0.5h)}{v2}
\fmffreeze
\fmfposition
\fmfipath{p[]}
\fmfiset{p1}{vpath1(__v1,__v2)}
\fmfiset{p2}{vpath2(__v2,__v1)}
\fmfiset{p3}{vpath3(__v1,__v2)}
\fmfi{fermion}{subpath (0,length(p1)/2) of p1}
\fmfi{fermion}{subpath (0,length(p2)/2) of p2}
\fmfi{fermion}{subpath (length(p2)/2,length(p2)) of p2}
\fmfi{fermion}{subpath (length(p1)/2,length(p1)) of p1}
\fmfi{epsscalar}{point length(p2)/2 of p1 .. point length(p2)/2 of p2}
\fmfi{phantom,label=\circled{1},label.dist=0}{point length(p2)/2 of p1 .. point length(p2)/2 of p2}
\fmfdot{v1,v2}
\myvert{point length(p1)/2 of p1}
\myvert{point length(p2)/2 of p2}
\end{fmfgraph*}
}}

\newcommand{\GRAPHNLOAop}{
 \parbox[c][20mm][c]{30mm}{
\centering
\begin{fmfgraph*}(30,30)
\fmfleft{i}
\fmfright{o}
\fmf{photon}{i,v1}
\fmf{photon}{v2,o}
\fmf{phantom,right,tag=1}{v1,v2}
\fmf{phantom,right,tag=2}{v2,v1}
\fmf{phantom,tag=3}{v1,v2}
\fmfforce{(0w,0.5h)}{i}
\fmfforce{(1w,0.5h)}{o}
\fmfforce{(0.25w,0.5h)}{v1}
\fmfforce{(0.75w,0.5h)}{v2}
\fmffreeze
\fmfposition
\fmfipath{p[]}
\fmfiset{p1}{vpath1(__v1,__v2)}
\fmfiset{p2}{vpath2(__v2,__v1)}
\fmfiset{p3}{vpath3(__v1,__v2)}
\fmfi{fermion}{subpath (0,length(p1)) of p1}
\fmfi{fermion}{subpath (0,length(p2)/4) of p2}
\fmfi{selectron}{subpath (length(p2)/4,3length(p2)/4) of p2}
\fmfi{fermion}{subpath (3length(p2)/4,length(p2)) of p2}
\fmfi{photino}{point length(p2)/4 of p2 .. point length(p3)/2 of p3 .. point 3length(p2)/4 of p2}
\fmfi{phantom,label=\circled{1},label.dist=0}{point length(p2)/4 of p2 .. point length(p3)/2 of p3 .. point 3length(p2)/4 of p2}
\fmfdot{v1,v2}
\myvert{point length(p2)/4 of p2}
\myvert{point 3length(p2)/4 of p2}
\end{fmfgraph*}
}}

\newcommand{\GRAPHNLOAqr}{
 \parbox[c][20mm][c]{30mm}{
\centering
\begin{fmfgraph*}(30,30)
\fmfleft{i}
\fmfright{o}
\fmf{photon}{i,v1}
\fmf{photon}{v2,o}
\fmf{phantom,right,tag=1}{v1,v2}
\fmf{phantom,right,tag=2}{v2,v1}
\fmf{phantom,tag=3}{v1,v2}
\fmfforce{(0w,0.5h)}{i}
\fmfforce{(1w,0.5h)}{o}
\fmfforce{(0.25w,0.5h)}{v1}
\fmfforce{(0.75w,0.5h)}{v2}
\fmffreeze
\fmfposition
\fmfipath{p[]}
\fmfiset{p1}{vpath1(__v1,__v2)}
\fmfiset{p2}{vpath2(__v2,__v1)}
\fmfiset{p3}{vpath3(__v1,__v2)}
\fmfi{selectron}{subpath (0,length(p1)) of p1}
\fmfi{selectron}{subpath (0,length(p2)/4) of p2}
\fmfi{fermion}{subpath (length(p2)/4,3length(p2)/4) of p2}
\fmfi{selectron}{subpath (3length(p2)/4,length(p2)) of p2}
\fmfi{photino}{point length(p2)/4 of p2 .. point length(p3)/2 of p3 .. point 3length(p2)/4 of p2}
\fmfi{phantom,label=\circled{1},label.dist=0}{point length(p2)/4 of p2 .. point length(p3)/2 of p3 .. point 3length(p2)/4 of p2}
\fmfdot{v1,v2}
\myvert{point length(p2)/4 of p2}
\myvert{point 3length(p2)/4 of p2}
\end{fmfgraph*}
}}

\newcommand{\GRAPHNLOAst}{
 \parbox[c][20mm][c]{30mm}{
\centering
\begin{fmfgraph*}(30,30)
\fmfleft{i}
\fmfright{o}
\fmf{photon}{i,v1}
\fmf{photon}{v2,o}
\fmf{phantom,right,tag=1}{v1,v2}
\fmf{phantom,right,tag=2}{v2,v1}
\fmf{phantom,tag=3}{v1,v2}
\fmfforce{(0w,0.5h)}{i}
\fmfforce{(1w,0.5h)}{o}
\fmfforce{(0.25w,0.5h)}{v1}
\fmfforce{(0.75w,0.5h)}{v2}
\fmffreeze
\fmfposition
\fmfipath{p[]}
\fmfiset{p1}{vpath1(__v1,__v2)}
\fmfiset{p2}{vpath2(__v2,__v1)}
\fmfiset{p3}{vpath3(__v1,__v2)}
\fmfi{fermion}{subpath (0,length(p1)/2) of p1}
\fmfi{selectron}{subpath (0,length(p2)/2) of p2}
\fmfi{fermion}{subpath (length(p2)/2,length(p2)) of p2}
\fmfi{selectron}{subpath (length(p1)/2,length(p1)) of p1}
\fmfi{photino}{point length(p2)/2 of p1 .. point length(p2)/2 of p2}
\fmfi{phantom,label=\circled{1},label.dist=0}{point length(p2)/2 of p1 .. point length(p2)/2 of p2}
\fmfdot{v1,v2}
\myvert{point length(p1)/2 of p1}
\myvert{point length(p2)/2 of p2}
\end{fmfgraph*}
}}

\newcommand{\GRAPHNLOepsa}{
\parbox[c][20mm][c]{30mm}{
\centering
\begin{fmfgraph*}(30,30)
\fmfleft{i}
\fmfright{o}
\fmf{epsscalar}{i,v1}
\fmf{epsscalar}{v2,o}
\fmf{selectron,right}{v2,v1}
\fmf{selectron,right}{v1,v2}
\fmf{epsscalar}{v1,v2}
\fmf{phantom,label=\circled{1},label.dist=0}{v1,v2}
\fmfforce{(0w,0.5h)}{i}
\fmfforce{(1w,0.5h)}{o}
\fmfforce{(0.25w,0.5h)}{v1}
\fmfforce{(0.75w,0.5h)}{v2}
\fmffreeze
\fmfdot{v1,v2}
\end{fmfgraph*}
}}

\newcommand{\GRAPHNLOepsbc}{
\parbox[c][20mm][c]{30mm}{
\centering
\begin{fmfgraph*}(30,30)
\fmfleft{i}
\fmfright{o}
\fmf{epsscalar}{i,v1}
\fmf{epsscalar}{v2,o}
\fmf{phantom,right,tag=1}{v1,v2}
\fmf{phantom,right,tag=2}{v2,v1}
\fmf{phantom,tag=3}{v1,v2}
\fmfforce{(0w,0.5h)}{i}
\fmfforce{(1w,0.5h)}{o}
\fmfforce{(0.25w,0.5h)}{v1}
\fmfforce{(0.75w,0.5h)}{v2}
\fmffreeze
\fmfposition
\fmfipath{p[]}
\fmfiset{p1}{vpath1(__v1,__v2)}
\fmfiset{p2}{vpath2(__v2,__v1)}
\fmfiset{p3}{vpath3(__v1,__v2)}
\fmfi{fermion}{subpath (0,length(p1)) of p1}
\fmfi{fermion}{subpath (0,length(p2)/4) of p2}
\fmfi{fermion}{subpath (length(p2)/4,3length(p2)/4) of p2}
\fmfi{fermion}{subpath (3length(p2)/4,length(p2)) of p2}
\fmfi{epsscalar}{point length(p2)/4 of p2 .. point length(p3)/2 of p3 .. point 3length(p2)/4 of p2}
\fmfi{phantom,label=\circled{1},label.dist=0}{point length(p2)/4 of p2 .. point length(p3)/2 of p3 .. point 3length(p2)/4 of p2}
\fmfdot{v1,v2}
\myvert{point length(p2)/4 of p2}
\myvert{point 3length(p2)/4 of p2}
\end{fmfgraph*}
}}

\newcommand{\GRAPHNLOepsd}{
\parbox[c][20mm][c]{30mm}{
\centering
\begin{fmfgraph*}(30,30)
\fmfleft{i}
\fmfright{o}
\fmf{epsscalar}{i,v1}
\fmf{epsscalar}{v2,o}
\fmf{phantom,right,tag=1}{v1,v2}
\fmf{phantom,right,tag=2}{v2,v1}
\fmf{phantom,tag=3}{v1,v2}
\fmfforce{(0w,0.5h)}{i}
\fmfforce{(1w,0.5h)}{o}
\fmfforce{(0.25w,0.5h)}{v1}
\fmfforce{(0.75w,0.5h)}{v2}
\fmffreeze
\fmfposition
\fmfipath{p[]}
\fmfiset{p1}{vpath1(__v1,__v2)}
\fmfiset{p2}{vpath2(__v2,__v1)}
\fmfiset{p3}{vpath3(__v1,__v2)}
\fmfi{fermion}{subpath (0,length(p1)/2) of p1}
\fmfi{fermion}{subpath (0,length(p2)/2) of p2}
\fmfi{fermion}{subpath (length(p2)/2,length(p2)) of p2}
\fmfi{fermion}{subpath (length(p1)/2,length(p1)) of p1}
\fmfi{epsscalar}{point length(p2)/2 of p1 .. point length(p2)/2 of p2}
\fmfi{phantom,label=\circled{1},label.dist=0}{point length(p2)/2 of p1 .. point length(p2)/2 of p2}
\fmfdot{v1,v2}
\myvert{point length(p1)/2 of p1}
\myvert{point length(p2)/2 of p2}
\end{fmfgraph*}
}}

\newcommand{\GRAPHNLOepsef}{
\parbox[c][20mm][c]{30mm}{
\centering
\begin{fmfgraph*}(30,30)
\fmfleft{i}
\fmfright{o}
\fmf{epsscalar}{i,v1}
\fmf{epsscalar}{v2,o}
\fmf{phantom,right,tag=1}{v1,v2}
\fmf{phantom,right,tag=2}{v2,v1}
\fmf{phantom,tag=3}{v1,v2}
\fmfforce{(0w,0.5h)}{i}
\fmfforce{(1w,0.5h)}{o}
\fmfforce{(0.25w,0.5h)}{v1}
\fmfforce{(0.75w,0.5h)}{v2}
\fmffreeze
\fmfposition
\fmfipath{p[]}
\fmfiset{p1}{vpath1(__v1,__v2)}
\fmfiset{p2}{vpath2(__v2,__v1)}
\fmfiset{p3}{vpath3(__v1,__v2)}
\fmfi{fermion}{subpath (0,length(p1)) of p1}
\fmfi{fermion}{subpath (0,length(p2)/4) of p2}
\fmfi{fermion}{subpath (length(p2)/4,3length(p2)/4) of p2}
\fmfi{fermion}{subpath (3length(p2)/4,length(p2)) of p2}
\fmfi{photon}{point length(p2)/4 of p2 .. point length(p3)/2 of p3 .. point 3length(p2)/4 of p2}
\fmfi{phantom,label=\circled{1},label.dist=0}{point length(p2)/4 of p2 .. point length(p3)/2 of p3 .. point 3length(p2)/4 of p2}
\fmfdot{v1,v2}
\myvert{point length(p2)/4 of p2}
\myvert{point 3length(p2)/4 of p2}
\end{fmfgraph*}
}}

\newcommand{\GRAPHNLOepsg}{
\parbox[c][20mm][c]{30mm}{
\centering
\begin{fmfgraph*}(30,30)
\fmfleft{i}
\fmfright{o}
\fmf{epsscalar}{i,v1}
\fmf{epsscalar}{v2,o}
\fmf{phantom,right,tag=1}{v1,v2}
\fmf{phantom,right,tag=2}{v2,v1}
\fmf{phantom,tag=3}{v1,v2}
\fmfforce{(0w,0.5h)}{i}
\fmfforce{(1w,0.5h)}{o}
\fmfforce{(0.25w,0.5h)}{v1}
\fmfforce{(0.75w,0.5h)}{v2}
\fmffreeze
\fmfposition
\fmfipath{p[]}
\fmfiset{p1}{vpath1(__v1,__v2)}
\fmfiset{p2}{vpath2(__v2,__v1)}
\fmfiset{p3}{vpath3(__v1,__v2)}
\fmfi{fermion}{subpath (0,length(p1)/2) of p1}
\fmfi{fermion}{subpath (0,length(p2)/2) of p2}
\fmfi{fermion}{subpath (length(p2)/2,length(p2)) of p2}
\fmfi{fermion}{subpath (length(p1)/2,length(p1)) of p1}
\fmfi{photon}{point length(p2)/2 of p1 .. point length(p2)/2 of p2}
\fmfi{phantom,label=\circled{1},label.dist=0}{point length(p2)/2 of p1 .. point length(p2)/2 of p2}
\fmfdot{v1,v2}
\myvert{point length(p1)/2 of p1}
\myvert{point length(p2)/2 of p2}
\end{fmfgraph*}
}}

\newcommand{\GRAPHNLOepshi}{
\parbox[c][20mm][c]{30mm}{
\centering
\begin{fmfgraph*}(30,30)
\fmfleft{i}
\fmfright{o}
\fmf{epsscalar}{i,v1}
\fmf{epsscalar}{v2,o}
\fmf{phantom,right,tag=1}{v1,v2}
\fmf{phantom,right,tag=2}{v2,v1}
\fmf{phantom,tag=3}{v1,v2}
\fmfforce{(0w,0.5h)}{i}
\fmfforce{(1w,0.5h)}{o}
\fmfforce{(0.25w,0.5h)}{v1}
\fmfforce{(0.75w,0.5h)}{v2}
\fmffreeze
\fmfposition
\fmfipath{p[]}
\fmfiset{p1}{vpath1(__v1,__v2)}
\fmfiset{p2}{vpath2(__v2,__v1)}
\fmfiset{p3}{vpath3(__v1,__v2)}
\fmfi{fermion}{subpath (0,length(p1)) of p1}
\fmfi{fermion}{subpath (0,length(p2)/4) of p2}
\fmfi{selectron}{subpath (length(p2)/4,3length(p2)/4) of p2}
\fmfi{fermion}{subpath (3length(p2)/4,length(p2)) of p2}
\fmfi{photino}{point length(p2)/4 of p2 .. point length(p3)/2 of p3 .. point 3length(p2)/4 of p2}
\fmfi{phantom,label=\circled{1},label.dist=0}{point length(p2)/4 of p2 .. point length(p3)/2 of p3 .. point 3length(p2)/4 of p2}
\fmfdot{v1,v2}
\myvert{point length(p2)/4 of p2}
\myvert{point 3length(p2)/4 of p2}
\end{fmfgraph*}
}}

\newcommand{\GRAPHNLOlambdaab}{
\parbox[c][20mm][c]{30mm}{
\centering
\begin{fmfgraph*}(30,30)
\fmfleft{i}
\fmfright{o}
\fmf{photino}{i,v1}
\fmf{photino}{v2,o}
\fmf{phantom,right,tag=1}{v1,v2}
\fmf{phantom,right,tag=2}{v2,v1}
\fmf{phantom,tag=3}{v1,v2}
\fmfforce{(0w,0.5h)}{i}
\fmfforce{(1w,0.5h)}{o}
\fmfforce{(0.25w,0.5h)}{v1}
\fmfforce{(0.75w,0.5h)}{v2}
\fmffreeze
\fmfposition
\fmfipath{p[]}
\fmfiset{p1}{vpath1(__v1,__v2)}
\fmfiset{p2}{vpath2(__v2,__v1)}
\fmfiset{p3}{vpath3(__v1,__v2)}
\fmfi{fermion}{subpath (0,length(p1)) of p1}
\fmfi{selectron}{subpath (0,length(p2)/4) of p2}
\fmfi{selectron}{subpath (length(p2)/4,3length(p2)/4) of p2}
\fmfi{selectron}{subpath (3length(p2)/4,length(p2)) of p2}
\fmfi{photon}{point length(p2)/4 of p2 .. point length(p3)/2 of p3 .. point 3length(p2)/4 of p2}
\fmfi{phantom,label=\circled{1},label.dist=0}{point length(p2)/4 of p2 .. point length(p3)/2 of p3 .. point 3length(p2)/4 of p2}
\fmfdot{v1,v2}
\myvert{point length(p2)/4 of p2}
\myvert{point 3length(p2)/4 of p2}
\end{fmfgraph*}
}}

\newcommand{\GRAPHNLOlambdacd}{
\parbox[c][20mm][c]{30mm}{
\centering
\begin{fmfgraph*}(30,30)
\fmfleft{i}
\fmfright{o}
\fmf{photino}{i,v1}
\fmf{photino}{v2,o}
\fmf{phantom,right,tag=1}{v1,v2}
\fmf{phantom,right,tag=2}{v2,v1}
\fmf{phantom,tag=3}{v1,v2}
\fmfforce{(0w,0.5h)}{i}
\fmfforce{(1w,0.5h)}{o}
\fmfforce{(0.25w,0.5h)}{v1}
\fmfforce{(0.75w,0.5h)}{v2}
\fmffreeze
\fmfposition
\fmfipath{p[]}
\fmfiset{p1}{vpath1(__v1,__v2)}
\fmfiset{p2}{vpath2(__v2,__v1)}
\fmfiset{p3}{vpath3(__v1,__v2)}
\fmfi{selectron}{subpath (0,length(p1)) of p1}
\fmfi{fermion}{subpath (0length(p2)/4,1length(p2)/4) of p2}
\fmfi{fermion}{subpath (1length(p2)/4,3length(p2)/4) of p2}
\fmfi{fermion}{subpath (3length(p2)/4,4length(p2)/4) of p2}
\fmfi{photon}{point length(p2)/4 of p2 .. point length(p3)/2 of p3 .. point 3length(p2)/4 of p2}
\fmfi{phantom,label=\circled{1},label.dist=0}{point length(p2)/4 of p2 .. point length(p3)/2 of p3 .. point 3length(p2)/4 of p2}
\fmfdot{v1,v2}
\myvert{point length(p2)/4 of p2}
\myvert{point 3length(p2)/4 of p2}
\end{fmfgraph*}
}}

\newcommand{\GRAPHNLOlambdaef}{
\parbox[c][20mm][c]{30mm}{
\centering
\begin{fmfgraph*}(30,30)
\fmfleft{i}
\fmfright{o}
\fmf{photino}{i,v1}
\fmf{photino}{v2,o}
\fmf{phantom,right,tag=1}{v1,v2}
\fmf{phantom,right,tag=2}{v2,v1}
\fmf{phantom,tag=3}{v1,v2}
\fmfforce{(0w,0.5h)}{i}
\fmfforce{(1w,0.5h)}{o}
\fmfforce{(0.25w,0.5h)}{v1}
\fmfforce{(0.75w,0.5h)}{v2}
\fmffreeze
\fmfposition
\fmfipath{p[]}
\fmfiset{p1}{vpath1(__v1,__v2)}
\fmfiset{p2}{vpath2(__v2,__v1)}
\fmfiset{p3}{vpath3(__v1,__v2)}
\fmfi{fermion}{subpath (0,length(p1)/2) of p1}
\fmfi{selectron}{subpath (0,length(p2)/2) of p2}
\fmfi{selectron}{subpath (length(p2)/2,length(p2)) of p2}
\fmfi{fermion}{subpath (length(p1)/2,length(p1)) of p1}
\fmfi{photon}{point length(p2)/2 of p1 .. point length(p2)/2 of p2}
\fmfi{phantom,label=\circled{1},label.dist=0}{point length(p2)/2 of p1 .. point length(p2)/2 of p2}
\fmfdot{v1,v2}
\myvert{point length(p1)/2 of p1}
\myvert{point length(p2)/2 of p2}
\end{fmfgraph*}
}}

\newcommand{\GRAPHNLOlambdagh}{
\parbox[c][20mm][c]{30mm}{
\centering
\begin{fmfgraph*}(30,30)
\fmfleft{i}
\fmfright{o}
\fmf{photino}{i,v1}
\fmf{photino}{v2,o}
\fmf{phantom,right,tag=1}{v1,v2}
\fmf{phantom,right,tag=2}{v2,v1}
\fmf{phantom,tag=3}{v1,v2}
\fmfforce{(0w,0.5h)}{i}
\fmfforce{(1w,0.5h)}{o}
\fmfforce{(0.25w,0.5h)}{v1}
\fmfforce{(0.75w,0.5h)}{v2}
\fmffreeze
\fmfposition
\fmfipath{p[]}
\fmfiset{p1}{vpath1(__v1,__v2)}
\fmfiset{p2}{vpath2(__v2,__v1)}
\fmfiset{p3}{vpath3(__v1,__v2)}
\fmfi{selectron}{subpath (0,length(p1)) of p1}
\fmfi{fermion}{subpath (0length(p2)/4,1length(p2)/4) of p2}
\fmfi{fermion}{subpath (1length(p2)/4,3length(p2)/4) of p2}
\fmfi{fermion}{subpath (3length(p2)/4,4length(p2)/4) of p2}
\fmfi{epsscalar}{point length(p2)/4 of p2 .. point length(p3)/2 of p3 .. point 3length(p2)/4 of p2}
\fmfi{phantom,label=\circled{1},label.dist=0}{point length(p2)/4 of p2 .. point length(p3)/2 of p3 .. point 3length(p2)/4 of p2}
\fmfdot{v1,v2}
\myvert{point length(p2)/4 of p2}
\myvert{point 3length(p2)/4 of p2}
\end{fmfgraph*}
}}

\newcommand{\GRAPHNLOlambdaij}{
\parbox[c][20mm][c]{30mm}{
\centering
\begin{fmfgraph*}(30,30)
\fmfleft{i}
\fmfright{o}
\fmf{photino}{i,v1}
\fmf{photino}{v2,o}
\fmf{phantom,right,tag=1}{v1,v2}
\fmf{phantom,right,tag=2}{v2,v1}
\fmf{phantom,tag=3}{v1,v2}
\fmfforce{(0w,0.5h)}{i}
\fmfforce{(1w,0.5h)}{o}
\fmfforce{(0.25w,0.5h)}{v1}
\fmfforce{(0.75w,0.5h)}{v2}
\fmffreeze
\fmfposition
\fmfipath{p[]}
\fmfiset{p1}{vpath1(__v1,__v2)}
\fmfiset{p2}{vpath2(__v2,__v1)}
\fmfiset{p3}{vpath3(__v1,__v2)}
\fmfi{fermion}{subpath (0,length(p1)) of p1}
\fmfi{selectron}{subpath (0length(p2)/4,1length(p2)/4) of p2}
\fmfi{fermion}{subpath (1length(p2)/4,3length(p2)/4) of p2}
\fmfi{selectron}{subpath (3length(p2)/4,4length(p2)/4) of p2}
\fmfi{photino}{point length(p2)/4 of p2 .. point length(p3)/2 of p3 .. point 3length(p2)/4 of p2}
\fmfi{phantom,label=\circled{1},label.dist=0}{point length(p2)/4 of p2 .. point length(p3)/2 of p3 .. point 3length(p2)/4 of p2}
\fmfdot{v1,v2}
\myvert{point length(p2)/4 of p2}
\myvert{point 3length(p2)/4 of p2}
\end{fmfgraph*}
}}

\newcommand{\GRAPHNLOlambdakl}{
\parbox[c][20mm][c]{30mm}{
\centering
\begin{fmfgraph*}(30,30)
\fmfleft{i}
\fmfright{o}
\fmf{photino}{i,v1}
\fmf{photino}{v2,o}
\fmf{phantom,right,tag=1}{v1,v2}
\fmf{phantom,right,tag=2}{v2,v1}
\fmf{phantom,tag=3}{v1,v2}
\fmfforce{(0w,0.5h)}{i}
\fmfforce{(1w,0.5h)}{o}
\fmfforce{(0.25w,0.5h)}{v1}
\fmfforce{(0.75w,0.5h)}{v2}
\fmffreeze
\fmfposition
\fmfipath{p[]}
\fmfiset{p1}{vpath1(__v1,__v2)}
\fmfiset{p2}{vpath2(__v2,__v1)}
\fmfiset{p3}{vpath3(__v1,__v2)}
\fmfi{selectron}{subpath (0,length(p1)) of p1}
\fmfi{fermion}{subpath (0length(p2)/4,1length(p2)/4) of p2}
\fmfi{selectron}{subpath (1length(p2)/4,3length(p2)/4) of p2}
\fmfi{fermion}{subpath (3length(p2)/4,4length(p2)/4) of p2}
\fmfi{photino}{point 3length(p2)/4 of p2 .. point length(p3)/2 of p3 .. point length(p2)/4 of p2}
\fmfi{phantom,label=\circled{1},label.dist=0}{point 3length(p2)/4 of p2 .. point length(p3)/2 of p3 .. point length(p2)/4 of p2}
\fmfdot{v1,v2}
\myvert{point length(p2)/4 of p2}
\myvert{point 3length(p2)/4 of p2}
\end{fmfgraph*}
}}

\newcommand{\GRAPHNLOlambdamn}{
\parbox[c][20mm][c]{30mm}{
\centering
\begin{fmfgraph*}(30,30)
\fmfleft{i}
\fmfright{o}
\fmf{photino}{i,v1}
\fmf{photino}{v2,o}
\fmf{phantom,right,tag=1}{v1,v2}
\fmf{phantom,right,tag=2}{v2,v1}
\fmf{phantom,tag=3}{v1,v2}
\fmfforce{(0w,0.5h)}{i}
\fmfforce{(1w,0.5h)}{o}
\fmfforce{(0.25w,0.5h)}{v1}
\fmfforce{(0.75w,0.5h)}{v2}
\fmffreeze
\fmfposition
\fmfipath{p[]}
\fmfiset{p1}{vpath1(__v1,__v2)}
\fmfiset{p2}{vpath2(__v2,__v1)}
\fmfiset{p3}{vpath3(__v1,__v2)}
\fmfi{fermion}{subpath (0length(p2)/2,1length(p1)/2) of p1}
\fmfi{fermion}{subpath (0length(p2)/2,1length(p2)/2) of p2}
\fmfi{selectron}{subpath (length(p2)/2,length(p2)) of p2}
\fmfi{selectron}{subpath (length(p1)/2,length(p1)) of p1}
\fmfi{photino}{point length(p2)/2 of p1 .. point length(p2)/2 of p2}
\fmfi{phantom,label=\circled{1},label.dist=0}{point length(p2)/2 of p1 .. point length(p2)/2 of p2}
\fmfdot{v1,v2}
\myvert{point length(p1)/2 of p1}
\myvert{point length(p2)/2 of p2}
\end{fmfgraph*}
}}

\newcommand{\GRAPHNLOtest}{
\parbox[c][20mm][c]{30mm}{
\centering
\begin{fmfgraph*}(30,30)
\fmfleft{i}
\fmfright{o}
\fmf{plain}{i,v1}
\fmf{plain}{v2,o}
\fmf{plain}{v1,v2}
\fmf{plain}{v2,v3}
\fmf{plain}{v3,v4}
\fmf{plain}{v4,v5}
\fmf{plain}{v5,o}
\fmfforce{(0w,0.5h)}{i}
\fmfforce{(1w,0.5h)}{o}
\fmfforce{(0.25w,0.5h)}{v1}
\fmfforce{(0.375w,0.5h)}{v2}
\fmfforce{(0.50w,0.5h)}{v3}
\fmfforce{(0.625w,0.5h)}{v4}
\fmfforce{(0.75w,0.5h)}{v5}
\fmffreeze
\fmfdot{v1,v2,v3,v4,v5}
\end{fmfgraph*}
}}

\newcommand{\GRAPHNLOphia}{
\parbox[c][20mm][c]{30mm}{
\centering
\begin{fmfgraph*}(30,30)
\fmfleft{i}
\fmfright{o}
\fmf{selectron}{i,v1}
\fmf{selectron}{v1,v5}
\fmf{selectron}{v5,o}
\fmf{photon,left}{v1,v5}
\fmf{photon,right}{v1,v5}
\fmf{phantom,label=\circled{1},label.dist=0,left}{v1,v5}
\fmf{phantom,label=\circled{1},label.dist=0,right}{v1,v5}
\fmfforce{(0w,0.5h)}{i}
\fmfforce{(1w,0.5h)}{o}
\fmfforce{(0.25w,0.5h)}{v1}
\fmfforce{(0.75w,0.5h)}{v5}
\fmffreeze
\fmfdot{v1,v5}
\end{fmfgraph*}
}}

\newcommand{\GRAPHNLOphib}{
\parbox[c][20mm][c]{30mm}{
\centering
\begin{fmfgraph*}(30,30)
\fmfleft{i}
\fmfright{o}
\fmf{selectron}{i,v1}
\fmf{selectron}{v1,v5}
\fmf{selectron}{v5,o}
\fmf{epsscalar,left}{v1,v5}
\fmf{epsscalar,right}{v1,v5}
\fmf{phantom,label=\circled{1},label.dist=0,left}{v1,v5}
\fmf{phantom,label=\circled{1},label.dist=0,right}{v1,v5}
\fmfforce{(0w,0.5h)}{i}
\fmfforce{(1w,0.5h)}{o}
\fmfforce{(0.25w,0.5h)}{v1}
\fmfforce{(0.75w,0.5h)}{v5}
\fmffreeze
\fmfdot{v1,v5}
\end{fmfgraph*}
}}

\newcommand{\GRAPHNLOphicd}{
\parbox[c][20mm][c]{30mm}{
\centering
\begin{fmfgraph*}(30,30)
\fmfleft{i}
\fmfright{o}
\fmf{selectron}{i,v1}
\fmf{selectron}{v1,v3}
\fmf{selectron}{v3,v5}
\fmf{selectron}{v5,o}
\fmf{photon,left}{v1,v5}
\fmf{photon,right}{v3,v5}
\fmf{phantom,label=\circled{1},label.dist=0,left}{v1,v5}
\fmf{phantom,label=\circled{1},label.dist=0,right}{v3,v5}
\fmfforce{(0w,0.5h)}{i}
\fmfforce{(1w,0.5h)}{o}
\fmfforce{(0.25w,0.5h)}{v1}
\fmfforce{(0.50w,0.5h)}{v3}
\fmfforce{(0.75w,0.5h)}{v5}
\fmffreeze
\fmfdot{v1,v3,v5}
\end{fmfgraph*}
}}

\newcommand{\GRAPHNLOphie}{
\parbox[c][20mm][c]{30mm}{
\centering
\begin{fmfgraph*}(30,30)
\fmfleft{i}
\fmfright{o}
\fmf{selectron}{i,v1}
\fmf{selectron}{v1,v3}
\fmf{selectron}{v3,v5}
\fmf{selectron}{v5,o}
\fmf{photon,left}{v1,v3}
\fmf{photon,right}{v3,v5}
\fmf{phantom,label=\circled{1},label.dist=0,left}{v1,v3}
\fmf{phantom,label=\circled{1},label.dist=0,right}{v3,v5}
\fmfforce{(0w,0.5h)}{i}
\fmfforce{(1w,0.5h)}{o}
\fmfforce{(0.15w,0.5h)}{v1}
\fmfforce{(0.50w,0.5h)}{v3}
\fmfforce{(0.85w,0.5h)}{v5}
\fmffreeze
\fmfdot{v1,v3,v5}
\end{fmfgraph*}
}}

\newcommand{\GRAPHNLOphif}{
\parbox[c][20mm][c]{30mm}{
\centering
\begin{fmfgraph*}(30,30)
\fmfleft{i}
\fmfright{o}
\fmf{selectron}{i,v1}
\fmf{selectron}{v1,v2}
\fmf{selectron}{v2,v4}
\fmf{selectron}{v4,v5}
\fmf{selectron}{v5,o}
\fmf{photon,left}{v1,v4}
\fmf{photon,right}{v2,v5}
\fmf{phantom,label=\circled{1},label.dist=0,left}{v1,v4}
\fmf{phantom,label=\circled{1},label.dist=0,right}{v2,v5}
\fmfforce{(0w,0.5h)}{i}
\fmfforce{(1w,0.5h)}{o}
\fmfforce{(0.15w,0.5h)}{v1}
\fmfforce{(0.383w,0.5h)}{v2}
\fmfforce{(0.617w,0.5h)}{v4}
\fmfforce{(0.85w,0.5h)}{v5}
\fmffreeze
\fmfdot{v1,v2,v4,v5}
\end{fmfgraph*}
}}

\newcommand{\GRAPHNLOphigh}{
\parbox[c][20mm][c]{30mm}{
\centering
\begin{fmfgraph*}(30,30)
\fmfleft{i}
\fmfright{o}
\fmf{selectron}{i,v1}
\fmf{selectron}{v1,v2}
\fmf{fermion}{v2,v4}
\fmf{fermion}{v4,v5}
\fmf{selectron}{v5,o}
\fmf{photon,left}{v1,v4}
\fmf{photino,left}{v5,v2}
\fmf{phantom,label=\circled{1},label.dist=0,left}{v1,v4}
\fmf{phantom,label=\circled{1},label.dist=0,left}{v5,v2}
\fmfforce{(0w,0.5h)}{i}
\fmfforce{(1w,0.5h)}{o}
\fmfforce{(0.15w,0.5h)}{v1}
\fmfforce{(0.383w,0.5h)}{v2}
\fmfforce{(0.617w,0.5h)}{v4}
\fmfforce{(0.85w,0.5h)}{v5}
\fmffreeze
\fmfdot{v1,v2,v4,v5}
\end{fmfgraph*}
}}

\newcommand{\GRAPHNLOphii}{
\parbox[c][20mm][c]{30mm}{
\centering
\begin{fmfgraph*}(30,30)
\fmfleft{i}
\fmfright{o}
\fmf{selectron}{i,v1}
\fmf{fermion}{v1,v2}
\fmf{selectron}{v2,v4}
\fmf{fermion}{v4,v5}
\fmf{selectron}{v5,o}
\fmf{photino,left}{v1,v4}
\fmf{photino,left}{v5,v2}
\fmf{phantom,label=\circled{1},label.dist=0,left}{v1,v4}
\fmf{phantom,label=\circled{1},label.dist=0,left}{v5,v2}
\fmfforce{(0w,0.5h)}{i}
\fmfforce{(1w,0.5h)}{o}
\fmfforce{(0.15w,0.5h)}{v1}
\fmfforce{(0.383w,0.5h)}{v2}
\fmfforce{(0.617w,0.5h)}{v4}
\fmfforce{(0.85w,0.5h)}{v5}
\fmffreeze
\fmfdot{v1,v2,v4,v5}
\end{fmfgraph*}
}}

\newcommand{\GRAPHNLOphij}{
\parbox[c][20mm][c]{30mm}{
\centering
\begin{fmfgraph*}(30,30)
\fmfleft{i}
\fmfright{o}
\fmf{selectron}{i,v1}
\fmf{selectron}{v1,v2}
\fmf{selectron}{v2,v4}
\fmf{selectron}{v4,v5}
\fmf{selectron}{v5,o}
\fmf{photon,left}{v1,v5}
\fmf{photon,left}{v2,v4}
\fmf{phantom,label=\circled{1},label.dist=0,left}{v1,v5}
\fmf{phantom,label=\circled{1},label.dist=0,left}{v2,v4}
\fmfforce{(0w,0.5h)}{i}
\fmfforce{(1w,0.5h)}{o}
\fmfforce{(0.15w,0.5h)}{v1}
\fmfforce{(0.325w,0.5h)}{v2}
\fmfforce{(0.675w,0.5h)}{v4}
\fmfforce{(0.85w,0.5h)}{v5}
\fmffreeze
\fmfdot{v1,v2,v4,v5}
\end{fmfgraph*}
}}

\newcommand{\GRAPHNLOphik}{
\parbox[c][20mm][c]{30mm}{
\centering
\begin{fmfgraph*}(30,30)
\fmfleft{i}
\fmfright{o}
\fmf{selectron}{i,v1}
\fmf{selectron}{v1,v2}
\fmf{selectron}{v4,v5}
\fmf{selectron}{v5,o}
\fmf{photino,right}{v4,v2}
\fmf{fermion}{v2,v4}
\fmf{photon,left}{v1,v5}
\fmf{phantom,label=\circled{1},label.dist=0,left}{v2,v4}
\fmf{phantom,label=\circled{1},label.dist=0,left}{v1,v5}
\fmfforce{(0w,0.5h)}{i}
\fmfforce{(1w,0.5h)}{o}
\fmfforce{(0.15w,0.5h)}{v1}
\fmfforce{(0.325w,0.5h)}{v2}
\fmfforce{(0.675w,0.5h)}{v4}
\fmfforce{(0.85w,0.5h)}{v5}
\fmffreeze
\fmfdot{v1,v2,v4,v5}
\end{fmfgraph*}
}}

\newcommand{\GRAPHNLOphil}{
\parbox[c][20mm][c]{30mm}{
\centering
\begin{fmfgraph*}(30,30)
\fmfleft{i}
\fmfright{o}
\fmf{selectron}{i,v1}
\fmf{fermion}{v1,v2}
\fmf{fermion}{v2,v4}
\fmf{fermion}{v4,v5}
\fmf{selectron}{v5,o}
\fmf{photino,right}{v5,v1}
\fmf{photon,left}{v2,v4}
\fmf{phantom,label=\circled{1},label.dist=0,right}{v5,v1}
\fmf{phantom,label=\circled{1},label.dist=0,left}{v2,v4}
\fmfforce{(0w,0.5h)}{i}
\fmfforce{(1w,0.5h)}{o}
\fmfforce{(0.15w,0.5h)}{v1}
\fmfforce{(0.325w,0.5h)}{v2}
\fmfforce{(0.675w,0.5h)}{v4}
\fmfforce{(0.85w,0.5h)}{v5}
\fmffreeze
\fmfdot{v1,v2,v4,v5}
\end{fmfgraph*}
}}

\newcommand{\GRAPHNLOphim}{
\parbox[c][20mm][c]{30mm}{
\centering
\begin{fmfgraph*}(30,30)
\fmfleft{i}
\fmfright{o}
\fmf{selectron}{i,v1}
\fmf{fermion}{v1,v2}
\fmf{fermion}{v2,v4}
\fmf{fermion}{v4,v5}
\fmf{selectron}{v5,o}
\fmf{photino,right}{v5,v1}
\fmf{epsscalar,left}{v2,v4}
\fmf{phantom,label=\circled{1},label.dist=0,right}{v5,v1}
\fmf{phantom,label=\circled{1},label.dist=0,left}{v2,v4}
\fmfforce{(0w,0.5h)}{i}
\fmfforce{(1w,0.5h)}{o}
\fmfforce{(0.15w,0.5h)}{v1}
\fmfforce{(0.325w,0.5h)}{v2}
\fmfforce{(0.675w,0.5h)}{v4}
\fmfforce{(0.85w,0.5h)}{v5}
\fmffreeze
\fmfdot{v1,v2,v4,v5}
\end{fmfgraph*}
}}

\newcommand{\GRAPHNLOphin}{
\parbox[c][20mm][c]{30mm}{
\centering
\begin{fmfgraph*}(30,30)
\fmfleft{i}
\fmfright{o}
\fmf{selectron}{i,v1}
\fmf{fermion}{v1,v2}
\fmf{selectron}{v2,v4}
\fmf{fermion}{v4,v5}
\fmf{selectron}{v5,o}
\fmf{photino,right}{v5,v1}
\fmf{photino,left}{v2,v4}
\fmf{phantom,label=\circled{1},label.dist=0,right}{v5,v1}
\fmf{phantom,label=\circled{1},label.dist=0,left}{v2,v4}
\fmfforce{(0w,0.5h)}{i}
\fmfforce{(1w,0.5h)}{o}
\fmfforce{(0.15w,0.5h)}{v1}
\fmfforce{(0.325w,0.5h)}{v2}
\fmfforce{(0.675w,0.5h)}{v4}
\fmfforce{(0.85w,0.5h)}{v5}
\fmffreeze
\fmfdot{v1,v2,v4,v5}
\end{fmfgraph*}
}}

\newcommand{\GRAPHNLOphio}{
\parbox[c][20mm][c]{30mm}{
\centering
\begin{fmfgraph*}(30,30)
\fmfleft{i}
\fmfright{o}
\fmf{selectron}{i,v1}
\fmf{selectron}{v2,o}
\fmf{selectron}{v1,v2}
\fmf{photon,left}{v1,v2}
\fmf{phantom,left,label=\circled{2},label.dist=0}{v1,v2}
\fmfforce{(0w,0.5h)}{i}
\fmfforce{(1w,0.5h)}{o}
\fmfforce{(0.20w,0.5h)}{v1}
\fmfforce{(0.80w,0.5h)}{v2}
\fmffreeze
\fmfdot{v1,v2}
\end{fmfgraph*}
}}

\newcommand{\GRAPHNLOphip}{
\parbox[c][20mm][c]{30mm}{
\centering
\begin{fmfgraph*}(30,30)
\fmfleft{i}
\fmfright{o}
\fmf{selectron}{i,v1}
\fmf{selectron}{v2,o}
\fmf{fermion}{v1,v2}
\fmf{photino,right}{v2,v1}
\fmf{phantom,label=\circled{2},label.dist=0,right}{v2,v1}
\fmfforce{(0w,0.5h)}{i}
\fmfforce{(1w,0.5h)}{o}
\fmfforce{(0.20w,0.5h)}{v1}
\fmfforce{(0.80w,0.5h)}{v2}
\fmffreeze
\fmfdot{v1,v2}
\end{fmfgraph*}
}}

\newcommand{\GRAPHNLOpsiab}{
\parbox[c][20mm][c]{30mm}{
\centering
\begin{fmfgraph*}(30,30)
\fmfleft{i}
\fmfright{o}
\fmf{fermion}{i,v1}
\fmf{selectron}{v1,v2}
\fmf{selectron}{v2,v4}
\fmf{fermion}{v4,v5}
\fmf{fermion}{v5,o}
\fmf{photino,left}{v1,v4}
\fmf{photon,right}{v2,v5}
\fmf{phantom,label=\circled{1},label.dist=0,left}{v1,v4}
\fmf{phantom,label=\circled{1},label.dist=0,right}{v2,v5}
\fmfforce{(0w,0.5h)}{i}
\fmfforce{(1w,0.5h)}{o}
\fmfforce{(0.15w,0.5h)}{v1}
\fmfforce{(0.383w,0.5h)}{v2}
\fmfforce{(0.617w,0.5h)}{v4}
\fmfforce{(0.85w,0.5h)}{v5}
\fmffreeze
\fmfdot{v1,v2,v4,v5}
\end{fmfgraph*}
}}

\newcommand{\GRAPHNLOpsic}{
\parbox[c][20mm][c]{30mm}{
\centering
\begin{fmfgraph*}(30,30)
\fmfleft{i}
\fmfright{o}
\fmf{fermion}{i,v1}
\fmf{fermion}{v1,v2}
\fmf{fermion}{v2,v4}
\fmf{fermion}{v4,v5}
\fmf{fermion}{v5,o}
\fmf{photon,left}{v1,v4}
\fmf{photon,right}{v2,v5}
\fmf{phantom,label=\circled{1},label.dist=0,left}{v1,v4}
\fmf{phantom,label=\circled{1},label.dist=0,right}{v2,v5}
\fmfforce{(0w,0.5h)}{i}
\fmfforce{(1w,0.5h)}{o}
\fmfforce{(0.15w,0.5h)}{v1}
\fmfforce{(0.383w,0.5h)}{v2}
\fmfforce{(0.617w,0.5h)}{v4}
\fmfforce{(0.85w,0.5h)}{v5}
\fmffreeze
\fmfdot{v1,v2,v4,v5}
\end{fmfgraph*}
}}

\newcommand{\GRAPHNLOpside}{
\parbox[c][20mm][c]{30mm}{
\centering
\begin{fmfgraph*}(30,30)
\fmfleft{i}
\fmfright{o}
\fmf{fermion}{i,v1}
\fmf{fermion}{v1,v2}
\fmf{fermion}{v2,v4}
\fmf{fermion}{v4,v5}
\fmf{fermion}{v5,o}
\fmf{photon,left}{v1,v4}
\fmf{epsscalar,right}{v2,v5}
\fmf{phantom,label=\circled{1},label.dist=0,left}{v1,v4}
\fmf{phantom,label=\circled{1},label.dist=0,right}{v2,v5}
\fmfforce{(0w,0.5h)}{i}
\fmfforce{(1w,0.5h)}{o}
\fmfforce{(0.15w,0.5h)}{v1}
\fmfforce{(0.383w,0.5h)}{v2}
\fmfforce{(0.617w,0.5h)}{v4}
\fmfforce{(0.85w,0.5h)}{v5}
\fmffreeze
\fmfdot{v1,v2,v4,v5}
\end{fmfgraph*}
}}

\newcommand{\GRAPHNLOpsif}{
\parbox[c][20mm][c]{30mm}{
\centering
\begin{fmfgraph*}(30,30)
\fmfleft{i}
\fmfright{o}
\fmf{fermion}{i,v1}
\fmf{fermion}{v1,v2}
\fmf{fermion}{v2,v4}
\fmf{fermion}{v4,v5}
\fmf{fermion}{v5,o}
\fmf{epsscalar,left}{v1,v4}
\fmf{epsscalar,right}{v2,v5}
\fmf{phantom,label=\circled{1},label.dist=0,left}{v1,v4}
\fmf{phantom,label=\circled{1},label.dist=0,right}{v2,v5}
\fmfforce{(0w,0.5h)}{i}
\fmfforce{(1w,0.5h)}{o}
\fmfforce{(0.15w,0.5h)}{v1}
\fmfforce{(0.383w,0.5h)}{v2}
\fmfforce{(0.617w,0.5h)}{v4}
\fmfforce{(0.85w,0.5h)}{v5}
\fmffreeze
\fmfdot{v1,v2,v4,v5}
\end{fmfgraph*}
}}

\newcommand{\GRAPHNLOpsig}{
\parbox[c][20mm][c]{30mm}{
\centering
\begin{fmfgraph*}(30,30)
\fmfleft{i}
\fmfright{o}
\fmf{fermion}{i,v1}
\fmf{selectron}{v1,v2}
\fmf{fermion}{v2,v4}
\fmf{selectron}{v4,v5}
\fmf{fermion}{v5,o}
\fmf{photino,left}{v1,v4}
\fmf{photino,right}{v2,v5}
\fmf{phantom,label=\circled{1},label.dist=0,left}{v1,v4}
\fmf{phantom,label=\circled{1},label.dist=0,right}{v2,v5}
\fmfforce{(0w,0.5h)}{i}
\fmfforce{(1w,0.5h)}{o}
\fmfforce{(0.15w,0.5h)}{v1}
\fmfforce{(0.383w,0.5h)}{v2}
\fmfforce{(0.617w,0.5h)}{v4}
\fmfforce{(0.85w,0.5h)}{v5}
\fmffreeze
\fmfdot{v1,v2,v4,v5}
\end{fmfgraph*}
}}

\newcommand{\GRAPHNLOpsih}{
\parbox[c][20mm][c]{30mm}{
\centering
\begin{fmfgraph*}(30,30)
\fmfleft{i}
\fmfright{o}
\fmf{fermion}{i,v1}
\fmf{fermion}{v1,v2}
\fmf{fermion}{v2,v4}
\fmf{fermion}{v4,v5}
\fmf{fermion}{v5,o}
\fmf{photon,right}{v5,v1}
\fmf{photon,left}{v2,v4}
\fmf{phantom,label=\circled{1},label.dist=0,right}{v5,v1}
\fmf{phantom,label=\circled{1},label.dist=0,left}{v2,v4}
\fmfforce{(0w,0.5h)}{i}
\fmfforce{(1w,0.5h)}{o}
\fmfforce{(0.15w,0.5h)}{v1}
\fmfforce{(0.325w,0.5h)}{v2}
\fmfforce{(0.675w,0.5h)}{v4}
\fmfforce{(0.85w,0.5h)}{v5}
\fmffreeze
\fmfdot{v1,v2,v4,v5}
\end{fmfgraph*}
}}

\newcommand{\GRAPHNLOpsii}{
\parbox[c][20mm][c]{30mm}{
\centering
\begin{fmfgraph*}(30,30)
\fmfleft{i}
\fmfright{o}
\fmf{fermion}{i,v1}
\fmf{fermion}{v1,v2}
\fmf{fermion}{v2,v4}
\fmf{fermion}{v4,v5}
\fmf{fermion}{v5,o}
\fmf{photon,right}{v5,v1}
\fmf{epsscalar,left}{v2,v4}
\fmf{phantom,label=\circled{1},label.dist=0,right}{v5,v1}
\fmf{phantom,label=\circled{1},label.dist=0,left}{v2,v4}
\fmfforce{(0w,0.5h)}{i}
\fmfforce{(1w,0.5h)}{o}
\fmfforce{(0.15w,0.5h)}{v1}
\fmfforce{(0.325w,0.5h)}{v2}
\fmfforce{(0.675w,0.5h)}{v4}
\fmfforce{(0.85w,0.5h)}{v5}
\fmffreeze
\fmfdot{v1,v2,v4,v5}
\end{fmfgraph*}
}}

\newcommand{\GRAPHNLOpsij}{
\parbox[c][20mm][c]{30mm}{
\centering
\begin{fmfgraph*}(30,30)
\fmfleft{i}
\fmfright{o}
\fmf{fermion}{i,v1}
\fmf{fermion}{v1,v2}
\fmf{selectron}{v2,v4}
\fmf{fermion}{v4,v5}
\fmf{fermion}{v5,o}
\fmf{photon,right}{v5,v1}
\fmf{photino,left}{v2,v4}
\fmf{phantom,label=\circled{1},label.dist=0,right}{v5,v1}
\fmf{phantom,label=\circled{1},label.dist=0,left}{v2,v4}
\fmfforce{(0w,0.5h)}{i}
\fmfforce{(1w,0.5h)}{o}
\fmfforce{(0.15w,0.5h)}{v1}
\fmfforce{(0.325w,0.5h)}{v2}
\fmfforce{(0.675w,0.5h)}{v4}
\fmfforce{(0.85w,0.5h)}{v5}
\fmffreeze
\fmfdot{v1,v2,v4,v5}
\end{fmfgraph*}
}}

\newcommand{\GRAPHNLOpsik}{
\parbox[c][20mm][c]{30mm}{
\centering
\begin{fmfgraph*}(30,30)
\fmfleft{i}
\fmfright{o}
\fmf{fermion}{i,v1}
\fmf{fermion}{v1,v2}
\fmf{fermion}{v2,v4}
\fmf{fermion}{v4,v5}
\fmf{fermion}{v5,o}
\fmf{epsscalar,right}{v5,v1}
\fmf{photon,left}{v2,v4}
\fmf{phantom,label=\circled{1},label.dist=0,right}{v5,v1}
\fmf{phantom,label=\circled{1},label.dist=0,left}{v2,v4}
\fmfforce{(0w,0.5h)}{i}
\fmfforce{(1w,0.5h)}{o}
\fmfforce{(0.15w,0.5h)}{v1}
\fmfforce{(0.325w,0.5h)}{v2}
\fmfforce{(0.675w,0.5h)}{v4}
\fmfforce{(0.85w,0.5h)}{v5}
\fmffreeze
\fmfdot{v1,v2,v4,v5}
\end{fmfgraph*}
}}

\newcommand{\GRAPHNLOpsil}{
\parbox[c][20mm][c]{30mm}{
\centering
\begin{fmfgraph*}(30,30)
\fmfleft{i}
\fmfright{o}
\fmf{fermion}{i,v1}
\fmf{fermion}{v1,v2}
\fmf{fermion}{v2,v4}
\fmf{fermion}{v4,v5}
\fmf{fermion}{v5,o}
\fmf{epsscalar,right}{v5,v1}
\fmf{epsscalar,left}{v2,v4}
\fmf{phantom,label=\circled{1},label.dist=0,right}{v5,v1}
\fmf{phantom,label=\circled{1},label.dist=0,left}{v2,v4}
\fmfforce{(0w,0.5h)}{i}
\fmfforce{(1w,0.5h)}{o}
\fmfforce{(0.15w,0.5h)}{v1}
\fmfforce{(0.325w,0.5h)}{v2}
\fmfforce{(0.675w,0.5h)}{v4}
\fmfforce{(0.85w,0.5h)}{v5}
\fmffreeze
\fmfdot{v1,v2,v4,v5}
\end{fmfgraph*}
}}

\newcommand{\GRAPHNLOpsim}{
\parbox[c][20mm][c]{30mm}{
\centering
\begin{fmfgraph*}(30,30)
\fmfleft{i}
\fmfright{o}
\fmf{fermion}{i,v1}
\fmf{fermion}{v1,v2}
\fmf{selectron}{v2,v4}
\fmf{fermion}{v4,v5}
\fmf{fermion}{v5,o}
\fmf{epsscalar,right}{v5,v1}
\fmf{photino,left}{v2,v4}
\fmf{phantom,label=\circled{1},label.dist=0,right}{v5,v1}
\fmf{phantom,label=\circled{1},label.dist=0,left}{v2,v4}
\fmfforce{(0w,0.5h)}{i}
\fmfforce{(1w,0.5h)}{o}
\fmfforce{(0.15w,0.5h)}{v1}
\fmfforce{(0.325w,0.5h)}{v2}
\fmfforce{(0.675w,0.5h)}{v4}
\fmfforce{(0.85w,0.5h)}{v5}
\fmffreeze
\fmfdot{v1,v2,v4,v5}
\end{fmfgraph*}
}}

\newcommand{\GRAPHNLOpsin}{
\parbox[c][20mm][c]{30mm}{
\centering
\begin{fmfgraph*}(30,30)
\fmfleft{i}
\fmfright{o}
\fmf{fermion}{i,v1}
\fmf{selectron}{v1,v2}
\fmf{selectron}{v2,v4}
\fmf{selectron}{v4,v5}
\fmf{fermion}{v5,o}
\fmf{photino,left}{v1,v5}
\fmf{photon,right}{v4,v2}
\fmf{phantom,label=\circled{1},label.dist=0,left}{v1,v5}
\fmf{phantom,label=\circled{1},label.dist=0,right}{v4,v2}
\fmfforce{(0w,0.5h)}{i}
\fmfforce{(1w,0.5h)}{o}
\fmfforce{(0.15w,0.5h)}{v1}
\fmfforce{(0.325w,0.5h)}{v2}
\fmfforce{(0.675w,0.5h)}{v4}
\fmfforce{(0.85w,0.5h)}{v5}
\fmffreeze
\fmfdot{v1,v2,v4,v5}
\end{fmfgraph*}
}}

\newcommand{\GRAPHNLOpsio}{
\parbox[c][20mm][c]{30mm}{
\centering
\begin{fmfgraph*}(30,30)
\fmfleft{i}
\fmfright{o}
\fmf{fermion}{i,v1}
\fmf{selectron}{v1,v2}
\fmf{fermion}{v2,v4}
\fmf{selectron}{v4,v5}
\fmf{fermion}{v5,o}
\fmf{photino,left}{v1,v5}
\fmf{photino,right}{v4,v2}
\fmf{phantom,label=\circled{1},label.dist=0,left}{v1,v5}
\fmf{phantom,label=\circled{1},label.dist=0,right}{v4,v2}
\fmfforce{(0w,0.5h)}{i}
\fmfforce{(1w,0.5h)}{o}
\fmfforce{(0.15w,0.5h)}{v1}
\fmfforce{(0.325w,0.5h)}{v2}
\fmfforce{(0.675w,0.5h)}{v4}
\fmfforce{(0.85w,0.5h)}{v5}
\fmffreeze
\fmfdot{v1,v2,v4,v5}
\end{fmfgraph*}
}}

\newcommand{\GRAPHNLOpsip}{
\parbox[c][20mm][c]{30mm}{
\centering
\begin{fmfgraph*}(30,30)
\fmfleft{i}
\fmfright{o}
\fmf{fermion}{i,v1}
\fmf{fermion}{v2,o}
\fmf{fermion}{v1,v2}
\fmf{photon,left}{v1,v2}
\fmf{phantom,label=\circled{2},label.dist=0,left}{v1,v2}
\fmfforce{(0w,0.5h)}{i}
\fmfforce{(1w,0.5h)}{o}
\fmfforce{(0.20w,0.5h)}{v1}
\fmfforce{(0.80w,0.5h)}{v2}
\fmffreeze
\fmfdot{v1,v2}
\end{fmfgraph*}
}}

\newcommand{\GRAPHNLOpsiq}{
\parbox[c][20mm][c]{30mm}{
\centering
\begin{fmfgraph*}(30,30)
\fmfleft{i}
\fmfright{o}
\fmf{fermion}{i,v1}
\fmf{fermion}{v2,o}
\fmf{fermion}{v1,v2}
\fmf{epsscalar,left}{v1,v2}
\fmf{phantom,label=\circled{2},label.dist=0,left}{v1,v2}
\fmfforce{(0w,0.5h)}{i}
\fmfforce{(1w,0.5h)}{o}
\fmfforce{(0.20w,0.5h)}{v1}
\fmfforce{(0.80w,0.5h)}{v2}
\fmffreeze
\fmfdot{v1,v2}
\end{fmfgraph*}
}}

\newcommand{\GRAPHNLOpsir}{
\parbox[c][20mm][c]{30mm}{
\centering
\begin{fmfgraph*}(30,30)
\fmfleft{i}
\fmfright{o}
\fmf{fermion}{i,v1}
\fmf{fermion}{v2,o}
\fmf{selectron}{v1,v2}
\fmf{photino,left}{v1,v2}
\fmf{phantom,label=\circled{2},label.dist=0,left}{v1,v2}
\fmfforce{(0w,0.5h)}{i}
\fmfforce{(1w,0.5h)}{o}
\fmfforce{(0.20w,0.5h)}{v1}
\fmfforce{(0.80w,0.5h)}{v2}
\fmffreeze
\fmfdot{v1,v2}
\end{fmfgraph*}
}}
\newcommand{\sqedbubble}[2]{
\parbox[c][10mm][c]{12mm}{
\centering
\begin{fmfgraph*}(10,10)
\fmfleft{i}
\fmfright{o}
\fmf{#1,left}{i,o}
\fmf{#2,left}{o,i}
\end{fmfgraph*}
}}

\newcommand{\GRAPHAprop}[1]{
\parbox[c][0mm][c]{25mm}{
\centering
\begin{fmfgraph*}(15,0)
\fmfleft{i}
\fmfright{o}
\fmf{photon}{i,o}
\fmf{phantom,label=\circled{$#1$},label.dist=0}{i,o}
\fmflabel{$\mu$}{i}
\fmflabel{$\nu$}{o}
\marrow{a}{down}{-90}{$p$}{i,o}
\end{fmfgraph*}
}}

\newcommand{\GRAPHApropSD}{
\parbox[c][0mm][c]{25mm}{
\centering
\begin{fmfgraph*}(15,0)
\fmfleft{i}
\fmfright{o}
\fmf{photonSD}{i,o}
\fmflabel{$\mu$}{i}
\fmflabel{$\nu$}{o}
\marrow{a}{down}{-90}{$p$}{i,o}
\end{fmfgraph*}
}}

\newcommand{\GRAPHApropbare}{
\parbox[c][0mm][c]{25mm}{
\centering
\begin{fmfgraph*}(15,0)
\fmfleft{i}
\fmfright{o}
\fmf{photon}{i,o}
\fmflabel{$\mu$}{i}
\fmflabel{$\nu$}{o}
\marrow{a}{down}{-90}{$p$}{i,o}
\end{fmfgraph*}
}}

\newcommand{\GRAPHepsprop}[1]{
\parbox[c][0mm][c]{25mm}{
\centering
\begin{fmfgraph*}(15,0)
\fmfleft{i}
\fmfright{o}
\fmf{epsscalar}{i,o}
\fmf{phantom,label=\circled{$#1$},label.dist=0}{i,o}
\fmflabel{$\mu$}{i}
\fmflabel{$\nu$}{o}
\marrow{a}{down}{-90}{$p$}{i,o}
\end{fmfgraph*}
}}

\newcommand{\GRAPHepspropbare}{
\parbox[c][0mm][c]{25mm}{
\centering
\begin{fmfgraph*}(15,0)
\fmfleft{i}
\fmfright{o}
\fmf{epsscalar}{i,o}
\fmflabel{$\mu$}{i}
\fmflabel{$\nu$}{o}
\marrow{a}{down}{-90}{$p$}{i,o}
\end{fmfgraph*}
}}

\newcommand{\GRAPHepspropSD}{
\parbox[c][0mm][c]{25mm}{
\centering
\begin{fmfgraph*}(15,0)
\fmfleft{i}
\fmfright{o}
\fmf{epsscalarSD}{i,o}
\fmflabel{$\mu$}{i}
\fmflabel{$\nu$}{o}
\marrow{a}{down}{-90}{$p$}{i,o}
\end{fmfgraph*}
}}

\newcommand{\GRAPHlambdaprop}[1]{
\parbox[c][0mm][c]{25mm}{
\centering
\begin{fmfgraph*}(18,0)
\fmfleft{i}
\fmfright{o}
\fmflabel{ }{i}
\fmflabel{ }{o}
\fmf{photino}{i,o}
\fmf{phantom,label=\circled{$#1$},label.dist=0}{i,o}
\marrow{a}{down}{-90}{$p$}{i,o}
\end{fmfgraph*}
}}

\newcommand{\GRAPHlambdapropbare}{
\parbox[c][0mm][c]{25mm}{
\centering
\begin{fmfgraph*}(18,0)
\fmfleft{i}
\fmfright{o}
\fmflabel{ }{i}
\fmflabel{ }{o}
\fmf{photino}{i,o}
\marrow{a}{down}{-90}{$p$}{i,o}
\end{fmfgraph*}
}}

\newcommand{\GRAPHlambdapropSD}{
\parbox[c][0mm][c]{25mm}{
\centering
\begin{fmfgraph*}(18,0)
\fmfleft{i}
\fmfright{o}
\fmflabel{ }{i}
\fmflabel{ }{o}
\fmf{photinoSD}{i,o}
\marrow{a}{down}{-90}{$p$}{i,o}
\end{fmfgraph*}
}}

\newcommand{\GRAPHlambdapropbareparam}[1]{
\parbox[c][0mm][c]{25mm}{
\centering
\begin{fmfgraph*}(18,0)
\fmfleft{i}
\fmfright{o}
\fmflabel{ }{i}
\fmflabel{ }{o}
\fmf{photino}{i,o}
\marrow{a}{down}{-90}{$p$}{#1}
\end{fmfgraph*}
}}

\newcommand{\GRAPHpsiprop}{
\parbox[c][0mm][c]{20mm}{
\centering
\begin{fmfgraph*}(18,0)
\fmfleft{i}
\fmfright{o}
\fmf{fermion}{i,o}
\marrow{a}{down}{-90}{$p$}{i,o}
\end{fmfgraph*}
}}

\newcommand{\GRAPHpsipropSD}{
\parbox[c][0mm][c]{20mm}{
\centering
\begin{fmfgraph*}(18,0)
\fmfleft{i}
\fmfright{o}
\fmf{double_arrow}{i,o}
\marrow{a}{down}{-90}{$p$}{i,o}
\end{fmfgraph*}
}}

\newcommand{\GRAPHphiprop}{
\parbox[c][0mm][c]{20mm}{
\centering
\begin{fmfgraph*}(18,0)
\fmfleft{i}
\fmfright{o}
\fmf{selectron}{i,o}
\marrow{a}{down}{-90}{$p$}{i,o}
\end{fmfgraph*}
}}

\newcommand{\GRAPHpsipropparam}[1]{
\parbox[c][0mm][c]{20mm}{
\centering
\begin{fmfgraph*}(18,0)
\fmfleft{i}
\fmfright{o}
\fmf{fermion}{i,o}
\marrow{a}{down}{-90}{$p$}{#1}
\end{fmfgraph*}
}}

\newcommand{\GRAPHphipropSD}{
\parbox[c][0mm][c]{20mm}{
\centering
\begin{fmfgraph*}(18,0)
\fmfleft{i}
\fmfright{o}
\fmf{dbl_dashes_arrow}{i,o}
\marrow{a}{down}{-90}{$p$}{i,o}
\end{fmfgraph*}
}}

\newcommand{\GRAPHGammaApsipsi}{
\parbox[c][15mm][c]{15mm}{ 
\centering
\hspace*{0.25cm}
\begin{fmfgraph*}(10,10)
\fmfleft{i}
\fmflabel{$\mu$}{i}
\fmfright{o1,o2}
\fmf{photon}{i,v}
\fmf{fermion}{o1,v,o2}
\fmfdot{v}
\fmfforce{(1w,0h)}{o1}
\fmfforce{(1w,1h)}{o2}
\fmfforce{(0w,0.5h)}{i}
\fmfforce{(0.5w,0.5h)}{v}
\end{fmfgraph*}
}}

\newcommand{\GRAPHGammaepspsipsi}{
\parbox[c][15mm][c]{15mm}{ 
\centering
\hspace*{0.25cm}
\begin{fmfgraph*}(10,10)
\fmfleft{i}
\fmflabel{$\mu$}{i}
\fmfright{o1,o2}
\fmf{epsscalar}{i,v}
\fmf{fermion}{o1,v,o2}
\fmfdot{v}
\fmfforce{(1w,0h)}{o1}
\fmfforce{(1w,1h)}{o2}
\fmfforce{(0w,0.5h)}{i}
\fmfforce{(0.5w,0.5h)}{v}
\end{fmfgraph*}
}}

\newcommand{\GRAPHGammaAAphiphi}{
\parbox[c][15mm][c]{15mm}{ 
\centering
\hspace*{0.25cm}
\begin{fmfgraph*}(10,10)
\fmfleft{i1,i2}
\fmfright{o1,o2}
\fmf{photon}{i1,v,i2}
\fmf{selectron}{o1,v,o2}
\fmfdot{v}
\fmflabel{$\mu$}{i1}
\fmflabel{$\nu$}{i2}
\fmfforce{(0w,0h)}{i1}
\fmfforce{(0w,1h)}{i2}
\fmfforce{(1w,0h)}{o1}
\fmfforce{(1w,1h)}{o2}
\fmfforce{(0.5w,0.5h)}{v}
\end{fmfgraph*}
}}

\newcommand{\GRAPHGammaepsepsphiphi}{
\parbox[c][15mm][c]{15mm}{ 
\centering
\hspace*{0.25cm}
\begin{fmfgraph*}(10,10)
\fmfleft{i1,i2}
\fmfright{o1,o2}
\fmf{epsscalar}{i1,v,i2}
\fmf{selectron}{o1,v,o2}
\fmfdot{v}
\fmflabel{$\mu$}{i1}
\fmflabel{$\nu$}{i2}
\fmfforce{(0w,0h)}{i1}
\fmfforce{(0w,1h)}{i2}
\fmfforce{(1w,0h)}{o1}
\fmfforce{(1w,1h)}{o2}
\fmfforce{(0.5w,0.5h)}{v}
\end{fmfgraph*}
}}

\newcommand{\GRAPHGammaAphiphi}{
\parbox[c][15mm][c]{15mm}{ 
\centering
\hspace*{0.25cm}
\begin{fmfgraph*}(10,10)
\fmfleft{i}
\fmflabel{$\mu$}{i}
\fmfright{o1,o2}
\fmf{photon}{i,v}
\fmf{selectron}{o1,v}
\fmf{selectron}{v,o2}
\marrow{b}{up}{135}{$k$}{v,o2}
\marrow{a}{down}{-135}{$p$}{o1,v}
\fmfforce{(1w,0h)}{o1}
\fmfforce{(1w,1h)}{o2}
\fmfforce{(0w,0.5h)}{i}
\fmfforce{(0.5w,0.5h)}{v}
\fmfdot{v}
\end{fmfgraph*}
}}

\newcommand{\GRAPHGammalambdaphipsi}{
\parbox[c][15mm][c]{15mm}{ 
\centering
\begin{fmfgraph*}(10,10)
\fmfleft{i}
\fmfright{o1,o2}
\fmf{photino}{i,v}
\fmf{fermion}{v,o1}
\fmf{selectron}{o2,v}
\fmffreeze
\fmfdot{v}
\fmfforce{(1w,0h)}{o1}
\fmfforce{(1w,1h)}{o2}
\fmfforce{(0w,0.5h)}{i}
\fmfforce{(0.5w,0.5h)}{v}
\end{fmfgraph*}
}}

\newcommand{\GRAPHGammalambdaphipsibis}{
\parbox[c][15mm][c]{15mm}{ 
\centering
\begin{fmfgraph*}(10,10)
\fmfleft{i}
\fmfright{o1,o2}
\fmf{photino}{i,v}
\fmf{fermion}{o1,v}
\fmf{selectron}{v,o2}
\fmffreeze
\fmfdot{v}
\fmfforce{(1w,0h)}{o1}
\fmfforce{(1w,1h)}{o2}
\fmfforce{(0w,0.5h)}{i}
\fmfforce{(0.5w,0.5h)}{v}
\end{fmfgraph*}
}}

\newcommand{\GRAPHLargeNa}{
\parbox[c][15mm][c]{20mm}{ 
\centering
\begin{fmfgraph*}(15,15)
\fmfleft{i}
\fmfright{o1,o2}
\fmf{selectron}{i,v}
\fmf{fermion}{o1,v}
\fmf{photino}{v,o2}
\fmffreeze
\fmfdot{v}
\fmfforce{(1w,0h)}{o1}
\fmfforce{(1w,1h)}{o2}
\fmfforce{(0w,0.5h)}{i}
\fmfforce{(0.5w,0.5h)}{v}
\end{fmfgraph*}
}}

\newcommand{\GRAPHLargeNb}{
\parbox[c][15mm][c]{20mm}{ 
\centering
\begin{fmfgraph*}(15,15)
\fmfleft{i}
\fmfright{o1,o2}
\fmf{selectron}{i,v}
\fmf{fermion}{o1,v}
\fmf{photino}{v,o2}
\fmffreeze
\fmfdot{v}
\fmfforce{(1w,0h)}{o1}
\fmfforce{(1w,1h)}{o2}
\fmfforce{(0w,0.5h)}{i}
\fmfforce{(0.5w,0.5h)}{v}
\end{fmfgraph*}
}}

\newcommand{\GRAPHLargeNc}{
\parbox[c][15mm][c]{20mm}{ 
\centering
\begin{fmfgraph*}(15,15)
\fmfleft{i}
\fmfright{o1,o2}
\fmf{selectron}{i,v}
\fmf{fermion}{o1,v}
\fmf{photino}{v,o2}
\fmffreeze
\fmfdot{v}
\fmfforce{(1w,0h)}{o1}
\fmfforce{(1w,1h)}{o2}
\fmfforce{(0w,0.5h)}{i}
\fmfforce{(0.5w,0.5h)}{v}
\end{fmfgraph*}
}}

\newcommand{\GRAPHpsipropfermionflow}[2]{
\parbox[c][10mm][c]{20mm}{
\centering
\begin{fmfgraph*}(20,5)
\fmfleft{i}
\fmfright{o}
\fmf{fermion}{i,o}
\marrow{a}{down}{-90}{$p$}{#1}
\marrow{b}{up}{-90}{}{#2}
\end{fmfgraph*}
}}

\newcommand{\GRAPHlambdapropfermionflow}[2]{
\parbox[c][10mm][c]{20mm}{
\centering
\begin{fmfgraph*}(20,5)
\fmfleft{i}
\fmfright{o}
\fmf{photino}{i,o}
\marrow{a}{down}{-90}{$p$}{#1}
\marrow{b}{up}{-90}{}{#2}
\end{fmfgraph*}
}}

\newcommand{\GRAPHLOAAA}{
\parbox[c][22mm][c]{22mm}{
\centering
\begin{fmfgraph*}(22,22)
\fmfleft{i}
\fmfright{o1,o2}
\fmf{photon}{i,v1}
\fmf{selectron,left}{v1,v2}
\fmf{selectron,left}{v2,v1}
\fmf{photon}{v2,o1}
\fmf{photon}{v2,o2}
\fmfforce{(0w,0.5h)}{i}
\fmfforce{(1w,0.75h)}{o1}
\fmfforce{(1w,0.25h)}{o2}
\fmfforce{(0.25w,0.5h)}{v1}
\fmfforce{(0.75w,0.5h)}{v2}
\fmfdot{v1,v2}
\end{fmfgraph*}
}}

\newcommand{\GRAPHLOAepseps}{
\parbox[c][22mm][c]{22mm}{
\centering
\begin{fmfgraph*}(22,22)
\fmfleft{i}
\fmfright{o1,o2}
\fmf{photon}{i,v1}
\fmf{selectron,left}{v1,v2}
\fmf{selectron,left}{v2,v1}
\fmf{epsscalar}{v2,o1}
\fmf{epsscalar}{v2,o2}
\fmfforce{(0w,0.5h)}{i}
\fmfforce{(1w,0.75h)}{o1}
\fmfforce{(1w,0.25h)}{o2}
\fmfforce{(0.25w,0.5h)}{v1}
\fmfforce{(0.75w,0.5h)}{v2}
\fmfdot{v1,v2}
\end{fmfgraph*}
}}

\newcommand{\GRAPHLOAeps}{
\parbox[c][22mm][c]{22mm}{
\centering
\begin{fmfgraph*}(22,22)
\fmfleft{i}
\fmfright{o}
\fmf{photon}{i,v1}
\fmf{fermion,left}{v1,v2}
\fmf{fermion,left}{v2,v1}
\fmf{epsscalar}{v2,o}
\fmfforce{(0w,0.5h)}{i}
\fmfforce{(1w,0.5h)}{o}
\fmfforce{(0.25w,0.5h)}{v1}
\fmfforce{(0.75w,0.5h)}{v2}
\fmfdot{v1,v2}
\end{fmfgraph*}
}}

\newcommand{\GRAPHLOtria}{
\parbox[c][20mm][c]{25mm}{
\centering
\begin{fmfgraph*}(20,20)
\fmfleft{i}
\fmfright{o1,o2}
\fmf{photon,tension=3}{i,v1}
\fmf{photon,tension=3}{v2,o1}
\fmf{photon,tension=3}{v3,o2}
\fmf{fermion}{v1,v2}
\fmf{fermion}{v2,v3}
\fmf{fermion}{v3,v1}
\fmfdot{v1,v2,v3}
\end{fmfgraph*}
}}

\newcommand{\GRAPHLOtriabis}{
\parbox[c][20mm][c]{25mm}{
\centering
\begin{fmfgraph*}(20,20)
\fmfleft{i}
\fmfright{o1,o2}
\fmf{photon,tension=3}{i,v1}
\fmf{photon,tension=3}{v2,o1}
\fmf{photon,tension=3}{v3,o2}
\fmf{fermion}{v2,v1}
\fmf{fermion}{v3,v2}
\fmf{fermion}{v1,v3}
\fmfdot{v1,v2,v3}
\end{fmfgraph*}
}}

\newcommand{\GRAPHLOtrib}{
\parbox[c][20mm][c]{25mm}{
\centering
\begin{fmfgraph*}(20,20)
\fmfleft{i}
\fmfright{o1,o2}
\fmf{photon,tension=3}{i,v1}
\fmf{photon,tension=3}{v2,o1}
\fmf{photon,tension=3}{v3,o2}
\fmf{selectron}{v1,v2}
\fmf{selectron}{v2,v3}
\fmf{selectron}{v3,v1}
\fmfdot{v1,v2,v3}
\end{fmfgraph*}
}}

\newcommand{\GRAPHLOtribbis}{
\parbox[c][20mm][c]{25mm}{
\centering
\begin{fmfgraph*}(20,20)
\fmfleft{i}
\fmfright{o1,o2}
\fmf{photon,tension=3}{i,v1}
\fmf{photon,tension=3}{v2,o1}
\fmf{photon,tension=3}{v3,o2}
\fmf{selectron}{v2,v1}
\fmf{selectron}{v3,v2}
\fmf{selectron}{v1,v3}
\fmfdot{v1,v2,v3}
\end{fmfgraph*}
}}

\newcommand{\GRAPHLOtric}{
\parbox[c][20mm][c]{25mm}{
\centering
\begin{fmfgraph*}(20,20)
\fmfleft{i}
\fmfright{o1,o2}
\fmf{epsscalar,tension=3}{i,v1}
\fmf{photon,tension=3}{v2,o1}
\fmf{photon,tension=3}{v3,o2}
\fmf{fermion}{v1,v2}
\fmf{fermion}{v2,v3}
\fmf{fermion}{v3,v1}
\fmfdot{v1,v2,v3}
\end{fmfgraph*}
}}

\newcommand{\GRAPHLOtrid}{
\parbox[c][20mm][c]{25mm}{
\centering
\begin{fmfgraph*}(20,20)
\fmfleft{i}
\fmfright{o1,o2}
\fmf{photon,tension=3}{i,v1}
\fmf{epsscalar,tension=3}{v2,o1}
\fmf{epsscalar,tension=3}{v3,o2}
\fmf{fermion}{v1,v2}
\fmf{fermion}{v2,v3}
\fmf{fermion}{v3,v1}
\fmfdot{v1,v2,v3}
\end{fmfgraph*}
}}

\newcommand{\GRAPHLOtridbis}{
\parbox[c][20mm][c]{25mm}{
\centering
\begin{fmfgraph*}(20,20)
\fmfleft{i}
\fmfright{o1,o2}
\fmf{photon,tension=3}{i,v1}
\fmf{epsscalar,tension=3}{v2,o1}
\fmf{epsscalar,tension=3}{v3,o2}
\fmf{fermion}{v2,v1}
\fmf{fermion}{v3,v2}
\fmf{fermion}{v1,v3}
\fmfdot{v1,v2,v3}
\end{fmfgraph*}
}}

\newcommand{\GRAPHLOtrie}{
\parbox[c][20mm][c]{25mm}{
\centering
\begin{fmfgraph*}(20,20)
\fmfleft{i}
\fmfright{o1,o2}
\fmf{photon,tension=3}{i,v1}
\fmf{photino,tension=3}{o2,v3}
\fmf{photino,tension=3}{v2,o1}
\fmf{fermion}{v1,v2}
\fmf{selectron}{v2,v3}
\fmf{fermion}{v3,v1}
\fmfdot{v1,v2,v3}
\end{fmfgraph*}
}}

\newcommand{\GRAPHLOtriebis}{
\parbox[c][20mm][c]{25mm}{
\centering
\begin{fmfgraph*}(20,20)
\fmfleft{i}
\fmfright{o1,o2}
\fmf{photon,tension=3}{i,v1}
\fmf{photino,tension=3}{o2,v3}
\fmf{photino,tension=3}{v2,o1}
\fmf{fermion}{v2,v1}
\fmf{selectron}{v3,v2}
\fmf{fermion}{v1,v3}
\fmfdot{v1,v2,v3}
\end{fmfgraph*}
}}

\newcommand{\GRAPHLOtrif}{
\parbox[c][20mm][c]{25mm}{
\centering
\begin{fmfgraph*}(20,20)
\fmfleft{i}
\fmfright{o1,o2}
\fmf{photon,tension=3}{i,v1}
\fmf{photino,tension=3}{o2,v3}
\fmf{photino,tension=3}{v2,o1}
\fmf{selectron}{v1,v2}
\fmf{fermion}{v2,v3}
\fmf{selectron}{v3,v1}
\fmfdot{v1,v2,v3}
\end{fmfgraph*}
}}

\newcommand{\GRAPHLOtrifbis}{
\parbox[c][20mm][c]{25mm}{
\centering
\begin{fmfgraph*}(20,20)
\fmfleft{i}
\fmfright{o1,o2}
\fmf{photon,tension=3}{i,v1}
\fmf{photino,tension=3}{o2,v3}
\fmf{photino,tension=3}{v2,o1}
\fmf{selectron}{v2,v1}
\fmf{fermion}{v3,v2}
\fmf{selectron}{v1,v3}
\fmfdot{v1,v2,v3}
\end{fmfgraph*}
}}

\newcommand{\GRAPHLOtrig}{
\parbox[c][20mm][c]{25mm}{
\centering
\begin{fmfgraph*}(20,20)
\fmfleft{i}
\fmfright{o1,o2}
\fmf{epsscalar,tension=3}{i,v1}
\fmf{photino,tension=3}{o2,v3}
\fmf{photino,tension=3}{v2,o1}
\fmf{fermion}{v1,v2}
\fmf{selectron}{v2,v3}
\fmf{fermion}{v3,v1}
\fmfdot{v1,v2,v3}
\end{fmfgraph*}
}}

\newcommand{\GRAPHLOtrih}{
\parbox[c][20mm][c]{25mm}{
\centering
\begin{fmfgraph*}(20,20)
\fmfleft{i}
\fmfright{o1,o2}
\fmf{epsscalar,tension=3}{i,v1}
\fmf{epsscalar,tension=3}{v2,o1}
\fmf{epsscalar,tension=3}{v3,o2}
\fmf{fermion}{v1,v2}
\fmf{fermion}{v2,v3}
\fmf{fermion}{v3,v1}
\fmfdot{v1,v2,v3}
\end{fmfgraph*}
}}

\newcommand{\GRAPHALone}{
\parbox[c][25mm][c]{40mm}{
\centering
\begin{fmfgraph*}(35,25)
\fmfleft{i}
\fmfright{o}
\fmf{photon}{i,v1}
\fmf{fermion}{v2,v1}
\fmf{fermion}{v3,v2}
\fmf{fermion}{v1,v3}
\fmf{photon}{v2,v4}
\fmf{photon}{v3,v6}
\fmf{fermion}{v5,v4}
\fmf{fermion}{v6,v5}
\fmf{fermion}{v4,v6}
\fmf{photon}{v5,o}
\fmfforce{(0w,0.5h)}{i}
\fmfforce{(1w,0.5h)}{o}
\fmfforce{(0.2w,0.5h)}{v1}
\fmfforce{(0.4w,0.75h)}{v2}
\fmfforce{(0.4w,0.25h)}{v3}
\fmfforce{(0.6w,0.75h)}{v4}
\fmfforce{(0.6w,0.25h)}{v6}
\fmfforce{(0.8w,0.5h)}{v5}
\fmfdot{v1,v2,v3,v4,v5,v6}
\end{fmfgraph*}
}}

\newcommand{\GRAPHALonebis}{
\parbox[c][25mm][c]{40mm}{
\centering
\begin{fmfgraph*}(35,25)
\fmfleft{i}
\fmfright{o}
\fmf{photon}{i,v1}
\fmf{fermion}{v2,v1}
\fmf{fermion}{v3,v2}
\fmf{fermion}{v1,v3}
\fmf{photon}{v2,v4}
\fmf{photon}{v3,v6}
\fmf{fermion}{v4,v5}
\fmf{fermion}{v5,v6}
\fmf{fermion}{v6,v4}
\fmf{photon}{v5,o}
\fmfforce{(0w,0.5h)}{i}
\fmfforce{(1w,0.5h)}{o}
\fmfforce{(0.2w,0.5h)}{v1}
\fmfforce{(0.4w,0.75h)}{v2}
\fmfforce{(0.4w,0.25h)}{v3}
\fmfforce{(0.6w,0.75h)}{v4}
\fmfforce{(0.6w,0.25h)}{v6}
\fmfforce{(0.8w,0.5h)}{v5}
\fmfdot{v1,v2,v3,v4,v5,v6}
\end{fmfgraph*}
}}

\newcommand{\GRAPHALtwo}{
\parbox[c][25mm][c]{40mm}{
\centering
\begin{fmfgraph*}(35,25)
\fmfleft{i}
\fmfright{o}
\fmf{zigzag}{i,v1}
\fmf{dots_arrow}{v2,v1}
\fmf{dots_arrow}{v3,v2}
\fmf{dots_arrow}{v1,v3}
\fmf{zigzag}{v2,v4}
\fmf{zigzag}{v3,v6}
\fmf{dots_arrow}{v5,v4}
\fmf{dots_arrow}{v6,v5}
\fmf{dots_arrow}{v4,v6}
\fmf{zigzag}{v5,o}
\fmfforce{(0w,0.5h)}{i}
\fmfforce{(1w,0.5h)}{o}
\fmfforce{(0.2w,0.5h)}{v1}
\fmfforce{(0.4w,0.75h)}{v2}
\fmfforce{(0.4w,0.25h)}{v3}
\fmfforce{(0.6w,0.75h)}{v4}
\fmfforce{(0.6w,0.25h)}{v6}
\fmfforce{(0.8w,0.5h)}{v5}
\fmfdot{v1,v2,v3,v4,v5,v6}
\end{fmfgraph*}
}}

\newcommand{\GRAPHALtwobis}{
\parbox[c][25mm][c]{40mm}{
\centering
\begin{fmfgraph*}(35,25)
\fmfleft{i}
\fmfright{o}
\fmf{zigzag}{i,v1}
\fmf{dots_arrow}{v2,v1}
\fmf{dots_arrow}{v3,v2}
\fmf{dots_arrow}{v1,v3}
\fmf{zigzag}{v2,v4}
\fmf{zigzag}{v3,v6}
\fmf{dots_arrow}{v4,v5}
\fmf{dots_arrow}{v5,v6}
\fmf{dots_arrow}{v6,v4}
\fmf{zigzag}{v5,o}
\fmfforce{(0w,0.5h)}{i}
\fmfforce{(1w,0.5h)}{o}
\fmfforce{(0.2w,0.5h)}{v1}
\fmfforce{(0.4w,0.75h)}{v2}
\fmfforce{(0.4w,0.25h)}{v3}
\fmfforce{(0.6w,0.75h)}{v4}
\fmfforce{(0.6w,0.25h)}{v6}
\fmfforce{(0.8w,0.5h)}{v5}
\fmfdot{v1,v2,v3,v4,v5,v6}
\end{fmfgraph*}
}}

\newcommand{\photon}{
\parbox[c][0mm][c]{7mm}{
\centering
\begin{fmfgraph*}(7,0)
\fmfleft{i}
\fmfright{o}
\fmf{photon}{i,o}
\end{fmfgraph*}
}}

\newcommand{\photontrunc}[1]{
\parbox[c][0mm][c]{14mm}{
\centering
\begin{fmfgraph*}(14,0)
\fmfleft{i}
\fmfright{o}
\fmf{photon}{i,o}
\fmf{phantom,label=\circled{#1},label.dist=0}{i,o}
\end{fmfgraph*}
}}

\newcommand{\photonSD}{
\parbox[c][0mm][c]{7mm}{
\centering
\begin{fmfgraph*}(7,0)
\fmfleft{i}
\fmfright{o}
\fmf{dbl_wiggly }{i,o}
\end{fmfgraph*}
}}

\newcommand{\bubble}{
\parbox[c][10mm][c]{8mm}{
\centering
\begin{fmfgraph*}(8,8)
\fmfleft{i}
\fmfright{o}
\fmf{plain,right}{i,o}
\fmf{plain,right}{o,i}
\fmfdot{i,o}
\end{fmfgraph*}
}}

\newcommand{\blob}{
\parbox[c][12mm][c]{10mm}{
\centering
\begin{fmfgraph*}(10,10)
\fmfleft{i}
\fmfright{o}
\fmf{phantom}{i,v1}
\fmf{phantom}{v1,o}
\fmfblob{1.0cm}{v1}
\fmfdot{i,o}
\end{fmfgraph*}
}}

\newcommand{\selfenergy}{
\parbox[c][15mm][c]{20mm}{
\centering
\begin{fmfgraph*}(20,20)
\fmfleft{i}
\fmfright{o}
\fmf{plain}{i,v1}
\fmf{plain}{v1,v2}
\fmf{photon,left}{v1,v2}
\fmf{phantom,label=\circled{1},label.dist=0,left}{v1,v2}
\fmf{plain}{v2,o}
\fmfforce{(0w,0.4h)}{i}
\fmfforce{(1w,0.4h)}{o}
\fmfforce{(0.2w,0.4h)}{v1}
\fmfforce{(0.8w,0.4h)}{v2}
\fmfdot{v1,v2}
\end{fmfgraph*}
}}

\newcommand{\fourpointsfotened}{
\parbox[c][15mm][c]{20mm}{
\centering
\begin{fmfgraph*}(20,10)
\fmfleft{i1}
\fmfleft{i2}
\fmfright{o1}
\fmfright{o2}
\fmf{plain}{i1,v1}
\fmf{plain}{i2,v1}
\fmf{photon}{v1,v2}
\fmf{phantom,label=\circled{1},label.dist=0}{v1,v2}
\fmf{plain}{v2,o1}
\fmf{plain}{v2,o2}
\fmfforce{(0w,0h)}{i1}
\fmfforce{(0w,1h)}{i2}
\fmfforce{(1w,1h)}{o1}
\fmfforce{(1w,0h)}{o2}
\fmfforce{(0.2w,0.5h)}{v1}
\fmfforce{(0.8w,0.5h)}{v2}
\fmfdot{v1,v2}
\end{fmfgraph*}
}}

\newcommand{\diamondd}{
\parbox[c][12mm][c]{10mm}{
\centering
\begin{fmfgraph*}(10,10)
\fmfleft{v1}
\fmfright{v2}
\fmf{phantom,right,tag=1}{v1,v2}
\fmf{phantom,right,tag=2}{v2,v1}
\fmf{phantom,tag=3}{v1,v2}
\fmffreeze
\fmfposition
\fmfipath{p[]}
\fmfiset{p1}{vpath1(__v1,__v2)}
\fmfiset{p2}{vpath2(__v2,__v1)}
\fmfiset{p3}{vpath3(__v1,__v2)}
\fmfi{plain}{subpath (0,length(p1)/2) of p1}
\fmfi{plain}{subpath (0,length(p2)/2) of p2}
\fmfi{plain}{subpath (length(p2)/2,length(p2)) of p2}
\fmfi{plain}{subpath (length(p1)/2,length(p1)) of p1}
\fmfi{photon}{point length(p2)/2 of p1 .. point length(p2)/2 of p2}
\fmfdot{v1,v2}
\myvert{point length(p1)/2 of p1}
\myvert{point length(p2)/2 of p2}
\end{fmfgraph*}
}}

\newcommand{\diamonddlab}[1]{
\parbox[c][12mm][c]{10mm}{
\centering
\begin{fmfgraph*}(10,10)
\fmfleft{v1}
\fmfright{v2}
\fmf{phantom,right,tag=1}{v1,v2}
\fmf{phantom,right,tag=2}{v2,v1}
\fmf{phantom,tag=3}{v1,v2}
\fmffreeze
\fmfposition
\fmfipath{p[]}
\fmfiset{p1}{vpath1(__v1,__v2)}
\fmfiset{p2}{vpath2(__v2,__v1)}
\fmfiset{p3}{vpath3(__v1,__v2)}
\fmfi{plain}{subpath (0,length(p1)/2) of p1}
\fmfi{plain}{subpath (0,length(p2)/2) of p2}
\fmfi{plain}{subpath (length(p2)/2,length(p2)) of p2}
\fmfi{plain}{subpath (length(p1)/2,length(p1)) of p1}
\fmfi{photon}{point length(p2)/2 of p1 .. point length(p2)/2 of p2}
\fmfi{phantom,label=\circled{#1},label.dist=0}{point length(p2)/2 of p1 .. point length(p2)/2 of p2}
\fmfdot{v1,v2}
\myvert{point length(p1)/2 of p1}
\myvert{point length(p2)/2 of p2}
\end{fmfgraph*}
}}

\newcommand{\eyee}{
 \parbox[c][12mm][c]{10mm}{
\centering
\begin{fmfgraph*}(10,10)
\fmfleft{v1}
\fmfright{v2}
\fmf{phantom,right,tag=1}{v1,v2}
\fmf{phantom,right,tag=2}{v2,v1}
\fmf{phantom,tag=3}{v1,v2}
\fmffreeze
\fmfposition
\fmfipath{p[]}
\fmfiset{p1}{vpath1(__v1,__v2)}
\fmfiset{p2}{vpath2(__v2,__v1)}
\fmfiset{p3}{vpath3(__v1,__v2)}
\fmfi{plain}{subpath (0,length(p1)) of p1}
\fmfi{plain}{subpath (0,length(p2)/4) of p2}
\fmfi{plain}{subpath (length(p2)/4,3length(p2)/4) of p2}
\fmfi{plain}{subpath (3length(p2)/4,length(p2)) of p2}
\fmfi{photon}{point length(p2)/4 of p2 .. point length(p3)/2 of p3 .. point 3length(p2)/4 of p2}
\fmfdot{v1,v2}
\myvert{point length(p2)/4 of p2}
\myvert{point 3length(p2)/4 of p2}
\end{fmfgraph*}
}}

\newcommand{\eyeelab}[1]{
 \parbox[c][12mm][c]{10mm}{
\centering
\begin{fmfgraph*}(10,10)
\fmfleft{v1}
\fmfright{v2}
\fmf{phantom,right,tag=1}{v1,v2}
\fmf{phantom,right,tag=2}{v2,v1}
\fmf{phantom,tag=3}{v1,v2}
\fmffreeze
\fmfposition
\fmfipath{p[]}
\fmfiset{p1}{vpath1(__v1,__v2)}
\fmfiset{p2}{vpath2(__v2,__v1)}
\fmfiset{p3}{vpath3(__v1,__v2)}
\fmfi{plain}{subpath (0,length(p1)) of p1}
\fmfi{plain}{subpath (0,length(p2)/4) of p2}
\fmfi{plain}{subpath (length(p2)/4,3length(p2)/4) of p2}
\fmfi{plain}{subpath (3length(p2)/4,length(p2)) of p2}
\fmfi{photon}{point length(p2)/4 of p2 .. point length(p3)/2 of p3 .. point 3length(p2)/4 of p2}
\fmfi{phantom,label=\circled{#1},label.dist=0}{point length(p2)/4 of p2 .. point length(p3)/2 of p3 .. point 3length(p2)/4 of p2}
\fmfdot{v1,v2}
\myvert{point length(p2)/4 of p2}
\myvert{point 3length(p2)/4 of p2}
\end{fmfgraph*}
}}

\newcommand{\rqedfplot}{
\begin{tikzpicture}[scale=1]
\begin{axis}[
ymode=log,
log ticks with fixed point,
xmin = 0, xmax = 5,
ymax = 50,
xtick distance = 1,
grid = both,
minor tick num = 1,
major grid style = {lightgray},
minor grid style = {lightgray!25},
width = 0.95\textwidth,
height = 0.85\textwidth,
xlabel = {$\Nf$},
ylabel = {$\alpha_r(\Nf)$},
legend cell align = {left},
x label style={at={(axis description cs:1.0,0.)},anchor=west},
y label style={at={(axis description cs:0.,1.0)},rotate=-90,anchor=east},
]

\addplot[
draw=none,
domain = 0:3.39,
samples = 10
] {1.27324/(3.04395-x)};

\addplot[
domain = 0:3.39,
samples = 200,
smooth,
thick,
black,
] {861.588/(787.506 - 43.2798*x - 35.7484 *x^2 - 2.44026 *x^3 - x^4)};

\addplot[
smooth,
thick,
black,
only marks,
mark=square
] file[skip first] {plots/plot3.dat};

\draw[dashed,thick] (3.3954,0.0001) -- (3.3954,50);

\node at (1.75,9)[text width=2cm] {\small$m_\text{dyn}\neq0 $ \mbox{(insulator)}};
\node at (4.5,3)[text width=2cm] {\small$m_\text{dyn}=0 $ \mbox{(metal)}};
\node at (2.3,0.6)[text width=2cm] {\small$m_\text{dyn}=0$ \mbox{(metal)}};

\end{axis}
\end{tikzpicture}
}

\newcommand{\srqedftplot}{
\begin{tikzpicture}[scale=1]
\begin{axis}[
ymode=log,
log ticks with fixed point,
xmin = 0, xmax = 5,
ymax = 50,
xtick distance = 1,
grid = both,
minor tick num = 1,
major grid style = {lightgray},
minor grid style = {lightgray!25},
width = 0.95\textwidth,
height = 0.85\textwidth,
xlabel = {$\Nf$},
ylabel = {$\alpha_r(\Nf)$},
legend cell align = {left},
x label style={at={(axis description cs:1.0,0.)},anchor=west},
y label style={at={(axis description cs:0.,1.0)},rotate=-90,anchor=east},
]

\addplot[
draw=none,
domain = 0:3.39,
samples = 10
] {1.27324/(3.04395-x)};

\addplot[
domain = 0:3.39,
samples = 200,
smooth,
thick,
black,
] {0.63662/(1.62114-x)};

\addplot[
smooth,
thick,
black,
only marks,
mark=square
] file[skip first] {plots/plot4.dat};

\draw[dashed,thick] (1.62114,0.0001) -- (1.62114,50);

\node at (1.1,9)[text width=2cm] {\small$m_\text{dyn}\neq0 $ \mbox{(insulator)}};
\node at (3.4,3)[text width=2cm] {\small$m_\text{dyn}=0 $ \mbox{(metal)}};
\node at (1.25,0.34)[text width=2cm] {\small$m_\text{dyn}=0$};

\end{axis}

\end{tikzpicture}
}

\maketitle

\begin{abstract}
We review several variants of three-dimensional quantum electrodynamics (QED$_3$) with $N_f$ fermion (or boson) flavors including fermionic (or spinorial) QED$_3$, bosonic (or scalar) QED$_3$, $\mathcal{N}=1$ supersymmetric QED and also models of reduced QED (supersymmetric or not). We begin with an introduction to these models and their flow to a stable infra-red fixed point in the large-$N_f$ limit. We then present detailed state-of-the-art computations of the critical exponents of these models within the dimensional regularization (and reduction) scheme(s), at the next-to-leading order in the $1/N_f$ expansion and in an arbitrary covariant gauge. We finally discuss dynamical (matter) mass generation and the current status of our understanding of the phase structure of these models.
\end{abstract}

\thispagestyle{empty}

\begin{fmffile}{fmf-main}

\newcommand{\marrow}[5]{%
\fmfcmd{style_def marrow#1
expr p = drawarrow subpath(1/4, 3/4)of p shifted 6 #2 withpen pencircle scaled 0.4;
LaTeX_text(point 0.5 of p shifted 10 #2, #3, "#4");
enddef;}
\fmf{marrow#1,tension=0}{#5}
}

\newcommand{\Marrow}[6]{%
\fmfcmd{style_def marrow#1
expr p = drawarrow subpath (1/4, 3/4) of p shifted #6 #2 withpen pencircle scaled 0.4;
label.#3(btex #4 etex, point 0.5 of p shifted #6 #2);
enddef;}
\fmf{marrow#1,tension=0}{#5}}

\def\myvert#1{\fmfiv{decor.shape=circle,decor.filled=full,decor.size=2thick}{#1}}

\fmfset{dash_len}{2mm} 
\fmfset{arrow_len}{2.5mm} 
\fmfset{arrow_ang}{19} 


\fmfcmd{%
style_def photino expr p =
draw_wiggly p;
draw_plain p;
enddef;}

\fmfcmd{%
style_def selectron expr p =
draw_dashes p;
cfill (marrow (p, .5));
enddef;}

\fmfcmd{%
style_def epsscalar expr p =
draw_dbl_dots p;
enddef;}

\fmfcmd{%
style_def photinoSD expr p =
draw_dbl_wiggly p;
draw_dbl_plain p;
enddef;}

\fmfcmd{ 
style_def epsscalarSD expr p =
draw_dbl_dots p shifted (0,1.5);
draw_dbl_dots p shifted (0,-1.5); 
enddef;}

\fmfcmd{%
style_def photonSD expr p =
draw_dbl_wiggly p;
enddef;}

\fmfcmd{%
style_def fermionSD expr p =
draw_double_arrow p;
enddef;}

\fmfcmd{%
style_def selectronSD expr p =
draw_dbl_dashes_arrow p;
cfill (marrow (p, .5));
enddef;}

\newcommand*\circled[1]{\tikz[baseline=(char.base)]{\node[thick, shape=circle,draw,fill=white,inner sep=2pt] (char) {\tiny #1};}}

\newcommand\custompoint[3]{
\node[] at (#1,#2) {|};
\fill[draw=none,fill=black,opacity=0.3] ($(0,#2)+(0,-0.1)$) rectangle ($(#1,#2)+(0,0.1)$);
\node[above=0.1cm,anchor=south west] at ($(#1,#2)+(-0.2,0)$) {#3};
\node[below=0.2cm] at (#1,#2) {#1};
}

\newcommand\customtick[2]{
\node[] at (#1,#2) {|};
\node[below=0.3cm] at (#1,#2) {#1};
}

\newcommand\fmfsetbaseline{
\fmf{plain}{base1,base2}
\fmfforce{(0w,0h)}{base1}
\fmfforce{(0w,1h)}{base2}
}

\renewcommand{\baselinestretch}{0.85}\normalsize
{\small\tableofcontents}
\renewcommand{\baselinestretch}{1.0}\normalsize

\section{Introduction}

\subsection{Fermionic and Bosonic QEDs}

Three-dimensional quantum electrodynamics (QED$_3$) is an archetypal Abelian gauge field theory model describing strongly interacting planar fermions. In a Minkowski space and for massless fermions, it is described by the following action:
\be
S_{\text{fQED}_3} = \int \D^3x \left[ \I \bar{\psi}_i\fsl{D}\psi^i -\frac{1}{4}F_{\mu\nu}F^{\mu \nu} {- \frac{1}{2\xi}} (\partial_\mu A^\mu)^2 \right] ,
\label{sqed:action:qed3}
\ee
where $\psi^i$ are the $2\Nf$ flavors ($i=1,\dots,2\Nf $) of two-component massless Dirac fermions (or, equivalently, the $\Nf$ flavors of four-component massless Dirac fermions, $\Psi^i$), $A^\mu$ is the three-dimensional gauge field, $D_\mu = \partial_\mu + \I e A_\mu$ {is the gauge covariant derivative} and $\xi$ is the gauge-fixing parameter. It has been attracting continuous interest for the past four decades. The original motivation \cite{Appelquist:1981vg,Appelquist:1981sf} came from the fact that QED$_3$ shares several important features of four-dimensional quantum chromodynamics (QCD$_4$), such as asymptotic freedom, confinement and dynamical chiral symmetry breaking induced by the radiative generation of a fermion mass \cite{Pisarski:1984dj,Appelquist:1986fd,Appelquist:1988sr,Nash:1989xx,Atkinson:1989fp,Pennington:1990bx,Kotikov:1989nm,Kotikov:2011kg,Gusynin:1995bb,Maris:1996zg,Gusynin:2003ww,Fischer:2004nq} (see recent progress in \cite{Gusynin:2016som,Kotikov:2016wrb,Kotikov:2016prf,Karthik:2019mrr,Kotikov:2020slw}). In the last three decades, a considerable revival of interest in QED$_3$ also arose from its applications to condensed matter physics systems with relativistic-like gapless quasiparticle excitations at low energies, such as high-$T_c$ superconductors \cite{PhysRevB.42.4748,Dorey:1991kp,Franz:2001zz,Herbut:2002yq}, planar antiferromagnets \cite{Farakos:1997hg} and graphene \cite{Semenoff:1984dq,Novoselov:2004} (for graphene, see reviews in refs.~\cite{Gusynin:2007ix,CastroNeto:2009zz,Kotov:2010yh,Teber:2018jdh}).

Besides the above fermionic QED$_3$ (fQED$_3$), early studies \cite{Appelquist:1981sf} focused on the so-called (massless) bosonic QED$_3$ (bQED$_3$), the action of which is given by the following:
\be
S_{\text{bQED}_3} = \int \D^3x \left[\left|D_\mu\phi_i\right|^2 -\frac{1}{4}F_{\mu\nu}F^{\mu \nu} {- \frac{1}{2\xi}} (\partial_\mu A^\mu)^2 \right] ,
\label{eq:sqed:bqedaction}
\ee
where $\phi_i$ are $2\Nf$ flavors ($i=1,\dots,2\Nf$) of the complex pseudo-scalar field, $\phi$. Note that this model does not contain a $\phi^4$-term and is sometimes referred to as pure scalar QED or tricritical bQED in the literature (see, e.g., \cite{Benvenuti:2018cwd,Khachatryan:2019veb,Khachatryan:2019ygx}). 

A salient feature of both fQED$_3$ and bQED$_3$ is that they are super-renormalizable with a dimensionful coupling constant ($e^2$ has a dimension of mass). Remarkably, early \mbox{studies \cite{Appelquist:1981vg,Appelquist:1981sf,Appelquist:1986fd}} realized that, within a $1/\Nf$ expansion, an interacting fixed point emerges in the low-energy limit, and both models become effectively renormalizable deep in the infra-red (IR) with a dimensionless coupling constant, $1/\Nf$. This led to the study of their critical properties with the help of large-$\Nf$ techniques (see \cite{Gracey:2018ame} for a review). Critical exponents such as the field and mass anomalous dimensions are particularly important. They encode the renormalization of the composite operator, $\bar{\psi} \psi$ \cite{Gracey:1993iu,Gracey:1993sn}, and play a crucial role in the study of fundamental quantum field theory mechanisms, such as dynamical symmetry breaking and electron mass generation. Precision calculations require the computation of higher-order corrections that often represent a major technical challenge. Beyond precision, these corrections sometimes reveal new physics that is missed by simple low-order estimates. They also allow the study of the stability of a non-trivial IR fixed point with respect to radiative corrections. Finally, while the above models are well behaved in the limit of a large number of flavors, the fate of IR singularities are ubiquitous in super-renormalizable models, and QED$_3$ often serves as a toy model for such studies \cite{Jackiw:1980kv,Templeton:1981yp,Guendelman:1983dt,Guendelman:1982fm,King:1985hr}
(see recent progress in \cite{Karthik:2017hol,Gusynin:2020cra,Pikelner:2020mga}).

\subsection{Supersymmetric QEDs}

Another interesting variant of QED$_3$ corresponds to (minimal) $\mathcal{N}=1$ supersymmetric three-dimensional QED (SQED$_3$). This model can be obtained naively by combining the fermionic and bosonic QED$_3$ models described above, together with a superpartner for the photon, the photino. Mathematically, the degrees of freedom of $\cN = 1$ SQED$_3$ are the $2\Nf$ matter multiplets $\{\phi^j, \psi^j, F^j\}$ and a gauge multiplet $\{A^\mu, \lambda\}$. In the matter multiplet, the complex pseudo-scalars, $\phi^j$, are the {superpartners} of the two-component Dirac spinors, $\psi^j$, and the $F^j$ are complex auxiliary scalar fields without any dynamics (they ensure the equality of the degrees of freedom in the matter and gauge multiplet). 
In the gauge multiplet, obtained after choosing the Wess--Zumino gauge \cite{Wess:1971yu}, the photino, $\lambda$, is a two-component Majorana field. The action of $\cN =1$ massless SQED$_3$ is then given by
\ba
S_{\text{SQED}_3} = \int \D^3 x \bigg[ &
\I \bar{\psi_i}\fsl{D}\psi^i 
- \frac{1}{4}F_{\mu\nu}F^{\mu \nu} 
- \frac{1}{2\xi} (\partial_\mu A^\mu)^2
+ |D_\mu \phi_i|^2 
+ \frac{\I}{2}\bar{\lambda} \fsl{\partial}\lambda 
- \I e(\bar{\psi}^i \lambda \phi_i 
- \bar{\lambda} \psi^i \phi_i^\ast)
+ |F_i|^2 
\bigg] ,
\label{eq:sqed:actionnodred}
\ea
which, similarly to the two previous models, is super-renormalizable and has a non-trivial fixed point deep in the IR, at which it becomes effectively renormalizable \cite{Koopmans:1989kv}, the more recent \cite{Benvenuti:2019ujm} and \cite{Khachatryan:2019ygx} {for a review}). 

Supersymmetric variants of QED$_3$ have attracted continuous interest through the last decades. This has been partly motivated by the fact that the enhanced symmetry may simplify the resolution and, perhaps, even lead to an exact solution. As a matter of fact, the case of (non-minimal) $\mathcal{N}=2$ SQED$_3$ has been studied in an early seminal paper of Pisarski \cite{Pisarski:1984dj} by dimensional reduction from the case of (minimal) $\mathcal{N}=1$ four-dimensional supersymmetric QED (SQED$_4$), with focus on dynamical electron mass generation along the lines of the non-supersymmetric case. Actually, in $\mathcal{N}=1$ SQED$_4$, a non-perturbative non-renormalization theorem forbids dynamical mass generation \cite{Clark:1988zw}, and it was then argued \mbox{in \cite{Koopmans:1989kv}} that it, therefore, extends by dimensional reduction to $\mathcal{N}=2$ SQED$_3$. Further evidence for the absence of dynamical mass generation in $\mathcal{N}=2$ SQED$_3$ came from numerical \mbox{simulations \cite{Walker:1999rs}} and a refined analytic treatment \cite{CampbellSmith:1999vt}.

The situation in $\mathcal{N}=1$ SQED$_3$ is more subtle because of the absence of non-renormaliza\-tion theorem in this case. The model was first considered by Koopmans and Steringa \cite{Koopmans:1989kv} 
along the lines set by Appelquist et al. for standard fQED$_3$ \cite{Appelquist:1988sr}. 
Their truncated (to leading order (LO) in $1/\Nf $-expansion) Schwinger--Dyson equations approach resulted in a critical fermion flavor number, $\Nc=1.62$. This implies that a dynamical (parity-invariant) mass generation may occur for $\Nf=1$, i.e., one four-component Dirac spinor. A decade later, additional evidence for the generation of dynamical electron mass in minimal SQED$_3$ was also found in \cite{CampbellSmith:1999aw}. There is, however, no rigorous statement for electron mass generation for minimally supersymmetric SQED$_3$ \cite{Koopmans:1989kv,CampbellSmith:1999aw}.

In the last two decades, $\mathcal{N}=1$ SQED$_3$ has attracted significant attention (together with other supersymmetric and non-supersymmetric gauge theories) in the context of the study of IR dualities and renormalization group flows (see, e.g., 
\cite{Gremm:1999su,Gukov:2002es,Bashmakov:2018wts,Benini:2018umh,Gaiotto:2018yjh,Benini:2018bhk}). Interestingly, it was argued in \cite{Benvenuti:2019ujm} that $\mathcal{N}=1$ SQED$_3$ at $\Nf=1$ is dual to a conformal field theory in the IR. This suggests that no dynamical mass for the electron should be generated in contrast to the previously mentioned early (leading order) calculations \cite{Koopmans:1989kv,CampbellSmith:1999aw}. In this review, we will present a 
refined, next-to-leading order (NLO) analysis. We will show that, at NLO, $\Nc=0.39$, which is strong evidence that no electron mass is radiatively generated in $\mathcal{N}=1$ SQED$_3$, which is in agreement with the analysis based on dualities \cite{Benvenuti:2019ujm}.

At the interface with condensed matter physics, there have also been proposals during the last years that SUSY may emerge in the low-energy limit of various lattice models (see, e.g., \cite{Lee:2006if,2013PhRvB87d1401R,Grover:2013rc,Ponte:2012ru,Jian:2014pca,Witczak-Krempa:2015jca,Jian:2016zll,Han:2019car}). To this day, there is still no evidence that SUSY is realized in nature, and an emergent SUSY should certainly be difficult to detect in the lab \cite{Zhao:2017bhw}. Nevertheless, computing critical exponents in the corresponding models is certainly valuable in order to assess the potential impact of supersymmetry on experimentally measurable observables.

\subsection{Reduced QEDs}

Another model corresponds to the so-called reduced QED (QED$_{d_\gamma,d_e}$) that describes relativistic fermions in $d_e$-dimensional space--time and interacting via the exchange of massless bosons in $d_\gamma$-dimensions ($d_e \leq d_\gamma$). In a Minkowski space, the QED$_{d_\gamma,d_e}$ \mbox{action \cite{Gorbar:2001qt,Teber:2012de,Kotikov:2013eha}} reads
\be
{S = \int \D^{d_e} x \I\bar{\psi}_i \gamma^{\mu_e} D_{\mu_e} \psi^i
+ \int \D^{d_\gamma} x \left[- \frac{1}{4} F^{\mu_\gamma \nu_\gamma} F_{\mu_\gamma \nu_\gamma}
-\frac{1}{2\xi}\left(\partial_{\mu_\gamma}A^{\mu_\gamma}\right)^2 \right] ,}
\label{actionRQED}
\ee
where {$\psi_i$} are the $2\Nf$ flavors of two-component Dirac spinors in $d_e$ dimensions ($\mu_e=0,1, \dots ,d_e-1$). In (\ref{actionRQED}), the coupling of the fermion to the gauge field is restricted to $d_e$-dimensional space--time such that the gauge field is free in the $d_\gamma - d_e$ co-dimensional space, i.e., $D_{\mu_e}={\partial_{\mu_e}}+\I e A_{\mu_e}$, where $A_{\mu_e}=A_{\mu_e}(z=0)$ such that $z$ is the collective coordinate in the $(d_\gamma-d_e)$-dimensional space. It was introduced in \cite{Gorbar:2001qt}, motivated by the study of dynamical chiral symmetry breaking in brane-world theories (see also \cite{Kaplan:2009kr}). Soon after, a first application was devoted to the especially important 
case of conformal QED$_{4,3}$ (also known as pseudo-QED from \cite{Marino:1992xi} and mixed-dimensional QED from the recent \cite{Son:2015xqa}) in relation to \mbox{graphene 
\cite{Gorbar:2002iw}}. More precisely, QED$_{4,3}$ 
describes graphene \cite{Semenoff:1984dq} at its infra-red (IR) Lorentz-invariant fixed point 
\cite{Gonzalez:1993uz}. Importantly, it has been shown in \cite{Kotikov:2016yrn} that there is mapping between QED$_{4,3}$ and fermionic QED$_3$ in a large-$\Nf$ limit.

Theoretically, there has been rather extensive studies on QED$_{4,3}$ during the last decade with primary applications
to graphene-like systems, e.g., their transport and spectral properties \cite{Teber:2012de,Kotikov:2013kcl,Herbut:2013kca,Kotikov:2013eha,Valenzuela:2014uia,2016IJMPB..3050084H,Teber:2018goo} and quantum Hall effect \cite{2013arXiv1309.5879M,Son:2015xqa} (in \cite{Son:2015xqa}, the model was invoked as an effective field theory describing half-filled fractional quantum Hall systems) 
and dynamical symmetry breaking \cite{Kotikov:2016yrn,2017arXiv170400381S}, on which we will focus in the following. From a more field theoretic point of view, the model was shown to be unitary 
\cite{Marino:2014oba}, and its properties were studied under Landau--Khalatnikov--Fradkin transformation
\cite{Ahmad:2016dsb,James:2019ctc} as well as under duality transformations \cite{Hsiao:2017lch}. Renewed
interest in the model and its formal properties was triggered by a study \cite{Herzog:2017xha} on
interacting boundary conformal field theories (see, 
e.g., \cite{Bashmakov:2017rko,Karch:2018uft,Dudal:2018pta,DiPietro:2019hqe,Giombi:2019enr}). 

Motivated by condensed matter applications, supersymmetric extensions of reduced QED were also constructed and analyzed in \cite{Herzog:2018lqz,Gupta:2019qlg} via superconformal techniques on the boundary for both $\mathcal{N}=1$ and $\mathcal{N}=2$ cases. In \cite{Gupta:2019qlg}, non-perturbative computations of transport properties in the $\mathcal{N}=2$ case were carried out with the help of localization techniques. Once again, the $\mathcal{N}=1$ case is more subtle, and no exact solution is, so far, known. It is this case that will be of interest to us in the following. In particular, the supersymmetric extension of QED$_{4,3}$ will be denoted as SQED$_{4,3}$. In this ultra-relativistic super-graphene model, the matter fields (electrons and selectrons) are localized on a $(2+1)$-dimensional plane, while the gauge fields (photons and photinos) are $(3+1)$-dimensional. The corresponding action reads
\ba 
S_{\text{SQED}_{4,3}}& = \int \D^{d_e} x \bigg[ 
\I \bar{\psi}_i\gamma^{\mu_e}D_{\mu_e}\psi^i
+ |D_{\mu_e} \phi_i|^2 
+ |F_i|^2 
- \I e (\bar{\psi}^i \lambda \phi_i - \bar{\lambda} \psi^i \phi_i^\ast) 
\bigg] \nonum \\
& + \int \D^{d_\gamma} x \bigg[
-\frac{1}{4} F_{\mu_\gamma\nu_\gamma} F^{\mu_\gamma\nu_\gamma} 
-\frac{1}{2\xi} (\partial_{\mu_\gamma} A^{\mu_\gamma})^2
+ \frac{\I}{2} \bar{\Lambda} \Gamma^{\mu_\gamma}\partial_{\mu_\gamma}\Lambda 
+ \frac{1}{2} D^2 
\bigg] ,
\label{super-graphene}
\ea
where $d_e=3$ and $d_\gamma=4$, $\Lambda$ denotes a four-component Majorana field, $\Gamma^{\mu_\gamma}$ are four $4~\times~4$ gamma matrices, $\gamma^{\mu_e}$ are three $2~\times~2$ gamma matrices and $D$ is a real auxiliary field without dynamics (see \cite{Herzog:2018lqz} for more details).
As will be seen in the following, just as for the non-SUSY case, there exists mapping between SQED$_3$ in the large-$\Nf$ expansion and SQED$_{4,3}$ in the small-coupling expansion. The mapping is very similar to the non-SUSY case up to a factor of two from SUSY.

Let us emphasize that, in this review, we always consider {\it suspended} (super)-graphene, as opposed to a model defined on the boundary, as considered in, e.g., \cite{Herzog:2018lqz}. In our case, the boundary is considered as a transparent interface, while the model of \cite{Herzog:2018lqz} considers a purely reflecting boundary (graphene on a substrate). Nevertheless, the two models can be related by doubling the interaction, $\al_{\text{bdry}} = \al/2$. 

\subsection{Outline of the Review}

In order to adopt a unifying approach, we will introduce in Section \ref{sec:sqed:generalmodel} a general model that encompasses all the above-described models. We will present the perturbative setup, consisting of the Feynman rules (including subtleties related to the presence of Majorana spinors) and our renormalization conventions. In Section \ref{sec:sqed:perturbativecalculations}, we will then perform the full LO and NLO computations in the large-$\Nf$ expansion of all the polarizations and self-energies and derive all the corresponding anomalous dimensions. This will lead us to briefly discuss a generalized version of the Furry theorem for these models. In Section \ref{sec:sqed:results}, we apply our general results to the models of interest, i.e., fQED$_3$, bQED$_3$, SQED$_3$ and, finally, QED$_{4,3}$ and SQED$_{4,3}$, where we compute the optical conductivity of the (super-)graphene material. In Section \ref{sec:sqed:crit}, we will discuss the criteria for dynamical mass generation and the related phase structure of these models. The conclusion will be given in Section \ref{sec:conclusion}. Let us finally note that this review represents an {extended} version of our previous (short) paper \cite{Metayer:2022wre}. 

\section{General QED Model and Conventions}
\label{sec:sqed:generalmodel}

\subsection{The General gQED\titlemath{_3} Model}

In the Introduction, we formally introduced a total of five models, namely, fQED$_3$, bQED$_3$, SQED$_3$ and also the reduced QED$_{4,3}$ and SQED$_{4,3}$. In this section, we introduce a general model that encompasses the first three of them with the help of additional parameters. The reduced models will be analyzed from the results of the general model with \mbox{mapping \cite{Kotikov:2016yrn}. }

The general model is denoted as gQED$_3$, and its corresponding action reads
{
\begin{empheq}[box=\widefbox]{align}
S_{\text{gQED}_3} = \int \D^3 x \bigg[ & 
\I \bar{\psi}_i\slashed{D}\psi^i
-\frac{1}{4} \hat{F}^{\mu\nu} \hat{F}_{\mu\nu}
-\frac{1}{2\xi}(\hat\partial_{\mu} \hat{A}^{\mu})^2 \nonum \\
& 
+ S|\hat{D}_{\mu} \phi_i|^2 
+\frac{\I S}{2}\bar{\lambda} \slashed{\partial}\lambda
- \I e S(\bar{\psi}^i \lambda \phi_i 
- \bar{\lambda} \psi^i \phi^\ast_i)
+S|F_i|^2 \nonum \\
& -S\EP\frac{1}{2} (\hat\partial_{\mu} \bar{A}_{\nu})^2
- e\EP S \bar{\psi}_i \bar{\gamma}^{\mu}\bar{A}_{\mu}\psi^i 
+ e^2 \EP S\bar{A}^2 \lvert\phi_i\rvert^2 
\bigg] .
\label{eq:sqed:actiongeneral}
\end{empheq}
}
In \eqref{eq:sqed:actiongeneral}, the first line corresponds to the fQED$_3$ action {\eqref{sqed:action:qed3}}. The second line is the $\mathcal{N}=1$ SUSY content, where each superpartner field is associated with a tracking factor $S\in\{0,1\}$ such that $\Phi\rightarrow S \Phi$, $\forall \Phi\in\{\phi ,\lambda ,\bar{A}^{\mu}\}$ and $S^2=S$. Hence, at any step of the calculation, we may turn on (respectively, off) SUSY by setting $S=1$ (respectively, $S=0$). This will highlight SUSY effects in our computations and allow us to check our expressions by recovering known results for the corresponding non-SUSY theories. The third line of \eqref{eq:sqed:actiongeneral} is due to the use of the dimensional reduction (DRED) scheme \cite{Siegel:1979wq,Siegel:1980qs,Capper:1979ns} (see also \cite{Jack:1997sr} for a review), which is the most convenient regularization scheme for practical calculations in supersymmetric theories. DRED allows for preserving SUSY at the perturbative level by the introduction of extra particles, the so-called $\eps$-scalars, carried by the field $\bar A$. These particles arise from the formal splitting of the gauge field as $A^\mu = \hat A^{\mu} + \bar A^{\mu}$. Here, we use the notations of the review \cite{Mihaila:2013wma}, where hatted (respectively, barred) quantities 
have $d$ (respectively, $3-d$) components. To better appreciate the effects of DRED during the computations, the $\eps$-scalar field will be associated with a tracking factor $\EP \in\{0,1\}$ such that $\bar{A}^{\mu}\rightarrow \EP \bar{A}^{\mu}$ and $\EP ^2=\EP $. Indeed, as we shall see in the following, although $\eps$-scalars affect only few quantities at NLO, their effect is crucial in order to ensure the validity of supersymmetric identities.

In addition, we will work with $2\Nf$ arbitrary $n$-component spinors. 
In the SQED$_3$ case, $n=2$ is necessary to ensure the equality of the matter and gauge degrees of freedom. 
Nevertheless, working with arbitrary $n$ component spinors will allow us to take the case of $n=0$-component spinors, i.e., no fermions, which corresponds to the case of bQED$_3$. Indeed, by killing the spinorial degrees of freedom with $n=0$, one exactly recovers the action of bQED$_3$, \eqref{eq:sqed:bqedaction}. In order to keep track of both cases while limiting the complexity of our formulas, one can notice that the identity $n^2S=2nS$ holds in both cases. We shall, therefore, use the constraint, $n(n-2)S=0$, to simplify our computations. 

The action \eqref{eq:sqed:actiongeneral} is the general model we will work with in the rest of this article. It completely describes $\mathcal{N}=1$ supersymmetric QED in the DRED scheme with suitable parameters ($S,~\EP,~n$) that allow the recovery of the subcases of fQED$_3$ and bQED$_3$ (as well as QED$_{4,3}$ and SQED$_{4,3}$ via mapping) as well as the study of the effect of DRED by turning it on (or off) with $\EP=1$ (or 0). These parametrizations are summarized in Table \ref{tab:modelparam}.

\begin{table}[h]
\renewcommand{\arraystretch}{1.2}
\centering
\begin{tabular}{|c||c|c|c|}
\hline
Model & $S$ & $n$ & $\EP$ \\
\hline
$\mathcal{N}=1$ SQED$_3$ & 1 & 2 & 1 \\
bosonic QED$_3$ & 1 & 0 & 0 \\ 
fermionic QED$_3$ & 0 & 2 & 0 \\ 
\hline
\end{tabular}
\caption{Parameter values used to recover the different large-$\Nf$ models from the gQED$_3$ action \eqref{eq:sqed:actiongeneral}.}
\label{tab:modelparam}
\end{table}

Let us remark that the action \eqref{eq:sqed:actiongeneral} is completely massless. In the following, in order to compute the mass anomalous dimensions of the model, we will introduce a mass term for the matter multiplet, i.e., for the electron and the selectron. Since we are interested in dynamical mass generation, we will focus on the parity-even mass terms, i.e., of the form,
\be
\mathcal{L}_m=
m_\psi \left(\sum_{i=1}^{\Nf}\bar\psi_i\psi^i - \hspace{-0.25cm} \sum_{i=1+\Nf}^{2\Nf} \hspace{-0.25cm} \bar\psi_i\psi^i\right) 
+ m_\phi^2 \left(\sum_{i=1}^{\Nf}|\phi_i|^2 + \hspace{-0.25cm} \sum_{i=1+\Nf}^{2\Nf} \hspace{-0.25cm} |\phi_i|^2\right) .
\label{eq:sqed:massterm}
\ee
Moreover, we will work within the limit of small masses, i.e.,
\be
m_x \ll \pE \ll e^2\Nf , \qquad (m_x=m_\psi,m_\phi) ,
\label{eq:sqed:IRlimits}
\ee
with $\pE$ the Euclidean momentum. This limit will have, as a main advantage, to remove all the tadpoles in the theory. In bQED$_3$ and SQED$_3$ (and, therefore, SQED$_{4,3}$), the four-point bosonic coupling indeed gives rise to tadpoles, as opposed to the case of fQED$_3$, where no tadpoles are present, since the theory only has three-point coupling. The masses will, therefore, enter the electron and selectron propagators as small IR regulative masses{, allowing} the computation of their corresponding mass anomalous dimensions. All our calculations will then be carried using massless techniques (see, e.g., the review in \cite{Kotikov:2018wxe}). We postpone to {Section} \ref{sec:sqed:crit} the study of a potential dynamical generation of small parity-even masses \eqref{eq:sqed:massterm} in the matter multiplet.

\subsection{Feynman Rules}

The gQED$_3$ model \eqref{eq:sqed:actiongeneral} contains both Dirac and Majorana fermions. Therefore, one has to be careful to properly define the Feynman rules of the model in order to avoid sign mistakes. In the following, we will use a method based on the conventions of \cite{Denner:1992me,Denner:1992vza}. We first derive the bare gauge-multiplet propagators from the general action \eqref{eq:sqed:actiongeneral}, {reading}
\bs
\label{eq:sqed:gaugeprop}
\ba
\text{Photon:~~} &\hat{D}_{0AA}^{\mu\nu}(p)= \langle \hat{A}^\mu(-p)\hat{A}^\nu(p)\rangle_0 = \GRAPHApropbare = \frac{-\I}{p^2}\hat{d}^{\mu\nu}(p) ,\\
\text{$\eps$-scalar:~~}&\bar{D}_{0AA}^{\mu\nu}(p)= \langle \bar{A}^\mu(-p)\bar{A}^{\nu}(p)\rangle_0 
= \GRAPHepspropbare = \frac{-\I S \mathcal{E}}{p^2}\bar{g}^{\mu\nu} ,\\
\text{Photino:~~} &D_{0\lambda\bar{\lambda}}(p)= \langle \lambda(-p)\bar{\lambda}(p)\rangle_0 = \GRAPHlambdapropbare = \frac{\I S}{\fsl{p}} ,
\label{eq:sqed:gaugeprop:photino}
\ea
\es
with $\hat{d}^{\mu\nu}(p)=\hat{g}^{\mu\nu} -(1 - \xi)(\hat{p}^\mu \hat{p}^\nu/p^2)$. It is important to remark that the photino ($\lambda$) Majorana line \eqref{eq:sqed:gaugeprop:photino} carries a fermion flow, but is not represented with a dedicated arrow. We also derive from \eqref{eq:sqed:actiongeneral} the bare matter-multiplet propagators, reading
\bs
\label{eq:sqed:matterprop}
\ba
\text{Electron:~~} &S_{0\psi {\bar{\psi}}}(p)= \langle \psi(-p)\bar{\psi}(p)\rangle_0 = \GRAPHpsiprop = \frac{\I}{\fsl{p}- m_\psi} ,
\label{eq:sqed:matterprop:electron} \\
\text{Selectron:~~} &S_{0\phi{\phi^*}}(p)= \langle \phi(-p){\phi^*} (p)\rangle_0 = \GRAPHphiprop = \frac{\I S}{p^2-m_\phi^2} .
\label{eq:sqed:matterprop:selectron}
\ea
\es
Note that the arrow on the Dirac fermion ($\psi$) and the pseudo-scalar ($\phi$) propagators indicates the charge flow or, equivalently, the matter flow. As for the photino \eqref{eq:sqed:gaugeprop:photino}, the fermion flow on the Dirac fermion line \eqref{eq:sqed:matterprop:electron} is not indicated.
Together with these gauge and matter propagators comes additional rules for the loops
\begin{itemize}
\item Each matter-field loop ($\psi$ and $\phi$ field charge flow) gives a factor of $2\Nf$, i.e., graphically 
\be
\sqedbubble{fermion}{fermion} \equiv 2\Nf , \qquad
\sqedbubble{selectron}{fermion} \equiv 2\Nf , \qquad
\sqedbubble{selectron}{selectron} \equiv 2\Nf . 
\label{eq:sqed:matterloops}
\ee
\item Each fermion {loop} ($\psi$ and $\lambda$ field fermion flow) gives a factor $(-1)$ and a trace over the spinorial indices, i.e., graphically 
\be
\sqedbubble{fermion}{fermion} \equiv -\Tr , \qquad
\sqedbubble{photino}{fermion} \equiv -\Tr , \qquad
\sqedbubble{photino}{photino} \equiv -\Tr .
\label{eq:sqed:fermionicloops}
\ee
\end{itemize}
Lastly, we provide all the vertices of the action \eqref{eq:sqed:actiongeneral}, yielding in graphical form
\bs
\label{eq:sqed:vertices}
\ba
& \hat{\Gamma}^{\mu}_{0A\psi\bar\psi}=\GRAPHGammaApsipsi=-\I e \hat{\gamma}^{\mu} , 
&& \label{eq:sqed:vertex:qed} \\
& \hat{\Gamma}^{\mu}_{0A\phi{\phi^*}}(p,k)=\GRAPHGammaAphiphi=-\I e S (\hat{p}+\hat{k})^{\mu} , 
&& \hat{\Gamma}^{\mu \nu}_{0AA\phi{\phi^*}}=\GRAPHGammaAAphiphi=2\I e^2 S \hat{g}^{\mu \nu} , 
\label{eq:sqed:vertex:bqed}\\[0.2cm] 
& \bar{\Gamma}^{\mu}_{0A\psi\bar\psi}=\GRAPHGammaepspsipsi=-\I e \EP S \bar{\gamma}^{\mu} , 
&& \bar{\Gamma}^{\mu \nu}_{0AA\phi{\phi^*}}=\GRAPHGammaepsepsphiphi=2\I e^2 \EP S \bar{g}^{\mu \nu} , 
\label{eq:sqed:vertex:epsqed}\\
& \Gamma_{0\bar\lambda\psi{\phi^*}}=\GRAPHGammalambdaphipsi=eS , 
&& \Gamma_{0{\bar\psi\lambda\phi}}=\GRAPHGammalambdaphipsibis=-eS .
\label{eq:sqed:vertex:sqed}
\ea
\es
Note that the first vertex \eqref{eq:sqed:vertex:qed} is purely of fQED origin, the second line \eqref{eq:sqed:vertex:bqed} are the vertices of bQED origin, then the vertices \eqref{eq:sqed:vertex:epsqed} come from the $\eps$-scalar contributions and, finally, the vertices \eqref{eq:sqed:vertex:sqed} are of pure SUSY origin.

In addition to all these rules, one should be careful about fermion flows when both Dirac and Majorana fermions are present. This usually results in a multitude of additional Feynman rules to cope with all the possible flow cases in order to obtain the correct signs everywhere. In the following, we will use the compact Feynman rules of \cite{Denner:1992me,Denner:1992vza} that are based on assigning an additional fermion flow line on diagrams (when necessary) along fermionic lines to obtain the correct signs. The additional Feynman rules are then written down by specifying the fermion flow (arrow above) and, for the fermionic propagators (recalling that the middle arrow is the charge/matter flow and the bottom arrow is the momentum), they read
\bs
\label{eq:sqed:advancedfeynrulesa}
\ba
&&& \GRAPHpsipropfermionflow{i,o}{i,o} = S_{0\psi\bar\psi}(p) , 
&& \GRAPHpsipropfermionflow{o,i}{i,o} = S_{0\psi\bar\psi}(-p) , && \\
&&& \GRAPHpsipropfermionflow{i,o}{o,i} = S_{0\psi\bar\psi}(-p) , 
&& \GRAPHpsipropfermionflow{o,i}{o,i} = S_{0\psi\bar\psi}(p) , && \\
&&& \GRAPHlambdapropfermionflow{i,o}{i,o} = D_{0\lambda\bar\lambda}(p) , 
&& \GRAPHlambdapropfermionflow{o,i}{i,o} = D_{0\lambda\bar\lambda}(-p) , &&
\ea
\es
which amounts to adding a minus sign on the flowing momentum for each opposite arrow. Similarly, for the fermionic Dirac vertices (fermion flow indicated with the arrow on the right), they read
\vspace{-6pt}
\bs
\label{eq:sqed:advancedfeynrulesb}
\ba
&&& 
\begin{tikzpicture}[baseline=-0.1cm]
\node (graph) [] {\GRAPHGammaApsipsi};
\draw[-latex,thick] ($(graph.east)+(0,-0.4)$) arc [start angle=230,end angle=130, radius=0.5cm];
\end{tikzpicture}
=-\I e \hat{\gamma}^{\mu} ,
&&
\begin{tikzpicture}[baseline=-0.1cm]
\node (graph) [] {\GRAPHGammaApsipsi};
\draw[-latex,thick] ($(graph.east)+(0,0.4)$) arc [start angle=130,end angle=230, radius=0.5cm];
\end{tikzpicture}
=+\I e \hat{\gamma}^{\mu} , && \\
&&& 
\begin{tikzpicture}[baseline=-0.1cm]
\node (graph) [] {\GRAPHGammaepspsipsi};
\draw[-latex,thick] ($(graph.east)+(0,-0.4)$) arc [start angle=230,end angle=130, radius=0.5cm];
\end{tikzpicture}
=-\I e \EP S \bar{\gamma}^{\mu} ,
&& 
\begin{tikzpicture}[baseline=-0.1cm]
\node (graph) [] {\GRAPHGammaepspsipsi};
\draw[-latex,thick] ($(graph.east)+(0,0.4)$) arc [start angle=130,end angle=230, radius=0.5cm];
\end{tikzpicture}
=+\I e \EP S \bar{\gamma}^{\mu} , &&
\ea
\es
which amounts to a complex conjugation (charge conjugation) of the vertex if the fermion flow goes backward with respect to the charge/matter flow. Note that the other vertices mixing both Majorana and Dirac fermion (see \eqref{eq:sqed:vertex:sqed}) are real and are, therefore, unchanged under the inversion of the fermion flow. 

Actually, in the vast majority of cases, the simple rules {\eqref{eq:sqed:gaugeprop} to \eqref{eq:sqed:vertices}}, i.e., without the additional fermion flow lines \eqref{eq:sqed:advancedfeynrulesa} and \eqref{eq:sqed:advancedfeynrulesb}, are sufficient. This comes from the fact that most of the diagrams that we consider are such that the charge flow can follow naturally the fermion flow, both being continuous and unidirectional, i.e., graphically
\vspace{-6pt}
\be
\parbox[c][15mm][c]{50mm}{
\centering
\begin{fmfgraph*}(40,10)
\fmfleft{i}
\fmfright{o}
\fmf{fermion}{i,v1}
\marrow{a}{down}{-90}{$p_1$}{i,v1}
\fmf{photino}{v1,v2}
\marrow{b}{down}{-90}{$p_2$}{v1,v2}
\fmf{fermion}{v2,o}
\marrow{c}{down}{-90}{$p_3$}{v2,o}
\fmf{selectron}{v1,u1}
\fmf{selectron}{u2,v2}
\fmf{phantom}{u1,u2}
\fmffreeze
\fmfforce{(0.0w,0.25h)}{i}
\fmfforce{(1.0w,0.25h)}{o}
\fmfforce{(0.33w,1.0h)}{u1}
\fmfforce{(0.66w,1.0h)}{u2}
\fmfforce{(0.33w,0.25h)}{v1}
\fmfforce{(0.66w,0.25h)}{v2}
\end{fmfgraph*}
} ~ ,
\ee
where the hidden fermion flow goes from left to right, i.e., through the Dirac fermion, then the Majorana fermion and then the Dirac fermion again so that all arrows are properly aligned (reversing any of the arrows in this diagram would generate non-trivial minus signs not accounted for in the simple Feynman rules above).
In such a case, provided that the momentum arrows follow the (hidden) fermion flow and the charge flow, one can safely use the simple Feynman rules without the sign corrections shown above, i.e., {\eqref{eq:sqed:gaugeprop} to \eqref{eq:sqed:vertices}}.

Nevertheless, the advanced Feynman rules \eqref{eq:sqed:advancedfeynrulesa} and \eqref{eq:sqed:advancedfeynrulesb} will be required for a few diagrams, where one encounters a configuration of the type
\vspace{-6pt}
\be
\parbox[c][15mm][c]{50mm}{
\centering
\begin{fmfgraph*}(40,10)
\fmfleft{i}
\fmfright{o}
\fmf{fermion}{i,v1}
\marrow{a}{down}{-90}{$p_1$}{i,v1}
\fmf{photino,label=?,l.s=right}{v1,v2}
\fmf{fermion}{o,v2}
\marrow{c}{down}{-90}{$p_3$}{o,v2}
\fmf{selectron}{v1,u1}
\fmf{selectron}{v2,u2}
\fmf{phantom}{u1,u2}
\fmffreeze
\fmfforce{(0.0w,0.25h)}{i}
\fmfforce{(1.0w,0.25h)}{o}
\fmfforce{(0.33w,1.0h)}{u1}
\fmfforce{(0.66w,1.0h)}{u2}
\fmfforce{(0.33w,0.25h)}{v1}
\fmfforce{(0.66w,0.25h)}{v2}
\end{fmfgraph*}
} ~~ .
\label{eq:chain}
\ee
In such a case, we are forced to use the advanced Feynman rules \eqref{eq:sqed:advancedfeynrulesa} and \eqref{eq:sqed:advancedfeynrulesb}. In the following, this will be the case for only one diagram,
which is the seventh (labeled $(g)$) diagram of the two-loop contribution to the fermion self-energy at NLO, i.e., $\Sigma_2^{\psi(g)}$ (see \eqref{eq:specialdiagram}).

Note that, in principle, these advanced Feynman rules are also needed for the computation of the photino polarization because of the Majorana external legs. However, these diagrams are always appearing in pairs (with respect to opposite charge flows) that are exactly equal, such that we can consider only the case where all arrows are aligned and double the result. See the discussion below \eqref{eq:sqed:photinooneloop} for an example. 

We conclude this section by a brief warning to the reader that would like to use the software Qgraf \cite{Nogueira:1993,Nogueira:2021wfp} (as we did) to generate the diagram expressions of any theory involving both Dirac and {Majorana} fermions. First, Qgraf does not seem to be able to provide the correct minus signs from the fermionic loops in \eqref{eq:sqed:fermionicloops}. The simplest solution we found is to include additional trivial delta functions, $\delta^{\alpha\beta}$, in the propagators for $\psi$ and $\lambda$, where $\alpha,\beta$ are the spinor indices, such that $\delta^{\alpha\alpha}=-1$. Similarly, one can implement in an automated way the inclusion of the factors $2\Nf$ for \eqref{eq:sqed:fermionicloops} with similar delta functions on the fields $\psi^i$ and $\phi^j$, i.e., $\delta^{ij}$ such that $\delta^{ii}=2\Nf$. Moreover, Qgraf may have trouble in generating diagram expressions with continuous and unidirectional fermion flows in rare cases. More specifically, the software seems to always generate the flow properly (i.e., the indices generated by Qgraf that we use to orient the charge and fermion flows are aligned with the momenta arrows), except when there is an isolated fermion between two Majorana or the reverse, i.e., a chain of the form \eqref{eq:chain}. In this particular case, we need additional routines to check the Qgraf output and possibly correct these fermionic flows by using the rules \eqref{eq:sqed:advancedfeynrulesa} and \eqref{eq:sqed:advancedfeynrulesb}. As advertised before, our routine has corrected only one diagram in the NLO computations, which is $\Sigma_2^{\psi(g)}$ (see \eqref{eq:specialdiagram}).

\subsection{Numerator Algebra}

We work in a three-dimensional Minkowski space with the metric $g^{\mu \nu} = \text{diag}(+,-,-)$.
The three $n\times n$ Dirac $\gamma$-matrices satisfy the usual Clifford algebra, 
$\{ \gamma^\mu, \gamma^\nu \}= 2 g^{\mu \nu} I_n$, where $\Tr(I_n)=n$. Since we work in the DRED scheme, the metric tensor and $\gamma$-matrices are decomposed as
\be
g^{\mu \nu} = \hat{g}^{\mu\nu} + \bar g^{\mu \nu}, \qquad \gamma^\mu = {\hat\gamma^{\mu} +\bar\gamma^{\mu}} ,
\ee
so that there are $d=3-2\eps$ matrices $\hat\gamma^\mu$ and $2\eps$ matrices $\bar\gamma^\mu$, in order to keep a total integer number of three matrices $\gamma^\mu$. All of these matrices are of arbitrary size $n\times n$ to be able to take the limits $n=0$ for bQED$_3$, as well as $n=2$ for SQED$_3$ and fQED$_3$. In the DRED scheme, the following intuitive properties hold
\bs
\ba
& g^\mu{}_\mu = 3 , 
&& \hat g^\mu{}_\mu = d=3-2\eps , 
&& \bar g^\mu{}_\mu = 2 \eps , \\
& \{ \gamma^\mu, \gamma^{\nu} \}= 2 g^{\mu \nu} I_n ,
&& \{\hat\gamma^{\mu}, \hat\gamma^{\nu} \}= 2 \hat g^{\mu\nu} I_n , 
&& \{ \bar\gamma^{\mu}, \bar\gamma^{\nu} \}= 2 \bar g^{\mu \nu} I_n , 
\label{eq:sqed:commutators}
\ea
\es
as well as the very important case of the mixed dimensional anticommutator
\be
\{ \hat\gamma^{\mu}, \bar\gamma^{\nu} \}= 0 . 
\label{eq:miweddimcliff}
\ee
As expected, the usual Dirac trace computations will be modified but in a somewhat trivial way thanks to the property \eqref{eq:miweddimcliff}. In the following, we will have to compute traces involving gamma matrices living in two different spacetimes, such as $\Tr(\hat\gamma^\mu\bar\gamma^\nu\hat\gamma^\rho\bar\gamma^\sigma)$. This requires some care. In practice, one first sorts out the matrices, e.g., gathers hatted matrices to the left and barred ones to the right. This can be conducted using repetitively the anticommutation of the hatted and barred matrices \eqref{eq:miweddimcliff}, i.e., 
\be
\Tr(\hat\gamma^\mu \cdots \bar\gamma^\nu\hat\gamma^\rho \cdots \bar\gamma^\sigma) 
= - \Tr(\hat\gamma^\mu\cdots {\hat\gamma^\rho \bar\gamma^\nu} \cdots \bar\gamma^\sigma) .
\ee
Once completely sorted, one splits the traces into two parts using the following crucial trace splitting formula
\be
\Tr(\hat\gamma^{\nu_1} \cdots \hat\gamma^{\nu_n} \bar\gamma^{\mu_1} \cdots \bar\gamma^{\mu_m})= \frac{1}{2} \Tr(\hat\gamma^{\nu_1} \cdots \hat\gamma^{\nu_n})\Tr(\bar\gamma^{\mu_1} \cdots \bar\gamma^{\mu_m}) ,
\ee
where all matrices on the left are hatted and all matrices on the right are barred. 
Once sorted and split, both traces can be computed using the usual algorithm 
\be
\Tr(\hat\gamma^{\mu_1}\hat\gamma^{\mu_2} \cdots \hat\gamma^{\mu_m})= \sum_{i=2}^m (-1)^i {\hat g^{\mu_1\mu_i}} \Tr(\xcancel{\hat\gamma^{\mu_1}} \hat\gamma^{\mu_2} \cdots \xcancel{\hat\gamma^{\mu_i}} \cdots \hat\gamma^{\mu_m}) , \qquad m>3 ,
\label{eq:sqed:gammaformula}
\ee
and {the same algorithm} for traces over only barred matrices, $\bar\gamma^{\mu_i}$. These recursive formulas allow us to reduce any {trace} methodically until reaching the fundamental ones 
\be
\Tr(I_n)=n , \qquad 
\Tr(\gamma^\mu)=0 , \qquad 
\Tr(\gamma^\mu\gamma^\nu)=ng^{\mu\nu} , \qquad 
\Tr(\gamma^\mu\gamma^\nu\gamma^\rho)={\I n} \mathcal{T}_n \eps^{\mu\nu\rho} .
\label{eq:sqed:tracesidentities}
\ee
At this point, we recall that in three-dimensional theories, the trace over three gammas may not be zero, depending on the choice of the representation for the $\gamma$ matrices, but proportional to the fully antisymmetric tensor, $\eps^{\mu\nu\rho}$. To this end, we introduce the additional parameter $\mathcal{T}_n$, such that $\mathcal{T}_2=1$, $\mathcal{T}_4=0$ and $\mathcal{T}_n^2=\mathcal{T}_n$. Anyway, in large-$\Nf$ massless three-dimensional QED$_3$ (fermionic, bosonic and supersymmetric), these odd traces do not contribute to any result, as expected from parity-even theory. We will explicitly check this fact by observing that $\mathcal{T}_n$ will never appear in the rest of this article, even if we perform all the computations taking it into account. 
In the DRED scheme, the trace identities \eqref{eq:sqed:tracesidentities} split into two copies with the following intuitive properties
\bs
\ba
& \Tr(\hat\gamma^\mu)=0 , \qquad 
\Tr(\hat\gamma^\mu\hat\gamma^\nu)=n \hat g^{\mu\nu} ,\qquad 
\Tr(\hat\gamma^\mu\hat\gamma^\nu\hat\gamma^\rho)={\I n} \mathcal{T}_n \hat\eps^{\mu\nu\rho} , \\
& \Tr(\bar\gamma^\mu)=0 , \qquad 
\Tr(\bar\gamma^\mu\bar\gamma^\nu)=n \bar g^{\mu\nu} , \qquad \Tr(\bar\gamma^\mu\bar\gamma^\nu\bar\gamma^\rho)=0 .
\ea
\es
Note that we take $\bar\eps^{\mu\nu\rho}=0,$ as it makes sense that the Levi-Civita tensor in $2\eps$ dimensions vanishes as $\eps\rightarrow0$. 
Using the (mixed dimensional) trace techniques described above allows for computing any fermionic trace in gQED$_3$ and its subcases.

\subsection{Renormalization Setup}

We now have sufficient background material to introduce the renormalization setup and conventions for the gQED$_3$ model. Upon turning on the interactions, the Feynman rules for the gauge multiplet \eqref{eq:sqed:gaugeprop} are dressed via their respective Dyson equations and read 
\bs
\label{propagators:dressed}
\ba
& \hat{D}_{AA}^{\mu \nu} (p) = \langle \hat{A}^\mu(-p)\hat{A}^\nu(p)\rangle = \GRAPHApropSD = \frac{-\I}{1 - \Pi_\gamma(p^2)} \frac{\hat{d}^{\mu\nu}}{p^2} ,
\label{propagators:dressed:photon}\\
& \bar{D}_{AA}^{\mu \nu}(p)= \langle \bar{A}^\mu(-p)\bar A^{\nu}(p)\rangle = \GRAPHepspropSD = \frac{-\I \EP S}{1 - \Pi_\eps(p^2)} \frac{\bar{g}^{ \mu\nu}}{p^2} , 
\label{propagators:dressed:eps-scalar}\\
& D_{\lambda {\bar{\lambda}}}\hspace{0.08cm}(p)= \langle \lambda(-p)\bar{\lambda}(p)\rangle = \GRAPHlambdapropSD = \frac{\I S}{1 - \Pi_\lambda(p^2)}\frac{1}{\slashed{p}} , 
\label{propagators:dressed:photino}
\ea
\es
where the polarizations, $\Pi_x$, for the photon ($\Pi_\gamma$), the $\eps$-scalar ($\Pi_\eps$) and the photino ($\Pi_\lambda$), are parameterized via the following projections
\bs
\label{eq:sqed:projectorsgauge}
\ba
&\hat{\Pi}^{\mu \nu}(p)=(p^2 \hat{g}^{\mu\nu} - \hat{p}^\mu \hat{p}^\nu)\Pi_\gamma(p^2)
&& \hspace{-1cm} \implies \Pi_\gamma(p^2)= \left.\frac{\hat{\Pi}^{\mu}{}_{\mu}(p)}{(d-1) p^2}\right|_{m_x=0} ,
\label{eq:sqed:projectorsgauge:photon}\\
&\bar{\Pi}^{\mu\nu}(p)= p^2 \bar{g}^{\mu\nu} \Pi_\eps(p^2)
&& \hspace{-1cm} \implies \Pi_\eps(p^2)= \left.\frac{\bar{\Pi}^{\mu}{}_{\mu}(p)}{2\eps p^2}\right|_{m_x=0} ,
\label{eq:sqed:projectorsgauge:eps}\\
&\Pi_\lambda(p)= \fsl{p} \Pi_\lambda(p^2)
&& \hspace{-1cm} \implies \Pi_\lambda(p^2)= \left.\frac{\Tr[\fsl{p} \Pi_\lambda(p)]}{n p^2}\right|_{m_x=0} ,
\label{eq:sqed:projectorsgauge:lambda}
\ea
\es
respectively.
Using this setup, all integrals can be carried out in the massless limit, i.e., $m_x\rightarrow 0$ for $x=\{\psi,\phi\}$, as an IR rearrangement. 

An important remark is that in \eqref{propagators:dressed:photon}, the tensorial structure still yields $\hat{d}^{\mu\nu}(p)=\hat{g}^{\mu\nu} -(1 - \xi)(\hat{p}^\mu \hat{p}^\nu/p^2)$ because we are using a non-local gauge, i.e., we take
\be
\xi \rightarrow \xi(p^2)=\frac{\xi}{1-\Pi_\gamma(p^2)} ,
\label{eq:sqed:nonlocalgauge}
\ee
where $\xi$ will still be considered as the gauge-fixing parameter in the following. This trick is widely used in the QED$_3$ literature to keep computations light (see, e.g., \cite{Gracey:1993iu}; see also, for the SUSY case, \cite{Wess:1974jb,Zumino:1959wt,Walker:2003kc}). We recall that the use of a non-local gauge \eqref{eq:sqed:nonlocalgauge} is possible without affecting the physical results because the gauge-fixing parameter, $\xi$, is a mathematical artifact that does not appear in physical results.

As we will prove explicitly in the next sections, all the polarizations \eqref{eq:sqed:projectorsgauge} are finite. Indeed, we recall that, in the large-$\Nf$ limit, SQED$_3$ \cite{Koopmans:1989kv}, similarly to bQED$_3$ \cite{Appelquist:1981vg,Appelquist:1981sf} and \mbox{fQED$_3$ \cite{Pisarski:1984dj,Appelquist:1986fd}}, is a non-running (``standing'') gauge theory, i.e., the coupling is not renormalized, implying finite polarizations and, therefore, vanishing beta functions. This leads to the triviality of the renormalization constants for the coupling, gauge-multiplet fields and gauge-fixing parameter, formally $Z_x=1$, and $\gamma_x=0$ with $x\in\{e, \gamma, \eps, \lambda, \xi\}$, which imply a trivial beta function for the running of the coupling $e$ reading $\beta = -2 \eps \bar\alpha$, where $\alpha=e^2/(4\pi)$ and $\bar\alpha=\alpha/(4\pi)$. In this case, the coupling trivially renormalizes as {$\alpha\rightarrow\mu^{2\eps}\alpha$}, where $\mu$ is the renormalization scale. In the following, we will work in the modified minimal subtraction scheme, where the renormalization scale is defined as $\bar\mu^2=4\pi e^{-\gamma_E}\mu^2$, and further, ($\overline{\text{MS}}$ scheme) subtracts $4\pi$ and $\gamma_E${, the Euler--Mascheroni constant}. We will refer to this modified version of the dimensional reduction scheme as $\overline{\textrm{DRED}}$. 

Now considering the matter multiplet, turning on the interactions leads to the following dressed propagators
\vspace{-6pt}
\bs
\ba
& S_{\psi {\bar{\psi}}}\hspace{0.05cm}(p) = \GRAPHpsipropSD = \frac{\I}{1-\Sigma_p^\psi(p^2)}\frac{1}{\slashed{p}} , \\
& S_{\phi {{\phi^*}}}\hspace{-0.05cm}(p) = \GRAPHphipropSD = \frac{\I S}{1-\Sigma_p^\phi(p^2)}\frac{1}{p^2} ,
\ea
\es
where the matter-multiplet self-energies are parameterized as
\bs
\label{eq:sqed:selfenergyparam}
\ba
& \Sigma^\psi(p)= \fsl{p} \Sigma_p^\psi(p^2)+ m_{\psi} \Sigma_m^\psi(p^2) , 
\label{eq:sqed:selfenergyparampsi}\\
& \Sigma^\phi(p)= p^2 \Sigma_p^\phi(p^2)+m_{\phi}^2 \Sigma_m^\phi(p^2) .
\label{eq:sqed:selfenergyparamphi}
\ea
\es
From these, the components $p$ and $m$ can be extracted with the following projectors
\bs
\label{eq:sqed:projectors}
\ba
& \Sigma_p^\psi(p^2)=\left.\frac{\Tr[\fsl{p} \Sigma^\psi(p)]}{n p^2}\right|_{m_x=0} , 
&& \hspace{-2cm} \Sigma_m^\psi(p^2)=\left.\frac{\Tr[\Sigma^\psi(p)]}{n m_\psi}\right|_{m_x=0} , 
\label{eq:sqed:projectorsp}\\
& \Sigma_{p}^\phi(p^2)=\left.\frac{\Sigma^\phi(p)}{p^2}\right|_{m_x=0} , 
&& \hspace{-2cm} \Sigma_m^\phi(p^2)=\left.\frac{\partial \Sigma^\phi(p)}{\partial m_{\phi}^2}\right|_{m_x=0} ,
\label{eq:sqed:projectorsm}
\ea
\es
where $m_x=\{m_\phi,m_\psi\}$. As for the gauge polarizations, using this setup allows all integrals to be computed in the $m_x\rightarrow 0$ limit, i.e., completely massless, as an IR rearrangement.

The renormalization conventions for the non-trivial renormalization constants are defined as
\be
\psi=Z_\psi^{1/2}\psi_r , \qquad 
\phi=Z_\phi^{1/2}\phi_r , \qquad 
m_\psi=Z_{m_\psi}m_{\psi r} , \qquad 
m_\phi=Z_{m_\phi}m_{\phi r} .
\label{eq:Zdef}
\ee
The renormalization constants can be extracted from the bare self-energies thanks to the expression of the renormalized self-energies
\bs
\label{eq:sqed:renidentities}
\ba
& \Sigma_{pr}^{\psi}=1-(1-\Sigma_p^\psi)Z_\psi ,
&& \Sigma_{mr}^{\psi}=1-(1+\Sigma_m^\psi)Z_\psi Z_{m_\psi} , \\
& \Sigma_{pr}^{\phi}=1-(1-\Sigma_p^\phi)Z_\phi ,
&& \Sigma_{mr}^{\phi}=1-(1+\Sigma_m^\phi)Z_\phi Z_{m_\phi}^2 , 
\ea
\es
leading to the following simple set of relations
\be
\hspace{-1.8cm}
\left(1-\Sigma_p^\psi\right)Z_\psi=\text{finite} , ~~
\left(1-\Sigma_p^\phi\right)Z_\phi=\text{finite} , ~~
\frac{1+\Sigma_m^\psi}{1-\Sigma_p^\psi}Z_{m_\psi}=\text{finite} , ~~
\frac{1+\Sigma_m^\phi}{1-\Sigma_p^\phi}Z^2_{m_\phi}=\text{finite} , \hspace{-1cm}
\label{eq:sqed:ZSigmadef}
\ee
where ``finite'' means of the order of $\eps^0$, so that no additional counter diagrams needs to be computed. Finally, the associated anomalous dimensions are defined as
\be
\gamma_x=\frac{\D\log Z_x}{\D\log \mu} ,~~~x\in\{\psi, \phi, m_\psi, m_\phi\} ,
\label{eq:gammadef}
\ee
and correspond to the critical exponents of the theory that we want to compute.

\subsection{The Large-\titlemath{\Nf} Expansion}

In this section, we briefly introduce graphically the idea of the large-$\Nf$ expansion (see, e.g., \cite{Gracey:2018ame} for complete a review). Let us consider fQED$_3$ for simplicity. We first recall that, in the loop expansion, the Dyson equation {\eqref{propagators:dressed:photon}} for the photon can be written {graphically} as
\ba
\photonSD\photonSD ~ & = ~ \photon\photon \nonum \\ 
& ~+~ \photon\bubble\photon \nonum \\
& ~+~ \photon\bubble\photon\bubble\photon ~+~ \photon\diamondd\photon ~+~ \photon\eyee\photon \nonum \\
& ~+~ \photon\bubble\photon\bubble\photon\bubble\photon ~+~ \photon\bubble\photon\diamondd\photon ~+~ \cdots
\label{eq:sqed:couplingexp}
\ea
In the loop expansion \eqref{eq:sqed:couplingexp}, the perturbative series is well defined in the small-coupling $e$ regime, just by vertex counting. When the coupling, $e$, is not suitable as the expansion parameter, such as in super-renormalizable theories like gQED$_3$, one can use the so-called large-$\Nf$ expansion technique. Naively, the series \eqref{eq:sqed:couplingexp} is not perturbative in this regime since each fermion loop gives a factor $\Nf$, thereby increasing with the complexity of the diagram. The trick to perform the $1/\Nf$ expansion is then to resum the infinite chain of simple matter loops in force field propagators. Hence, considering the first term of each line in \eqref{eq:sqed:couplingexp}, i.e., the simple bubble chains, we can define the new propagator 
\ba
\photontrunc{1} ~ & = ~ \photon\photon ~+~ \photon\bubble\photon ~+~ \photon\bubble\photon\bubble\photon ~ + \cdots \nonum \\
& = ~ \photon\photon ~ \times ~ \left(1+ ~ \bubble\photon ~ + \left(\bubble\photon\right)^2 + \cdots \right) \nonum \\
& = ~ \photon\photon ~ \times ~ \frac{1}{1- ~\bubble\photon} .
\label{eq:sqed:softenedproprconstruct}
\ea

Going further, we recall that the bare photon propagator has momentum dependence, $\sim p^{-2}$, and the fermionic simple bubble diagram reads {$\sim e^2 \Nf \pE$} (with $\pE$ the Euclidean momentum). Therefore, the new propagator \eqref{eq:sqed:softenedproprconstruct}, in the large-$\Nf$ limit, reads
\be
\photontrunc{1} ~ {\sim ~ \left(~\bubble~\right)^{-1}} \sim ~ \frac{1}{e^2\Nf\pE} .
\ee
This new photon propagator is then said to be softened \cite{Appelquist:1981vg,Appelquist:1981sf} since its behavior in the infra-red is attenuated. Using this softened propagator, the first contribution to the electron self-energy is, therefore,
\be
\selfenergy ~ \sim ~ \frac{1}{\Nf} ,
\vspace{-0.25cm}
\ee
where the (dimensionful) coupling, $e^2$, drops in favor of $1/\Nf$. Therefore, in the large-$\Nf$ limit (that takes into account an infinite number of diagrams), fQED$_3$ becomes renormalizable with dimensionless coupling $1/\Nf$. Moreover, the expansion for the dressed photon propagator \eqref{eq:sqed:couplingexp} can be rewritten as
\ba
\hspace{-1.5cm}
\photonSD\photonSD ~ = ~ \photontrunc{1} ~ & + ~ \photontrunc{1}\diamonddlab{1}\photontrunc{1} ~+~ \photontrunc{1}\eyeelab{1}\photontrunc{1} ~+~ \cdots \hspace{-1cm}
\label{eq:sqed:largeNexp}
\ea
which now behave perturbatively in $1/\Nf$. 

Similarly, at the next-to-leading order (NLO), one can resum the two-loop contributions, yielding a new propagator, $\photontrunc{2}~\sim 1/\Nf^2$, in the IR limit, which allows computing NLO corrections to the electron self-energy at $1/\Nf^2$, etc. So, the strategy goes as follows. At leading order (LO): 
\begin{itemize}
\setlength\itemsep{0em}
\item[(1)] Compute the one-loop polarization using bare Feynman rules and compute the LO-softened photon by resumming the one-loop polarization.
\item[(2)] Compute the other diagrams of the theory at $\Ord(1/\Nf)$ using the LO-softened photon only. 
\end{itemize}

{Then,} at next-to-leading order (NLO): 
\begin{itemize}
\setlength\itemsep{0em}
\item[(3)] Compute the two-loop polarizations using the LO-softened photon propagator and compute the new NLO-softened photon propagator by resumming the two-loop polarization.
\item[(4)] Compute the other diagrams of the theory at NLO, i.e., $\Ord(1/\Nf^2)$ using both the LO and NLO-softened photon propagators. 
\end{itemize}
and pursue similarly at NNLO if desired, which goes beyond the scope of this review. 

This reasoning can be easily extended for the full gQED$_3$ model by resumming all polarizations of the gauge multiplet. In general, large-$\Nf$ techniques are expected to be very powerful as they resum an infinite number of diagrams. Moreover, since the new coupling of the theory is $1/\Nf$, the value of $\alpha=e^2/(4\pi)$ can be arbitrarily large, which is extremely useful to study the critical properties of the corresponding field theories that originate from non-perturbative effects.

\newpage

\section{Perturbative Calculations up to NLO in gQED\titlemath{_3}}
\label{sec:sqed:perturbativecalculations}

\subsection{Gauge-Multiplet Polarizations at LO}

In this first section, we compute in detail the first correction to the polarizations of the gauge multiplet, i.e., for the photon, the $\eps$-scalar and the photino, at LO in the $1/\Nf$ expansion, i.e., at $\Ord(\Nf)$.

\subsubsection{Photon Polarization at LO}

We first consider the photon propagator \eqref{propagators:dressed:photon} and compute the LO photon polarization operator, which consists of the following two contributions
\be
\hat\Pi_{1}^{\mu\nu}(p) = \hat\Pi_{1(a)}^{\mu\nu}(p) + \hat\Pi_{1(b)}^{\mu\nu}(p) .
\label{LO:photon:1l:tot}
\ee
Graphically, the corresponding two diagrams read
\bs
\label{LO:photon:def}
\ba
\I \hat\Pi_{1(a)}^{\mu\nu}(p) = \GRAPHLOAa & = - \mu^{2\ve} 2\Nf \int[\D^d k] \Tr \left[\hat\Gamma^{\mu}_{0A\psi\bar\psi} S_{0\psi\bar\psi}(k-p)\hat\Gamma^{\nu}_{0A\psi\bar\psi} S_{0\psi\bar\psi}(k) \right] ,
\label{LO:photon:def:a} \\
\I \hat\Pi_{1(b)}^{\mu\nu}(p) = \GRAPHLOAb & = \mu^{2\ve} 2\Nf \int [\D^d k] \hat\Gamma_{0A\phi{\phi^*}}^{\mu}(k-p,k) S_{0\phi{\phi^*}}(k-p) \hat\Gamma_{0A\phi{\phi^*}}^{\nu}(k,k-p) S_{0\phi{\phi^*}}(k) . \nonum \\[-0.5cm]
\label{LO:photon:def:b}
\ea
\es
Note that the first diagram (a) is of pure fermionic (QED$_3$) origin, while the second one (b) is of pure bosonic (bQED$_3$) origin. Therefore, at this order, the SQED$_3$ photon polarization directly appears as a simple sum of the fermionic (spinorial) and bosonic (scalar) results.
Using the Feynman rules for the vertices \eqref{eq:sqed:vertices} and the matter (electrons and selectrons) propagators \eqref{eq:sqed:matterprop} leads to the following expression
\bs
\ba
\I \hat\Pi_{1(a)}^{\mu\nu}(p) 
& = -\mu^{2\ve} 2\Nf e^2 \int [\D^d k] \frac{\Tr[\hat\gamma^\mu(\slashed{k}-\slashed{p}+m_\psi) \hat\gamma^\nu (\slashed{k}+m_\psi)]}{((k-p)^2-m_\psi^2)(k^2-m_\psi^2)} , \\
\I \hat\Pi_{1(b)}^{\mu\nu}(p) 
& = \mu^{2\ve} 2\Nf S e^2 \int [\D^d k] \frac{(2\hat k - \hat p )^\mu(2\hat k - \hat p )^\nu }{((k-p)^2-m_\phi^2)(k^2-m_\phi^2)} .
\ea
\es
Using the photonic polarization projector \eqref{eq:sqed:projectorsgauge:photon} and performing the trace on the $d=3-2\eps$ (hatted) space using the recursive Formula \eqref{eq:sqed:gammaformula} gives the following expressions
\bs
\ba
\Pi_{1\gamma}^{(a)}(p^2) &= -\I\mu^{2\ve} \frac{\Nf e^2}{p^2}\frac{2(d-2)n}{d-1} \int [\D^d k] \frac{k^2-k \cdot p}{k^2(k-p)^2} , \\
\Pi_{1\gamma}^{(b)}(p^2) &= -\I\mu^{2\ve} \frac{\Nf e^2}{p^2}\frac{2S}{d-1} \int [\D^d k] \frac{(2 k - p )^2}{k^2(k-p)^2} .  
\ea
\es
These integrals, once wick rotated to the Euclidean space, are then straightforward to compute using the results of Appendix \ref{chap:multiloop} and yield in the $\overline{\textrm{DRED}}$ scheme
\bs
\label{LO:photon:res}
\ba
\Pi_{1\gamma}^{(a)}(p^2) & = -\frac{\Nf e^2}{(4\pi)^{3/2} \pE} \left(\frac{\overline{\mu}^{ 2}}{-p^2} \right)^\ep \frac{(d-2)n}{d-1} e^{\gamma_E \ep} G(d,1,1) ,  
\label{LO:photon:res:a} \\
\Pi_{1\gamma}^{(b)}(p^2) & = -\frac{\Nf e^2}{(4\pi)^{3/2} \pE} \left(\frac{\overline{\mu}^{ 2}}{-p^2} \right)^\ep \frac{2S}{d-1} e^{\gamma_E \ep} G(d,1,1) ,  
\label{LO:photon:res:b}
\ea
\es
where $G(d,\al,\beta)$ is known exactly and defined in Appendix \ref{chap:multiloop}. 
Performing the $\ep$-expansion yields
\bs
\label{LO:photon:exp}
\ba
\Pi_{1\gamma}^{(a)}(p^2) &= -\frac{n\Nf e^2}{16 \pE} \bigg(1 - (1 - 2 \log 2 + L_p) \ep + \Ord(\ep^2) \bigg) ,  
\label{LO:photon:exp:a} \\
\Pi_{1\gamma}^{(b)}(p^2) &= -\frac{S\Nf e^2}{8 \pE} \bigg(1 + (1 + 2 \log 2 - L_p) \ep + \Ord(\ep^2) \bigg) ,  
\label{LO:photon:exp:b}
\ea
\es
where $L_p = \log(-p^2/\overline{\mu}^{ 2})$. As expected, in the fQED$_3$ case ($S=0$, $n=2$), only the first diagram, which is purely fermionic, contributes. In contrast, in the bQED$_3$ case ($S=1$, $n=0$), only the second diagram, which is purely bosonic, contributes. The total photon polarization function is, therefore, given by
\ba
\Pi_{1\gamma}(p^2) =\Pi_{1\gamma}^{(a)}(p^2) + \Pi_{1\gamma}^{(b)}(p^2) = -\frac{\Nf e^2}{(4\pi)^{3/2} \pE} \left(\frac{\overline{\mu}^{ 2}}{-p^2} \right)^\ep\frac{(d-2)n+2 S}{(d-1)} e^{\gamma_E \ep} G(d,1,1) ,  
\label{LO:photon:tot}
\ea
and since it is finite in $d=3$, its exact expression in this dimensionality reads
\ba
\Pi_{1\gamma}(p^2) = - \frac{(n+2S)\Nf e^2}{16\pE} .  
\label{LO:photon:tot:d=3}
\ea
%
Interestingly, in the cases of SQED$_3$ ($S=1$, $n=2$), fQED$_3$ ($S=0$, $n=2$) and bQED$_3$ ($S=1$, $n=0$), we deduce the following results 
\ba
\Pi_{1\gamma}^{\text{SQED}_3}(p^2) = - \frac{\Nf e^2}{4\pE} ,
\qquad 
\Pi_{1\gamma}^{\text{fQED}_3}(p^2) = - \frac{\Nf e^2}{8\pE} ,
\qquad 
\Pi_{1\gamma}^{\text{bQED}_3}(p^2) = - \frac{\Nf e^2}{8\pE} .
\label{eq:sqed:LOphoton}
\ea
In this very simple case, the SQED$_3$ photon polarization is simply the sum of the fermionic and bosonic parts since there is no one-loop diagram involving a mixture of both. Therefore, the SQED$_3$ photon polarization is twice the value found for fQED$_3$, which was first obtained in \cite{Appelquist:1981vg,Appelquist:1981sf}. 
Note that our result for SQED$_3$ coincides with the earlier one-loop result given in ref.~\cite{Koopmans:1989kv} but now obtained in the dimensional reduction scheme. 

\subsubsection{$\eps$-Scalar Polarization at LO}

Next, we proceed similarly for the $\ep$-scalar propagator \eqref{propagators:dressed:eps-scalar} 
and compute the LO $\ep$-scalar polarization function, which consists of a single non-vanishing diagram, defined as
\be
\hspace{-1.7cm}
\I \bar\Pi_{1}^{\mu\nu}(p) = \GRAPHLOeps = -\mu^{2\ve} 2\Nf \int [\D^d k] \Tr\left[\bar\Gamma^{\mu}_{0{A}\psi\bar\psi} S_{0\psi\bar\psi}(k-p)\bar\Gamma^{\nu}_{0{A}\psi\bar\psi} S_{0\psi\bar\psi}(k)\right] . \hspace{-1cm}
\ee
Using the Feynman rules for the vertices \eqref{eq:sqed:vertices} and the matter (electron and selectron) propagators \eqref{eq:sqed:matterprop} leads to the following expression
\ba
\I \bar\Pi_{1}^{\mu\nu}(p) 
& = -\mu^{2\ve} 2\Nf \EP S e^2 \int [\D^d k]\frac{\Tr[\bar\gamma^\mu(\slashed{k}-\slashed{p}+m_\psi) \bar\gamma^\nu (\slashed{k}+m_\psi)]}{((k-p)^2-m_\psi^2)(k^2-m_\psi^2)} . 
\label{LO:eps-scalar:def}
\ea
Using the projector defined in \eqref{eq:sqed:projectorsgauge:eps} and performing the trace in the $2\eps$-dimensional (barred) space with the help of the recursive formula \eqref{eq:sqed:gammaformula} yields
\ba
\Pi_{1\eps}(p^2) = -4\I\mu^{2\eps}\frac{\Nf e^2}{p^2} \EP S \int [\D^d k] \frac{k^2-k\cdot p}{k^2(k-p)^2} .  
\ea
After wick rotation and using the results of Appendix \ref{chap:multiloop}, we have
\ba
\Pi_{1\eps}(p^2) = -\frac{\Nf e^2}{(4\pi)^{3/2} \pE} \left(\frac{\overline{\mu}^{ 2}}{-p^2} \right)^\ep 2 \EP S e^{\gamma_E \ep} G(d,1,1) .
\label{LO:eps-scalar:tot}
\ea
Since this result is again finite in $d=3$, it can be written as
\ba
\Pi_{1\ep}(p^2) = - \frac{\EP S\Nf e^2}{4\pE} .
\label{LO:eps-scalar:tot:d=3}
\ea
Let us note that in the case of bQED$_3$ ($S=1$, $n=0$, $\EP=0$), as well as in the case of fQED$_3$ ($S=0$, $n=2$, $\EP=0$), this polarization is obviously zero. Indeed, the $\eps$-scalars are relevant only in the case of SQED$_3$ ($S=1$, $n=2$, $\EP=1$), yielding
\be
\Pi^{\text{SQED}_3}_{1\eps}(p^2)=-\frac{\Nf e^2}{4\pE} ,
\ee
which is exactly equal to the polarization of the photon in the same case, $\Pi^{\text{SQED}_3}_{1\gamma}(p^2)$ calculated in \eqref{eq:sqed:LOphoton}. As we will comment later on, such an equality is expected from SUSY.

\subsubsection{Photino Polarization at LO}

Lastly, we proceed in the same way for the photino propagator \eqref{propagators:dressed:photino} and compute LO photino self-energy, which consists of two non-vanishing diagrams with opposite charge flows. Since it is a photino (Majorana) polarization, in principle, we need to follow the advanced Feynman rules \eqref{eq:sqed:advancedfeynrulesa} and \eqref{eq:sqed:advancedfeynrulesb}, leading to
\vspace*{-0.75cm}\\
\bs
\label{eq:sqed:photinooneloop}
\ba
-\I \Pi_{1\lambda}^{(a)}(p) & = \GRAPHLOlambda = \mu^{2\ve} 2\Nf \int [\D^d k] \Gamma_{0{\bar\psi\lambda\phi}} S_{0\psi\bar\psi}(k) \Gamma_{0{\bar\lambda\psi\phi^*}} S_{0\phi{\phi^*}}(k-p) , \\[-0.1cm]
-\I \Pi_{1\lambda}^{(b)}(p) & = \GRAPHLOlambdabis = \mu^{2\ve} 2\Nf \int [\D^d k] \Gamma_{0{\bar\lambda\psi\phi^*}} S_{0\psi\bar\psi}(k) \Gamma_{0{\bar\psi\lambda\phi}} S_{0\phi{\phi^*}}(k-p) , 
\ea
\es
\vspace*{-0.5cm}\\
where we assigned a continuous and unidirectional fermion flow that goes from left to right.
In the case of {the diagram} $(a)$, all flows are in the same direction so that there is no need for the advanced Feynman rules. In the case of diagram $(b)$, the charge flow and the fermion flow are opposite to each others. However, the charge flow is also opposite to the momentum flow so that the momentum obtains an additional minus sign. All together, the Dirac propagator remains unchanged, and diagram $(b)$ is equal to diagram $(a)$. The resulting contribution is then defined as twice the configuration where all flows are aligned
\be
-\I \Pi_{1\lambda}(p) = 2\times \mu^{2\ve} 2\Nf \int [\D^d k] \Gamma_{0{\bar\psi\lambda\phi}} S_{0\psi\bar\psi}(k) \Gamma_{0{\bar\lambda\psi\phi^*}} S_{0\phi{\phi^*}}(k-p) .
\ee
It turns out that this reasoning will apply to all the photino polarization diagrams at higher orders. Therefore, for automation purposes, one can always consider only the configuration where all flows are aligned and simply multiply it by two so that the advanced Feynman rules with fermion flow specification are almost never needed (see discussion below \eqref{eq:chain}). Using now the simple Feynman rules \eqref{eq:sqed:matterloops} and \eqref{eq:sqed:vertices}, its expression reads
\ba
-\I \Pi_{1\lambda}(p) 
& = 4 \mu^{2\ve} \Nf e^2 S \int [\D^d k] \frac{\slashed{k}+m_\psi}{(k^2-m_\psi^2)((k-p)^2-m_\phi^2)} .
\label{LO:photino:def}
\ea
Then, using the projector \eqref{eq:sqed:projectorsgauge:lambda} and performing the fermionic trace, we have
\be
\Pi_{1\lambda}(p^2) = 4\I \mu^{2\eps} \frac{\Nf e^2}{p^2} S \int [\D^d k] \frac{k \cdot p }{k^2 (k-p)^2} ,  
\ee
and, after wick rotation, using the results of Appendix \ref{chap:multiloop}, it yields
\ba
\Pi_{1\lambda}(p^2) = -\frac{\Nf e^2}{(4\pi)^{3/2} \pE} \left(\frac{\overline{\mu}^{ 2}}{-p^2} \right)^\ep 2 S e^{\gamma_E \ep} G(d,1,1) .  
\label{LO:photino:tot}
\ea
Since this result is again finite, we set it exactly in $d=3$ and obtain
\ba
\Pi_{1\lambda}(p^2) = - \frac{S\Nf e^2}{4\pE} .  
\label{LO:lambda:tot:d=3}
\ea
Note that this result is relevant only in the SQED$_3$ case ($S=1$, $n=2$), reading
\ba
\Pi_{1\lambda}^{\text{SQED}_3}(p^2) = - \frac{\Nf e^2}{4\pE} ,  
\label{LO:photino:tot:d=3:sqed}
\ea
which is exactly equal to both the one-loop photon \eqref{eq:sqed:LOphoton} and $\ep$-scalar \eqref{LO:eps-scalar:tot:d=3} polarization functions. Summarizing, we find that for SQED$_3$, the photon, $\ep$-scalar and photino self-energies are all equal and finite at the LO of the $1/\Nf $-expansion, reading
\be
\Pi_{1\gamma}^{\text{SQED}_3}(p^2) = \Pi_{1\eps}^{\text{SQED}_3}(p^2) = \Pi_{1\lambda}^{\text{SQED}_3}(p^2) = - \frac{\Nf e^2}{4\pE} ,
\label{LO:photon+eps+photino:tot:d=3}
\ee
which is first-order perturbative proof that the polarizations are all equal in the gauge multiplet, as expected from SUSY. 
Moreover, the finiteness of the polarizations is a first-order perturbative proof that the theory has no anomalous dimensions for the gauge fields in accordance with the fact that it is a standing gauge theory, as previously advertised.

\subsubsection{IR-Softened Gauge Multiplet at LO}

We are now in a position to compute the softened gauge propagators at the leading order of the $1/\Nf$ expansion (see \eqref{eq:sqed:softenedproprconstruct}). By substituting the one-loop (LO) results obtained for the polarization of the photon \eqref{LO:photon:tot:d=3}, the $\eps$-scalar \eqref{LO:eps-scalar:tot} and the photino \eqref{LO:lambda:tot:d=3} into the definition of the dressed gauge propagators \eqref{propagators:dressed}, the propagators soften in the large-$\Nf $ limit, i.e., $\pE \ll \Nf e^2$, and read
\vspace{-6pt}
\bs
\label{eq:sqed:LOprop}
\ba
&\hat{D}_{1AA}^{\mu \nu}(p) = \GRAPHAprop{1} = {\frac{16\I}{(n+2S)\Nf e^2}}\frac{\hat{d}^{\mu\nu}(p)}{\pE} , \label{eq:sqed:LOprop:photon} \\
& \bar{D}_{1AA}^{\mu \nu}(p) = \GRAPHepsprop{1} = \frac{4\I\EP S}{\Nf e^2}\frac{\bar{g}^{ {\mu}{\nu}}}{\pE} , \\
& D_{1\lambda {\bar{\lambda}}}(p) = \GRAPHlambdaprop{1} =\frac{-4\I S}{\Nf e^2}\frac{\slashed{p}}{\pE} ,
\ea
\es
where the tensorial structure of the photon is still given by $\hat{d}^{\mu\nu}(p)=\hat{g}^{\mu\nu} -(1 - \xi)(\hat{p}^\mu \hat{p}^\nu/p^2)$ thanks to the use of the non-local gauge (see \eqref{eq:sqed:nonlocalgauge}). These new softened propagators can then be used to compute the LO self-energies of both the electron and its superpartner.

\subsection{Matter-Multiplet Self-Energies at LO}

In this section, we compute in detail the first correction to the self-energies of the matter multiplet, i.e., for the electron and the selectron, at the LO in the $1/\Nf$ expansion, i.e., at $\Ord(1/\Nf)$.

\subsubsection{Electron Self-Energy at LO}

We start with the electron propagator \eqref{eq:sqed:selfenergyparam} and compute its LO correction, which consists of three contributions 
\be
\Sigma_{1}^{\psi}(p) = \Sigma_{1}^{\psi(a)}(p) + \Sigma_{1}^{\psi(b)}(p) + \Sigma_{1}^{\psi(c)}(p) ,
\label{LO:fermion:1l:tot}
\ee
one for each gauge interaction, which are defined {as } 
\bs
\vspace*{0.5cm}
\label{LO:fermion:def}
\ba
\hspace{-1.7cm}
-\I \Sigma_{1}^{\psi(a)}(p) = \GRAPHLOpsia & = \mu^{2\ve} \int [\D^d k] \hat\Gamma_{0A\psi\bar\psi}^{\mu} S_{0\psi\bar\psi}(k) \hat\Gamma_{0A\psi\bar\psi}^{\nu} \hat D_{1AA,\mu\nu} ({p-k}) ,
\label{LO:fermion:def:a} \hspace{-1cm} \\
\hspace{-1.7cm}
-\I \Sigma_{1}^{\psi(b)}(p) = \GRAPHLOpsib & = \mu^{2\ve} \int [\D^d k] \bar\Gamma_{0A\psi\bar\psi}^\mu S_{0\psi\bar\psi}(k) \bar\Gamma_{0A\psi\bar\psi}^\nu \bar D_{1AA,\mu\nu}(p-k) ,
\label{LO:fermion:def:b} \hspace{-1cm} \\
\hspace{-1.7cm}
-\I \Sigma_{1}^{\psi(c)}(p) = \GRAPHLOpsic & = \mu^{2\ve} \int [\D^d k] \Gamma_{0{\bar\lambda\psi\phi^*}}S_{0\phi{\phi^*}}(k) \Gamma_{0{\bar\psi\lambda\phi}} 
D_{1\lambda\bar\lambda}(p-k) , \hspace{-1cm}
\label{LO:fermion:def:c}
\ea
\vspace*{-0.5cm}\\
\es
where the photon, $\ve$-scalar and photino propagators are indeed the IR-softened ones at first order \eqref{eq:sqed:LOprop}. Using the simple Feynman rules here is enough, i.e., using Equations \eqref{eq:sqed:gaugeprop}--\eqref{eq:sqed:vertices}, and we obtain
\bs
\label{LO:fermion:def:bis}
\ba
-\I \Sigma_{1}^{\psi(a)}(p)
& =-\frac{16\I}{(n+2S)}\frac{\mu^{2\ve}}{\Nf } \int [\D^d k] \frac{\hat\gamma^\mu(\slashed{k}+m_\psi)\hat\gamma^\nu \hat d_{\mu\nu}({p-k})}{(k^2-m_\psi^2)\abs{{p-k}}} ,
\label{LO:fermion:def:a:bis} \\
-\I \Sigma_{1}^{\psi(b)}(p)
& = - 4\I\EP S\frac{\mu^{2\ve}}{\Nf } \int [\D^d k] \frac{\bar\gamma^\mu(\slashed{k}-m_\psi)\bar\gamma^\mu}{(k^2-m_\psi^2)\abs{{p-k}}} ,
\label{LO:fermion:def:b:bis} \\
-\I \Sigma_{1}^{\psi(c)}(p)
& = \I 4 S \frac{\mu^{2\ve}}{\Nf} \int [\D^d k] \frac{\slashed{{p}}-\slashed{{k}}}{(k^2-m_\phi^{{2}})\abs{{p-k}}} .
\ea
\es
Note that in \eqref{LO:fermion:def:bis}, the (dimensionful) electric constant, $e$, drops out in favor of the new coupling $1/\Nf $ thanks to the softening of the gauge-multiplet propagators. 
These diagrams are then split into the part proportional to the external momentum, $p$, (also called the vectorial part since
it is proportional to $\slashed{p}$) and the one proportional to the mass, $m_\psi$, (also called the scalar part) using the projectors \eqref{eq:sqed:projectors}. First, focusing on the vectorial part, using the projector \eqref{eq:sqed:projectorsp} and computing these three diagrams with projection, trace calculation, wick rotation, integral evaluation and wick rotate back, we find the following exact results
\bs
\label{LO:fermion:res}
\ba
\Sigma_{1p}^{\psi(a)}(p^2) &= \frac{4}{(4\pi)^{3/2} \Nf } \left(\frac{\overline{\mu}^{ 2}}{-p^2} \right)^\ep \frac{4(d-2)}{n+2S} \left(\frac{d-1}{2d-3} - \xi \right) e^{\gamma_E \ep} G(d,1,1/2) , 
\label{LO:fermion:res:a} \\
\Sigma_{1p}^{\psi(b)}(p^2) &= \frac{4}{(4\pi)^{3/2} \Nf } \left(\frac{\overline{\mu}^{ 2}}{-p^2} \right)^\ep \frac{2(d-3)(d-2)\EP S}{n(2d-3)} e^{\gamma_E \ep} G(d,1,1/2) ,  
\label{LO:fermion:res:b} \\
\Sigma_{1p}^{\psi(c)}(p^2) &= -\frac{4}{(4\pi)^{3/2} \Nf } \left(\frac{\overline{\mu}^{ 2}}{-p^2} \right)^\ep \frac{(d-1)S}{2d-3} e^{\gamma_E \ep} G(d,1,1/2) ,  
\label{LO:fermion:res:c}
\ea
\es
where $\Sigma_{1p}^{\psi(b)}$ is finite due to the $\eps$-scalar, while the two other contributions are singular in the limit $d \to 3$.
Secondly, focusing on the scalar part, using the projector \eqref{eq:sqed:projectorsm} and computing these three diagrams with the same approach yields the following exact results
\bs
\label{LO:fermion:res:bis}
\ba
\Sigma_{1m}^{\psi(a)}(p^2) &= \frac{4}{(4\pi)^{3/2} \Nf } \left(\frac{\overline{\mu}^{ 2}}{-p^2} \right)^\ep \frac{4(d-1+\xi)}{n+2S} e^{\gamma_E \ep} G(d,1,1/2) , \\
\Sigma_{1m}^{\psi(b)}(p^2) &= -\frac{4}{(4\pi)^{3/2} \Nf } \left(\frac{\overline{\mu}^{ 2}}{-p^2} \right)^\ep \frac{2 (d-3)\EP S }{n} e^{\gamma_E \ep} G(d,1,1/2) , \\
\Sigma_{1m}^{\psi(c)}(p^2) &= 0 ,  
\ea
\es
where the first contribution is singular in the limit $d\to3$, while the second diagram vanishes in $d=3$, and the last graph $(c)$ is exactly zero because of the gamma matrix trace. 
Summing all the contributions, the total vectorial and scalar electron self-energies are therefore, given, expanded in $d=3-2\eps$, by
\bs
\ba
& \hspace{-1.7cm}
\Sigma_{1p}^{\psi}(p^2) = \frac{4}{3(n+2S)\pi^2 \Nf } \left(\frac{\overline{\mu}^{ 2}}{-4p^2} \right)^\ep \left(\frac{2 - 3 \xi- 2 S }{\eps} +\frac{2}{3}\Big(7-(13+3\EP)-9\xi\Big) + \Ord(\ep) \right) , \hspace{-1cm} \\
& \hspace{-1.7cm} \Sigma_{1m}^{\psi}(p^2) = \frac{4}{3(n+2S)\pi^2 \Nf } \left(\frac{\overline{\mu}^{ 2}}{-4p^2} \right)^\ep \left(\frac{3(2+\xi)}{\eps} +6(3+\EP S +2\xi) + \Ord(\ep) \right) . \hspace{-1cm}
\ea
\label{LO:fermion:exp}
\es
Note that some $\log(2)$ are resummed by adding a 4 next to the momentum $p^2$. 
From this result, we extract straightforwardly, with \eqref{eq:sqed:ZSigmadef}, the LO electron wave function and mass renormalization
\be
Z_{\psi} =1{+}\frac{{4} (2-3 \xi -2 S)}{3 (n+2S) \pi^2 \Nf \eps } +\Ord(1/\Nf^2) , \qquad 
Z_{m} =1 + \frac{{8} (S-4)}{3 (n+2S) \pi^2 \Nf \eps }+\Ord(1/\Nf^2) .
\ee
As expected, the general mass renormalization factor is completely gauge-invariant, which is a strong check on our results. From the definition of the anomalous dimension \eqref{eq:gammadef}, we derive the general anomalous dimensions
\be
\gamma_{\psi} = {-}\frac{{8} (2-3 \xi -2 S)}{3 (n+2S) \pi^2 \Nf } + \Ord(1/\Nf ^2) , \qquad 
\gamma_{m_\psi} = \frac{{16 (4-S)}}{3 (n+2S) \pi^2 \Nf } + \Ord(1/\Nf ^2) .
\label{LO:fermion:gammapsi:LO}
\ee
In the relevant cases of SQED$_3$ ($S=1$, $n=2$) and fQED$_3$ ($S=0$, $n=2$), the general results simplify as
\bs
\label{eq:sqed:resonelooppsi}
\ba
&& & \gamma_\psi^{\text{SQED}_3} = \frac{2\xi}{\pi^2\Nf}+ \Ord(1/\Nf ^2) , 
&& \gamma_{m_\psi}^{\text{SQED}_3} = \frac{4}{\pi^2\Nf} + \Ord(1/\Nf ^2) , && 
\label{eq:sqed:resonelooppsisqed}\\
&& & \gamma_\psi^{\text{fQED}_3} = -\frac{4(2-3\xi)}{3\pi^2\Nf} + \Ord(1/\Nf ^2) , 
&& \gamma_{m_\psi}^{\text{fQED}_3} = \frac{32}{3\pi^2\Nf}+ \Ord(1/\Nf ^2) . &&
\label{eq:sqed:resonelooppsifqed}
\ea
\es
Note that $\gamma_{\psi}^{\text{SQED}_3}$ vanishes in the Landau gauge ($\xi=0$), which is then the so-called ``good gauge'' at LO. This is to be contrasted with the non-supersymmetric case, $\gamma_{\psi}^{\text{fQED}_3}$, (first obtained in \cite{Gracey:1993iu}) that vanishes in the so-called Nash gauge \cite{Nash:1989xx}, $\xi=2/3$. Note also that the bQED$_3$ case is obviously irrelevant here since we consider the anomalous dimension of the electron field and mass. We will further discuss the quantities \eqref{eq:sqed:resonelooppsi} once we obtain their supersymmetric counterpart in the next section and their NLO correction after that. 

\subsubsection{Selectron Self-Energy at LO}

We proceed similarly for the electron superpartner, the selectron, which is the scalar propagator and compute its LO scalar self-energy, which consists of the sum of the two diagrams 
\be
\Sigma_{1}^\phi(p) = \Sigma_{1}^{\phi(a)}(p) + \Sigma_{1}^{\phi(b)}(p) ,
\label{LO:scalar:1l:tot}
\ee
that are defined as
\bs
\vspace*{0.25cm}
\label{LO:scalar:def}
\ba
& \hspace{-1.1cm} -\I \Sigma_{1}^{\phi(a)}(p) = \GRAPHLOphia = 
\mu^{2\ve} \int [\D^d k] \hat\Gamma_{0A\phi{\phi^*}}^\mu(k,p) S_{0\phi{\phi^*}}(k) \hat\Gamma_{0A\phi{\phi^*}}^\nu(p,k) \hat D_{1AA,\mu\nu}(k-p) ,
\label{LO:scalar:def:a} \hspace{-1cm} \\
& \hspace{-1.1cm} -\I \Sigma_{1}^{\phi(b)}(p) = \GRAPHLOphib = -\mu^{2\ve} \int [\D^d k] \Tr \Big[\Gamma_{0{\bar\psi\lambda\phi}} S_{0\psi\bar\psi}(k) \Gamma_{0{\bar\lambda\psi\phi^*}} D_{1\lambda\bar\lambda}(k-p) \Big] , \hspace{-1cm}
\label{LO:scalar:def:b} 
\ea
\vspace*{-0.5cm}\\
\es
where the photon and photino propagators are indeed the IR-softened ones \eqref{eq:sqed:LOprop} and $\Sigma_{1b}^\phi$ contains a hybrid (Dirac/Majorana) fermion loop. Note that, for the diagram (b), we can assign a counter-clockwise fermion flow and momentum flows that follows the fermion loop consisting of the Dirac and Majorana fermions. Therefore, using the simple Feynman rules given by \eqref{eq:sqed:gaugeprop} to \eqref{eq:sqed:vertices} is indeed enough and leads to
\bs
\label{LO:scalar:def:bis}
\ba
-\I \Sigma_{1}^{\phi(a)}(p) & = -\frac{16\I S}{n+2S} \frac{\mu^{2\eps}}{\Nf} \int [\D^d k] \frac{(\hat k+\hat p)^\mu(\hat k+\hat p)^\nu \hat d_{\mu\nu}(k-p)}{(k^2-m_\phi^2)\abs{k-p}} , \label{LO:scalar:def:a:bis} \\ 
-\I \Sigma_{1}^{\phi(b)}(p) & = -4\I S\frac{\mu^{2\eps}}{\Nf} \int [\D^d k] \frac{\Tr \big[(\slashed{k}+m_\psi)(\slashed{k}-\slashed{p})\big]}{(k^2-m_\psi^2)\abs{k-p}} .
\label{LO:scalar:def:b:bis} 
\ea
\es
Performing the traces and using the projection \eqref{eq:sqed:projectorsm} yields the two LO contributions to the momentum part of the selectron self-energy
\bs
\label{LO:scalar:res}
\ba
\Sigma_{1p}^{\phi(a)}(p^2) &= \frac{4e^{\gamma_E \ep}}{(4\pi)^{3/2} \Nf } \left(\frac{\overline{\mu}^{ 2}}{-p^2} \right)^\ep {\frac{4 S}{(n+2S)}} \left(\frac{4(d-1)(d-2)}{2d-3} - (2d-5) \xi \right) G(d,1,1/2) ,
\label{LO:scalar:res:a} \\
\Sigma_{1p}^{\phi(b)}(p^2) &= -\frac{4e^{\gamma_E \ep}}{(4\pi)^{3/2} \Nf } \left(\frac{\overline{\mu}^{ 2}}{-p^2} \right)^\ep \frac{n(d-2)S}{2d-3} G(d,1,1/2) ,
\label{LO:scalar:res:b} 
\ea
\es
as well as the two LO contributions to the mass part of the selectron self-energy
\bs
\label{LO:scalar:res:m}
\ba
\hspace{-0.5cm} \Sigma_{1m}^{\phi(a)}(p^2) &=- \frac{4e^{\gamma_E \ep}}{(4\pi)^{3/2} \Nf } \left(\frac{\overline{\mu}^{ 2}}{-p^2} \right)^\ep \frac{2S}{n+2S} \Bigg(4(d-3)(d-1) -\frac{2d-5}{d-4}\left(2 d^2-13 d+19\right) \xi \Bigg)\nonum \\[-0.25cm]
& \hspace{9cm} \times G(d,1,1/2) ,
\label{LO:scalar:res:a:m} \\[-0.25cm]
\hspace{-0.5cm} \Sigma_{1m}^{\phi(b)}(p^2) &= \frac{4e^{\gamma_E \ep}}{(4\pi)^{3/2} \Nf } \left(\frac{\overline{\mu}^{ 2}}{-p^2} \right)^\ep \frac{n(d-1)S}{2} G(d,1,1/2) .
\label{LO:scalar:res:b:m} 
\ea
\es
The total selectron momentum and mass self-energy, in $\ep$-expanded form, yields 
\bs
\ba
& \Sigma_{1p}^{\phi}(p^2) = \frac{4{S}}{3(n+2S)\pi^2 \Nf } \left(\frac{\overline{\mu}^{ 2}}{-4p^2} \right)^\ep \left(\frac{8-3\xi-n}{\eps} + \frac{2}{3}\Big(28-5n\Big) + \Ord(\ep) \right) , \\
& \Sigma_{1m}^{\phi}(p^2) = \frac{4{S}}{3(n+2S)\pi^2 \Nf } \left(\frac{\overline{\mu}^{ 2}}{-4p^2} \right)^\ep \left(\frac{3(\xi+n)}{\eps} + 3 (8-3\xi+3n) + \Ord(\ep) \right) .
\label{LO:scalar:exp}
\ea
\es
From these results, we extract the LO scalar wave function and mass renormalization
\be
Z_{\phi} =1+ \frac{4 (8-3\xi-n)S}{3 (n+2) \pi^2 \Nf \eps} +\Ord(1/\Nf^2) , \qquad 
Z_{m_\phi} =1 -\frac{4 (n+4) S}{3 (n+2) \pi ^2 \Nf \eps } +\Ord(1/\Nf^2) . 
\ee
As expected from such a physical quantity, the general mass renormalization factor is completely gauge-invariant. We can now derive the anomalous dimensions for the selectron field and mass using the definition \eqref{eq:gammadef}, yielding
\be
\gamma_{\phi} =-\frac{8 (8-3\xi-n)S}{3 (n+2) \pi^2 \Nf} + \Ord(1/\Nf ^2) , \qquad
\gamma_{m_\phi} = \frac{8 (n+4) S}{3 (n+2) \pi ^2 \Nf} + \Ord(1/\Nf ^2) ,
\label{LO:scalar:gammapsi}
\ee
Note that in the relevant cases of SQED$_3$ ($S=1$, $n=2$) and bQED$_3$ ($S=1$, $n=0$), the above general results simplify as
\bs
\ba
&&& \gamma_{\phi}^{\text{SQED}_3} =-\frac{2 (2-\xi)}{\pi^2 \Nf} + \Ord(1/\Nf ^2) , \qquad
&&\gamma_{m_\phi}^{\text{SQED}_3} = \frac{4}{\pi ^2 \Nf} + \Ord(1/\Nf ^2) , && 
\label{eq:sqed:resoneloopphisqed}\\
&&& \gamma_{\phi}^{\text{bQED}_3} =-\frac{4 (8-3\xi)}{3\pi^2 \Nf} + \Ord(1/\Nf ^2) , \qquad
&&\gamma_{m_\phi}^{\text{bQED}_3} = \frac{16}{3\pi^2 \Nf} + \Ord(1/\Nf ^2) . &&
\label{eq:sqed:resoneloopphibqed}
\ea
\es
A few remarks are necessary here. 
First, we observe that for SQED$_3$, the mass anomalous dimension for the selectron \eqref{eq:sqed:resoneloopphisqed} is identical to the one of the electron \eqref{eq:sqed:resonelooppsisqed}, i.e., 
\be
\gamma_{m_\psi}^{\text{SQED}_3}=\gamma_{m_\phi}^{\text{SQED}_3} = \frac{4}{\pi ^2 \Nf} + \Ord(1/\Nf ^2) ,
\ee
as expected from supersymmetry.
In striking contrast, the field anomalous dimensions for the selectron \eqref{eq:sqed:resoneloopphisqed} and the electron \eqref{eq:sqed:resonelooppsisqed} are different. This is due to the use of a gauge-fixing term that breaks supersymmetry (Wess--Zumino gauge). This is not an issue since the breaking of SUSY will occur only for gauge-dependent quantities that are, by definition, non-physical. Secondly, let us remark that in the SQED$_3$ case, the field anomalous dimension of the selectron vanishes for $\xi=2$. Since for the fermionic part of SQED$_3$, the good gauge was the Landau gauge $\xi=0$, and it is, therefore, not possible to cancel both matter-field anomalous dimensions at the same time. This may cause trouble for computations of the critical properties of the model using the Schwinger--Dyson equations (see the devoted Section \ref{sec:sqed:crit}). As for the bQED case \eqref{eq:sqed:resoneloopphibqed}, we see that the good gauge is then $\xi=8/3$. We will further discuss these results again after improving them to NLO.

\subsection{Vanishing Contributions and Generalized Furry Theorem}
\label{sec:sqed:vanishingcontributions}

Before going to higher orders and computing any NLO diagrams, we first need to discuss some additional diagrams that may enter the incoming NLO computations as subdiagrams. These LO diagrams are made of matter bubbles and triangles and are of uttermost interest because a lot of them are vanishing, either exactly or in pairs. On the one hand, this will tremendously reduce the number of diagrams to be computed at NLO but will also ensure that matter bubbles are connected to each other in a way suitable for the large-$\Nf$ expansion.

\begin{figure}[h]
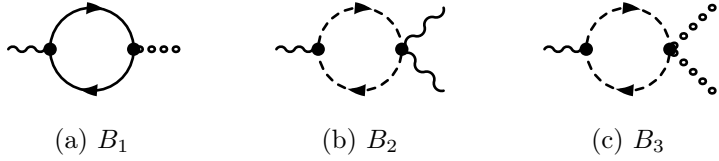

\centering
\begin{subfigure}[b]{0.2\textwidth}
\centering
\GRAPHLOAeps
\vspace{-0.3cm}
\caption{$B_1$}
\label{fig:sqed:vanish1}
\end{subfigure}
\begin{subfigure}[b]{0.2\textwidth}
\centering
\GRAPHLOAAA
\vspace{-0.3cm}
\caption{$B_2$}
\label{fig:sqed:vanish2}
\end{subfigure}
\begin{subfigure}[b]{0.2\textwidth}
\centering
\GRAPHLOAepseps
\vspace{-0.3cm}
\caption{$B_3$}
\label{fig:sqed:vanish3}
\end{subfigure}
\caption{Exactly vanishing one-loop bubble diagrams}
\label{fig:sqed:vanish}
\end{figure}

We first focus on three exactly-vanishing one-loop diagrams made of a matter bubble, as shown in Figure \ref{fig:sqed:vanish}.

\newpage

\noindent The first vanishing bubble contribution is the mixed polarization $B_1$ (ongoing photon and outgoing $\eps$-scalar), as displayed in Figure \ref{fig:sqed:vanish}a. Since this diagram is proportional to $\Tr(\bar\gamma^{\mu})=0$, it reads $B_1=0$. More generally, we conjecture that every diagram with an odd number of $\eps$-scalar external lines is exactly zero. The two other contributions, given by Figure \ref{fig:sqed:vanish}b,c, are also exactly vanishing by parity on the internal momentum integral, i.e., $B_2=B_3=0$.

\begin{figure}[h!]
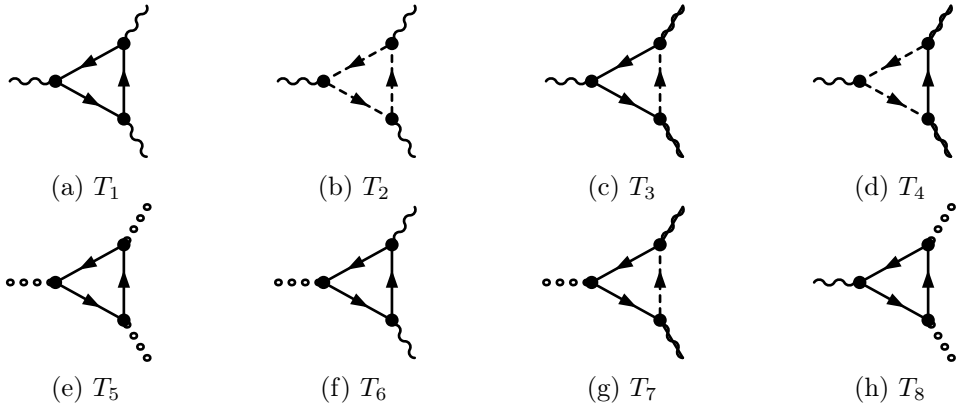

\centering
\begin{subfigure}[b]{0.20\textwidth}
\centering
\GRAPHLOtria
\caption{$T_1$}
\end{subfigure}
\begin{subfigure}[b]{0.20\textwidth}
\centering
\GRAPHLOtrib
\caption{$T_2$}
\end{subfigure}
\begin{subfigure}[b]{0.20\textwidth}
\centering
\GRAPHLOtrie
\caption{$T_3$}
\end{subfigure}
\begin{subfigure}[b]{0.20\textwidth}
\centering
\GRAPHLOtrif
\caption{$T_4$}
\end{subfigure}
\\
\begin{subfigure}[b]{0.20\textwidth}
\centering
\GRAPHLOtrih
\caption{$T_5$}
\end{subfigure}
\begin{subfigure}[b]{0.20\textwidth}
\centering
\GRAPHLOtric
\caption{$T_6$}
\end{subfigure}
\begin{subfigure}[b]{0.20\textwidth}
\centering
\GRAPHLOtrig
\caption{$T_7$}
\end{subfigure}
\begin{subfigure}[b]{0.20\textwidth}
\centering
\GRAPHLOtrid
\caption{$T_8$}
\end{subfigure}
\caption{All vanishing one-loop matter triangles in gQED$_3$.}
\label{fig:sqed:mattertrianglevanish}
\end{figure}

Multiple other vanishing contributions come from matter triangles. These are built from triangles of electrons and selectrons together with external legs of any allowed kinds, i.e., taken in the gauge multiplet. In total, there are 8 triangles (disregarding the possible charge flows), as shown in Figure \ref{fig:sqed:mattertrianglevanish}. In the following, we will briefly describe why all of these diagrams (that are proportional to $\Nf$) are vanishing.

The first one is the pure fermionic QED matter triangle diagram, $T_1$, and is vanishing because it always appears paired up with its mirror conjugate diagram with opposite charge/matter flow. An explicit computation is a check at the first order of the Furry theorem in QED \cite{Furry:1937zz}, which is the all-order proof that in QED, and any diagram with an odd number of photon legs can be discarded since they will cancel with their opposite flow diagrams, as a direct consequence of the conservation of energy and charge conjugation symmetry. In the following, we will discuss how the Furry theorem, generalized for gQED$_3$, holds at least at the leading order, i.e., for diagrams made of matter triangles and three external legs taken in the gauge multiplet. The second vanishing contribution is the pure bQED diagram, $T_2$, and also vanishes with its opposite charge flow counterpart. It, therefore, generalizes Furry's theorem to the case of bQED$_3$. The third and fourth vanishing diagrams are the supersymmetric triangles made of mixed electron and selectrons, thereby generalizing Furry's theorem to SQED$_3$ without $\eps$-scalars. Finally, as for the diagrams containing $\eps$-scalars, we have the three contributions $T_5$, $T_6$ and $T_7$ that are exactly zero for any momentum or charge flow direction. This is because they contain an odd number of $\eps$-scalar external legs, i.e., they are ultimately related to $\Tr(\bar\gamma^\mu)=0$. We are left with a last triangle, $T_8$, made of an electron loop together with one photon plus two $\eps$-scalars external legs. This diagram is different since it is not vanishing because of $\eps$-scalars, as in this case it is proportional to $\Tr(\bar\gamma^\mu\bar\gamma^\nu) \neq 0$. We then have to consider the diagrams with the two opposite charge flows, that also vanishes upon explicit computation.

We have to then check explicitly that every matter triangle does indeed vanish, either exactly or with respect to their (reversed matter flow) twin diagrams. This completes the perturbative leading order proof that the generalized Furry theorem holds in SQED$_3$ withing the DRED scheme, and, therefore, as subcases, also in fQED$_3$ and bQED$_3$. This means that every diagram containing a matter triangle can be set to zero in gQED$_3$, i.e., in SQED$_3$, fQED$_3$ and bQED$_3$. Some prominent examples of diagrams that we can drop are the large number of Aslamazov--Larkin-type diagrams. Taking into account these various vanishing contributions tremendously reduces the number of diagrams that has to be computed at NLO. Indeed, as we shall see in the following, it ensures that up to NLO, not a single diagram of three-loop type needs to be computed. This is crucial because the three-loop master integrals with half integer indices are still unknown and are a big challenge to compute due to the inherent branch-cut structure of the integrals, which results in intricate hypergeometric functions and transcendental numbers (Catalan number, Clausen function, etc.; see, e.g., \cite{Kazakov:1983,Kotikov:1996}). Moreover, the generalized Furry theorem at LO also guarantees that matter loops are connected by simple chains of force field propagators, like in the simpler fQED$_3$ case, in accordance with our starting assumption, ensuring that the large-$\Nf$ expansion is reliable. We can now go forward and proceed with the NLO computations.

\subsection{Gauge-Multiplet Polarizations at NLO}

In this section, we compute the NLO polarizations of the gauge multiplet, i.e., for the photon, the $\eps$-scalar and the photino at NLO in the $1/\Nf$ expansion, i.e., at $\Ord(1/\Nf^0)$. We will show that all of these polarizations are finite and gauge-invariant for gQED$_3$. In the following, we shall use shorthand notation for the polarization results
\be
\Pi_{2X}^{(y)}(p^2)= \frac{e^2}{2(n+2S)\pE} \left(\frac{\overline{\mu}^{2}}{-p^2} \right)^{2\eps} \tilde{\Pi}_{2X}^{(y)} , \quad \forall X\in\{\gamma,\eps,\lambda\} .
\ee

\subsubsection{Photon Polarization at NLO }

We first consider the NLO correction to the photon polarization that consists of $20$ Feynman diagrams labeled $(a,b,\dots,t)$.
Taking into account the fact that mirror conjugate graphs take the same value, we are left with 
$11$ distinct graphs to evaluate. This can be conducted exactly for all the diagrams, following the same procedure as for the one-loop case. Their expressions read
\bs
\label{eq:sqed:NLO:photon}
\ba
\frac{1}{2}\times\GRAPHNLOAa &~:~\tilde\Pi_{2\gamma}^{(a)} = -\frac{4S(2+\xi)}{\pi^2}+\Ord(\eps) ,  \\
4\times\GRAPHNLOAbcde &~:~\tilde\Pi_{2\gamma}^{(bcde)} = \frac{16S}{3\pi^2}\left(\frac{1}{\eps}+\frac{19}{3}+\frac{3\xi}{2}\right)+\Ord(\eps) , \\
2\times\GRAPHNLOAfg &~:~\tilde\Pi_{2\gamma}^{(fg)} = -\frac{2S}{3\pi^2}\left(\frac{8-3\xi}{\eps}+\frac{128}{3}-9\xi\right)+\Ord(\eps) ,  \\
\GRAPHNLOAh &~:~\tilde\Pi_{2\gamma}^{(h)} = -\frac{2S}{\pi^2}\left(\frac{\xi}{\eps}+\frac{70}{9}-3\zeta_2+5\xi\right)+\Ord(\eps) , \\
2\times\GRAPHNLOAij &~:~\tilde\Pi_{2\gamma}^{(ij)} = -\frac{n}{3\pi^2}\left(\frac{2-3\xi}{\eps}+\frac{14}{3}-6\xi\right)+\Ord(\eps) ,  \\
\GRAPHNLOAk &~:~\tilde\Pi_{2\gamma}^{(k)} = {\frac{n}{3\pi^2}}\left(\frac{2-3\xi}{\eps}-\frac{32}{3}+9\zeta_2-6\xi\right)+\Ord(\eps) , \\
2\times\GRAPHNLOAlm &~:~\tilde\Pi_{2\gamma}^{(lm)} = \frac{4S\mathcal{E}}{3\pi^2}+\Ord(\eps) ,  \\
\GRAPHNLOAn &~:~\tilde\Pi_{2\gamma}^{(n)} = -\frac{4S\mathcal{E}}{3\pi^2}+\Ord(\eps) , \\
2\times\GRAPHNLOAop &~:~\tilde\Pi_{2\gamma}^{(op)} = \frac{2nS}{3\pi^2}\left(\frac{1}{\eps}+\frac{13}{3}\right)+\Ord(\eps) , \\
2\times\GRAPHNLOAqr &~:~\tilde\Pi_{2\gamma}^{(qr)} = \frac{2nS}{3\pi^2}\left(\frac{1}{\eps}+\frac{19}{3}\right)+\Ord(\eps) , \\
2\times\GRAPHNLOAst &~:~\tilde\Pi_{2\gamma}^{(st)} = -\frac{4nS}{3\pi^2}\left(\frac{1}{\eps}+\frac{11}{3}\right)+\Ord(\eps) .  
\ea
\es
Summing all the contributions \eqref{eq:sqed:NLO:photon}, all poles cancel, and the final result is finite, reading
\be
\tilde\Pi_{2\gamma}={\frac{1}{18 \pi ^2}} \Big(\left(40 S-92+9 \pi ^2\right)n-2 \left(164-9 \pi ^2\right) S\Big)+\Ord(\eps) .
\ee
Several remarks are in order here. First, the $\eps$-scalars do not contribute here (the corresponding tracking factor, $\mathcal{E}$, is absent) because the two contributions, $\tilde\Pi_{2lm}^\gamma$ and $\tilde\Pi_{2n}^\gamma$, cancel each other. Second, the result is completely gauge-invariant, which provides a strong check on our result. Lastly, the finiteness of the results ensures that the theory is still standing at NLO, i.e., the coupling does not renormalize. 

Since the NLO result is finite and has the same form as the LO one \eqref{LO:photon:tot:d=3}, we can write the photon polarization in the form
\be
\Pi_\gamma(p^2) = \Pi_{1\gamma}(p^2) \bigg[ 1 + \frac{C_\gamma}{\Nf} + \Ord\big(1/\Nf^2\big) \bigg] , 
\quad \text{recalling} \quad \Pi_{1\gamma}(p^2)= -\frac{(n+2S)\Nf e^2}{16\pE} ,
\ee
and where the interaction correction coefficient to the photon polarization reads 
\be
C_\gamma= \frac{4n(92-9\pi^2)}{9(n+2S)^2\pi^2}+\frac{8(164-20n-9\pi^2)S}{9(n+2)^2\pi^2} .
\ee
In the different cases of interest, i.e., SQED$_3$ ($S=1$, $n=2$), fQED$_3$ ($S=0$, $n=2$) and bQED$_3$ ($S=1$, $n=0$), it yields the correction coefficients
\be
C_\gamma^{\text{SQED}_3} = \frac{12-\pi^2}{\pi^2} , \qquad 
C_\gamma^{\text{fQED}_3} = \frac{2(92-9\pi^2)}{9\pi^2} , \qquad 
C_\gamma^{\text{bQED}_3} = \frac{2(164-9\pi^2)}{9\pi^2} . \qquad 
\label{eq:Cgamma}
\ee

As advertised in the previous Section 
\ref{sec:sqed:vanishingcontributions}, it turns out that all the diagrams considered are two-loop. Indeed, since we are in the large-$\Nf$ expansion, higher loop diagrams could have contributions at the same order in $1/\Nf$. However, this is fortunately not the case. As proof, we have explicitly checked that, up to NLO, no three-loop diagram contributes to the photon polarization, either because they contain a vanishing contribution (see Section \ref{sec:sqed:vanishingcontributions}) or because they are of order $1/\Nf^2$ or higher. This requires the check of 361 diagrams in an automated way. This is conducted by generating the expressions for each diagram and then computing only what 
is in the order of $1/\Nf$, as well detecting subdiagram expressions that vanish because of the generalized Furry's theorem.

\subsubsection{$\eps$-Scalar Polarization at NLO}

We now consider the NLO correction to the $\eps$-scalar polarization that consists of nine Feynman diagrams labeled $(a,b,\dots,i)$.
Taking into account the fact that mirror conjugate graphs take the same value, we are left with 
six distinct graphs to evaluate. This can be conducted exactly for all the diagrams and reads 
\bs
\ba
\frac{1}{2}\times\GRAPHNLOepsa &~:~\tilde\Pi_{2\eps}^{(a)} = -\frac{8S\mathcal{E}}{\pi^2}+\Ord(\eps) ,  \\
2\times\GRAPHNLOepsbc &~:~\tilde\Pi_{2\eps}^{(bc)} = \frac{8S\mathcal{E}}{3\pi^2}+\Ord(\eps) ,  \\
\GRAPHNLOepsd &~:~\tilde\Pi_{2\eps}^{(d)} = \frac{8S\mathcal{E}}{\pi^2}\left(\frac{1}{\eps}+2\right)+\Ord(\eps) , \\
2\times\GRAPHNLOepsef &~:~\tilde\Pi_{2\eps}^{(ef)} = -\frac{4S\mathcal{E}}{3\pi^2}\left(\frac{2-3\xi}{\eps}+\frac{20}{3}-9\xi\right)+\Ord(\eps) , \\
\GRAPHNLOepsg &~:~\tilde\Pi_{2\eps}^{(g)} = -\frac{4S\mathcal{E}}{\pi^2}\left(\frac{2+\xi}{\eps}+10-3\zeta_2+3\xi\right)+\Ord(\eps) , \\
2\times\GRAPHNLOepshi &~:~\tilde\Pi_{2\eps}^{(hi)} = \frac{8S\mathcal{E}}{3\pi^2}\left(\frac{1}{\eps}+\frac{16}{3}\right)+\Ord(\eps) . 
\ea
\es
Summing all the contributions yields the complete result
\be
\tilde\Pi_{2\eps}=-\frac{2(12-\pi^2)S\mathcal{E}}{\pi^2}+\Ord(\eps) ,
\ee
which is, as expected, completely gauge-invariant and finite, providing a strong check on our result. Similar to the photon case, we can rewrite the LO + NLO $\eps$-scalar polarization as
\be
\Pi_\eps(p^2) = \Pi_{1\eps}(p^2) \bigg[ 1 + \frac{C_\eps}{\Nf} + \Ord\big(1/\Nf^2\big) \bigg] ,
\qquad \text{recalling} \qquad
\Pi_{1\ep}(p^2) = - \frac{\EP S\Nf e^2}{4\pE} ,
\ee
and where the interaction correction coefficient to the $\eps$-scalar polarization reads
\be
C_\eps=\frac{(12-\pi^2)S\EP}{\pi^2} .
\label{eq:Ceps}
\ee
Note that in the only case of interest here, SQED$_3$ ($S=1$, $n=2$), this result trivially simplifies as
\be
C_\eps^{\text{SQED}_3}=\frac{12-\pi^2}{\pi^2} ,
\ee
which is exactly the same result as the photon correction coefficient in the SQED$_3$ case, as shown in \eqref{eq:Cgamma}, as expected from such a supersymmetric gauge-invariant quantity. 

Again, we also have explicitly checked that none of the 147 three-loop diagrams contributes to the $\eps$-scalar polarization thanks to the generalized Furry theorem and the already resummed one-loop contributions.

\subsubsection{Photino Polarization at NLO}

The last polarization to consider is the NLO correction to the photino polarization that consists of $14$ Feynman diagrams labeled $(a,b,\dots,n)$. Taking into account the fact that mirror conjugate graphs take the same value, we are left with seven distinct graphs to evaluate. This can be conducted exactly for all the diagrams and reads 
\bs
\ba
2\times\GRAPHNLOlambdaab &~:~\tilde\Pi_{2\lambda}^{(ab)} = -\frac{2S}{3\pi^2}\left(\frac{8-3\xi}{\eps}+\frac{80}{3}-3\xi\right)+\Ord(\eps) , \\
2\times\GRAPHNLOlambdacd &~:~\tilde\Pi_{2\lambda}^{(cd)} = -\frac{2S}{3\pi^2}\left(\frac{2-3\xi}{\eps}+\frac{8}{3}-3\xi\right)+\Ord(\eps) , \\
2\times\GRAPHNLOlambdaef &~:~\tilde\Pi_{2\lambda}^{(ef)} = -\frac{4S}{\pi^2}\left(\frac{\xi}{\eps}+6-3\zeta_2+\xi\right)+\Ord(\eps) , \\
2\times\GRAPHNLOlambdagh &~:~\tilde\Pi_{2\lambda}^{(gh)} = \frac{4S\mathcal{E}}{3\pi^2}+\Ord(\eps) , \\
2\times\GRAPHNLOlambdaij &~:~\tilde\Pi_{2\lambda}^{(ij)} = \frac{4S}{3\pi^2}\left(\frac{1}{\eps}+\frac{13}{3}\right)+\Ord(\eps) , \\
2\times\GRAPHNLOlambdakl &~:~\tilde\Pi_{2\lambda}^{(kl)} = \frac{4S}{3\pi^2}\left(\frac{1}{\eps}+\frac{10}{3}\right)+\Ord(\eps) , \\
2\times\GRAPHNLOlambdamn &~:~\tilde\Pi_{2\lambda}^{(mn)} = \frac{4S}{\pi^2}\left(\frac{1}{\eps}+2\right)+\Ord(\eps) . 
\ea
\es
Summing all the contributions yields the gauge-invariant and finite result
\be
\tilde\Pi_{2\lambda}=\frac{2S(3\pi^2-38+2\mathcal{E})}{3\pi^2}+\Ord(\eps) .
\ee
Note that this result depends non-trivially on the parameter $\EP$, which implies that the $\eps$-scalars are crucial here to ensure that the result is correct, as we will see in the following. The LO + NLO result for the photino polarization can again be written in the form
\be
\Pi_\lambda(p^2) = \Pi_{1\lambda}(p^2) \bigg[ 1 + \frac{C_\lambda}{\Nf} + \Ord\big(1/\Nf^2\big) \bigg] ,
\qquad \text{recalling} \qquad
\Pi_{1\lambda}(p^2)= - \frac{S\Nf e^2}{4\pE} ,
\ee
where the interaction coefficient to the photino polarization reads
\be
C_\lambda=\frac{(38-2\EP -3\pi^2)S}{3\pi^2} .
\label{eq:Clambda}
\ee
Note that this result is only of interest in the case of SQED$_3$ ($S=1$, $n=2$), where it reduces to
\be
C_\lambda^{\text{SQED}_3}=\frac{12-\pi^2}{\pi^2} ,
\ee
provided that we allow for the $\eps$-scalars ($\EP=1$). This is again exactly the same result as for the photon and the $\eps$-scalar correction coefficient in SQED$_3$. Therefore, we have explicitly checked that, up to NLO,
\be
\Pi_\gamma^{\text{SQED}_3}=\Pi_\eps^{\text{SQED}_3}=\Pi_\lambda^{\text{SQED}_3} ,
\ee
meaning that all polarizations of the gauge multiplet are equal, as expected from supersymmetry for such gauge-invariant and finite quantities.

Again, we also have explicitly checked that none of the 234 three-loop diagrams contribute to the photino polarization thanks to the generalized Furry theorem and the already resummed one-loop contributions.

\subsubsection{IR-Softened Gauge Multiplet at NLO}

We are now in a position to compute the NLO-softened propagators, i.e., of order $1/\Nf^2$. Their expressions read
\bs
\label{eq:sqed:NLOprop}
\ba
&\hat{D}_{2AA}^{\mu \nu}(p) = \GRAPHAprop{2} = {\frac{-16\I C_\gamma}{(n+2S)\Nf^2 e^2}}\frac{\hat{d}^{\mu\nu}(p)}{\pE} , \\
& \bar{D}_{2AA}^{\mu \nu}(p) = \GRAPHepsprop{2} = \frac{-4\I\EP S C_\eps}{\Nf^2 e^2}\frac{\bar{g}^{ {\mu}{\nu}}}{\pE} , \\
& D_{2\lambda {\bar{\lambda}}}(p) = \GRAPHlambdaprop{2} =\frac{4\I S C_\lambda}{\Nf^2 e^2}\frac{\slashed{p}}{\pE} ,
\ea
\es
where we take the infra-red limit $\pE\ll e^2 \Nf$, as advertised in \eqref{eq:sqed:IRlimits}. 
Interestingly, we observe the nice property that the LO \eqref{eq:sqed:LOprop} and NLO \eqref{eq:sqed:NLOprop}-softened gauge-multiplet propagators are simply related via their polarization correction coefficients, i.e.,
\bs
\label{eq:sqed:LONLOrelationgeugeprop}
\ba
&\hat{D}_{2AA}^{\mu \nu}(p) = -C_\gamma \times \hat{D}_{1AA}^{\mu \nu}(p){/\Nf} , \label{eq:sqed:LONLOrelationgeugeprop:photon}\\
& \bar{D}_{2AA}^{\mu \nu}(p) = -C_\eps \times \bar{D}_{1AA}^{\mu \nu}(p){/\Nf} , \\
& D_{2\lambda {\bar{\lambda}}}(p) = -C_\lambda \times D_{1\lambda {\bar{\lambda}}}(p){/\Nf} ,
\ea
\es
where the tensorial structure of the photon is still given by $\hat{d}^{\mu\nu}(p)=\hat{g}^{\mu\nu} -(1 - \xi)(\hat{p}^\mu \hat{p}^\nu/p^2)$ thanks to the use of the non-local gauge (see \eqref{eq:sqed:nonlocalgauge}).

\subsection{Matter-Multiplet Self-Energies at NLO}

In this section, we compute the NLO self-energies of the matter multiplet, i.e., for the electron and the selectron at NLO in the $1/\Nf$ expansion, i.e., at $\Ord(1/\Nf^2)$ in gQED$_3$. In the following, we shall use shorthand notation for the self-energies
\be
\Sigma_{2 z}^{X(y)}(p^2)= \frac{4}{(n+2S)^2\Nf^2} \left(\frac{\overline{\mu}^{ 2}}{-4p^2} \right)^{2\eps} \tilde{\Sigma}_{2 z}^{X(y)} , \quad \forall X\in\{\psi,\phi\},~ z\in\{p,m\} .
\ee

\subsubsection{Electron Self-Energy at NLO}

We first consider the NLO correction to the electron self-energy that consists of $15$ two-loop and $3$ one-loop Feynman diagrams, altogether labeled $(a,b,\dots,r)$. Indeed, contributions of the same order in $\Nf$ with different loop orders are possible now that we have at our disposal both the LO \eqref{eq:sqed:LOprop} and NLO \eqref{eq:sqed:NLOprop}-softened propagators. 
Taking into account the fact that mirror conjugate graphs take the same value, we are left with a total of $16$ distinct graphs to evaluate. For each one of them, we extract both the momentum and mass parts using parametrization \eqref{eq:sqed:selfenergyparampsi}. With all computations conducted, we obtain the results 
\bs
\label{eq:sqed:NLOelec}
\ba
2\times\GRAPHNLOpsiab &~:
\begin{cases}
\tilde\Sigma^{\psi(ab)}_{2p}&\hspace{-0.3cm}= \cfrac{16S}{9\pi^4}\left(\cfrac{1-3\xi}{\eps^2}+\cfrac{13-60\xi}{3\eps}\right)+\Ord(\eps^0)\\
\tilde\Sigma^{\psi(ab)}_{2m}&\hspace{-0.3cm}= \cfrac{8S}{3\pi^4}\left(\cfrac{2+\xi}{\eps^2}+\cfrac{50+31\xi}{3\eps}\right)+\Ord(\eps^0)\\
\end{cases} ,\\
\GRAPHNLOpsic &~:
\begin{cases}
\tilde\Sigma^{\psi(c)}_{2p}&\hspace{-0.3cm}=-\cfrac{4}{9\pi^4}\left(\cfrac{(2-3\xi)^2}{\eps^2}+\cfrac{64-3\xi(56-39\xi)}{3\eps}\right)+\Ord(\eps^0)\\
\tilde\Sigma^{\psi(c)}_{2m}&\hspace{-0.3cm}=-\cfrac{4}{3\pi^4}\left(\cfrac{(2+\xi)(2-3\xi)}{\eps^2}+\cfrac{112-\xi(40+63\xi)}{3\eps}\right)+\Ord(\eps^0)
\end{cases} ,\\
2\times\GRAPHNLOpside &~:
\begin{cases}
\tilde\Sigma^{\psi(de)}_{2p}&\hspace{-0.3cm}= -\cfrac{16S\mathcal{E}(2+3\xi)}{9\pi^4\eps}+\Ord(\eps^0)\\
\tilde\Sigma^{\psi(de)}_{2m}&\hspace{-0.3cm}= \cfrac{32S\mathcal{E}(2+\xi)}{3\pi^4\eps}+\Ord(\eps^0)
\end{cases} ,\\
\GRAPHNLOpsif &~:
\begin{cases}
\tilde\Sigma^{\psi(f)}_{2p}&\hspace{-0.3cm}= \cfrac{16S\mathcal{E}}{3\pi^4\eps}+\Ord(\eps^0)\\
\tilde\Sigma^{\psi(f)}_{2m}&\hspace{-0.3cm}= -\cfrac{16S\mathcal{E}}{\pi^4\eps}+\Ord(\eps^0)
\end{cases} ,\\
\GRAPHNLOpsig &~:
\begin{cases}
\tilde\Sigma^{\psi(g)}_{2p}&\hspace{-0.3cm}= \cfrac{4S}{3\pi^4}\left(\cfrac{2}{\eps^2}+\cfrac{17}{\eps}\right)+\Ord(\eps^0)\\
\tilde\Sigma^{\psi(g)}_{2m}&\hspace{-0.3cm}= -\cfrac{4S}{\pi^4\eps}+\Ord(\eps^0)
\end{cases} ,
\label{eq:specialdiagram}\\
\GRAPHNLOpsih &~:
\begin{cases}
\tilde\Sigma^{\psi(h)}_{2p}&\hspace{-0.3cm}= \cfrac{2}{9\pi^4}\left(\cfrac{(2-3\xi)^2}{\eps^2}+\cfrac{2(2-3\xi)(16-21\xi)}{3\eps}\right)+\Ord(\eps^0)\\
\tilde\Sigma^{\psi(h)}_{2m}&\hspace{-0.3cm}= \cfrac{2}{3\pi^4}\left(\cfrac{(2+\xi)(10-3\xi)}{\eps^2}+\cfrac{2 (232 +86 \xi - 27 \xi^2)}{3\eps}\right)+\Ord(\eps^0)
\end{cases} ,\\
\GRAPHNLOpsii &~:
\begin{cases}
\tilde\Sigma^{\psi(i)}_{2p}&\hspace{-0.3cm}= -\cfrac{4S\mathcal{E}(2-3\xi)}{9\pi^4\eps}+\Ord(\eps^0)\\
\tilde\Sigma^{\psi(i)}_{2m}&\hspace{-0.3cm}= \cfrac{4S\mathcal{E}(2+\xi)}{3\pi^4\eps}+\Ord(\eps^0)
\end{cases} ,\\
\GRAPHNLOpsij &~:
\begin{cases}
\tilde\Sigma^{\psi(j)}_{2p}&\hspace{-0.3cm}= -\cfrac{4S}{9\pi^4}\left(\cfrac{2-3\xi}{\eps^2}+\cfrac{44-63\xi}{3\eps}\right)+\Ord(\eps^0)\\
\tilde\Sigma^{\psi(j)}_{2m}&\hspace{-0.3cm}= -\cfrac{8S}{3\pi^4}\left(\cfrac{2+\xi}{\eps^2}+\cfrac{56+31\xi}{3\eps}\right)+\Ord(\eps^0)
\end{cases} ,\\
\GRAPHNLOpsik &~:
\begin{cases}
\tilde\Sigma^{\psi(k)}_{2p}&\hspace{-0.3cm}= -\cfrac{4S\mathcal{E}(2-3\xi)}{9\pi^4\eps}+\Ord(\eps^0)\\
\tilde\Sigma^{\psi(k)}_{2m}&\hspace{-0.3cm}= \cfrac{4S\mathcal{E}(10-3\xi)}{3\pi^4\eps}+\Ord(\eps^0)
\end{cases} ,\\
\GRAPHNLOpsil &~:
\begin{cases}
\tilde\Sigma^{\psi(l)}_{2p}&\hspace{-0.3cm}= \cfrac{8\EP S}{9\pi^4} + \Ord(\eps^1)\\
\tilde\Sigma^{\psi(l)}_{2m}&\hspace{-0.3cm}= \cfrac{8\EP S}{3\pi^4} + \Ord(\eps^1)
\end{cases} ,\\
\GRAPHNLOpsim &~:
\begin{cases}
\tilde\Sigma^{\psi(m)}_{2p}&\hspace{-0.3cm}= \cfrac{8S\mathcal{E}}{9\pi^4\eps}+\Ord(\eps^0)\\
\tilde\Sigma^{\psi(m)}_{2m}&\hspace{-0.3cm}= -\cfrac{16S\mathcal{E}}{3\pi^4\eps}+\Ord(\eps^0)
\end{cases} ,\\
\GRAPHNLOpsin &~:
\begin{cases}
\tilde\Sigma^{\psi(n)}_{2p}&\hspace{-0.3cm}= -\cfrac{4S}{9\pi^2}\left(\cfrac{8-3\xi}{\eps^2}+\cfrac{7(32-9\xi)}{3\eps}\right)+\Ord(\eps^0)\\
\tilde\Sigma^{\psi(n)}_{2m}&\hspace{-0.3cm}= 0
\end{cases} ,\\
\GRAPHNLOpsio &~:
\begin{cases}
\tilde\Sigma^{\psi(o)}_{2p}&\hspace{-0.3cm}= \cfrac{8S}{9\pi^4}\left(\cfrac{1}{\eps^2}+\cfrac{31}{3\eps}\right)+\Ord(\eps^0)\\
\tilde\Sigma^{\psi(o)}_{2m}&\hspace{-0.3cm}= \cfrac{}{} 0
\end{cases} ,\\
\GRAPHNLOpsip &~:
\begin{cases}
\tilde\Sigma^{\psi(p)}_{2p}&\hspace{-0.3cm}= -\cfrac{2(2-3\xi)}{\pi^4\eps}(n+2S)\zeta_2C_\gamma+\Ord(\eps^0)\\
\tilde\Sigma^{\psi(p)}_{2m}&\hspace{-0.3cm}= -\cfrac{6(2+\xi)}{\pi^4\eps}(n+2S)\zeta_2C_\gamma+\Ord(\eps^0)
\end{cases} ,\label{eq:sqed:NLOselecCgamma}\\
\GRAPHNLOpsiq &~:
\begin{cases}
\tilde\Sigma^{\psi(q)}_{2p}&\hspace{-0.3cm}= \cfrac{4\EP S }{\pi^4} \zeta_2 C_\eps+ \Ord(\eps^1)\\
\tilde\Sigma^{\psi(q)}_{2m}&\hspace{-0.3cm}= -\cfrac{12\EP S }{\pi^4}\zeta_2 C_\eps + \Ord(\eps^1)
\end{cases} ,\label{eq:sqed:NLOselecCeps}\\
\GRAPHNLOpsir &~:
\begin{cases}
\tilde\Sigma^{\psi(r)}_{2p}&\hspace{-0.3cm}= \cfrac{4S}{\pi^4\eps}\zeta_2C_\lambda+\Ord(\eps^0) \\
\tilde\Sigma^{\psi(r)}_{2m}&\hspace{-0.3cm}= \cfrac{}{} 0
\end{cases} .\label{eq:sqed:NLOselecClambda}
\ea
\es
Note that the computation of the three last diagrams leads to the trivial result that they are simply their LO equivalents times their corresponding coefficient with a sign, {$-C_x/\Nf$}, thanks to the equality \eqref{eq:sqed:LONLOrelationgeugeprop}. Moreover, $C_\eps$ will not contribute to anomalous dimensions because the diagram \eqref{eq:sqed:NLOselecCeps}, $\tilde\Sigma^{\psi(q)}_{p2}$, is finite. Similarly, $C_\lambda$ does not contribute to the mass anomalous dimension because the diagram, $\tilde\Sigma^{\psi(r)}_{m2}$, is exactly zero. Again, we also have explicitly checked that none of the 390 three-loop diagrams contributes to the electron self-energy at NLO.

Summing all the NLO contributions \eqref{eq:sqed:NLOelec}, yields the following results
\bs
\label{eq:Sigmam}
\ba
& \hspace{-0.2cm}\Sigma^\psi_p =-\frac{2(S-\xib)}{R\Nf\eps}-\frac{2(S-\xib)^2}{R^2\Nf^2\eps^2}-\frac{1}{3R^2\Nf^2\eps}\bigg[4+(77+6\EP)S+4\big(1-(19+3\EP)S+6\xib\big)\xib-6R\big(SC_\lambda-\xib C_\gamma\big)\bigg]+\Ord(\eps^0) , \\[-0.2cm]
& \hspace{-0.2cm} \Sigma^\psi_m = \frac{3(2+\xi)}{R\Nf\eps}+\frac{9(2+\xi)^2}{2R^2\Nf^2\eps^2}+\frac{1}{R^2\Nf^2\eps}\bigg[220-21S-4(29-4\xib)\xib +3(2+\xi)\big(6\EP S-RC_\gamma\big)\bigg]+\Ord(\eps^0) ,
\ea
\es
where we introduced the useful notation
\be
\bar{\xi}=(2-3\xi)/2 , \qquad 
R=A(-4p^2/\bar\mu^2)^{\eps} , \qquad 
A=3\pi^2(n+2S)/4 .
\ee
We can now compute the renormalization functions up to NLO for the electron, reading
\bs
\label{eq:sqed:Zpsi+Zmpsi}
\ba
& Z_\psi= 1 - \frac{2 (S-\bar\xi )}{A \mu^{2\eps} \Nf \eps } 
+ \frac{2 \left(S-2 \bar\xi S+\bar\xi^2\right)}{A^2 \mu^{4\eps} \Nf^2 \eps^2} 
- \frac{4+(29-6\EP){S}-6 A \mu^{2\eps}(S C_\lambda-\bar\xi C_\gamma)}{3 A^2 \mu^{4\eps}\Nf^2 \ep } +\Ord(1/\Nf^3) , \\
& Z_{m_{\psi}} = 1- \frac{2(4-S)}{A \mu^{2\eps}\Nf \eps} 
+ \frac{2(16-7S)}{A^2 \mu^{4\eps}\Nf^2 \eps^2} 
- \frac{2(16-(46-3\EP){S}+3A\mu^{2\eps}(S C_\lambda-4C_\gamma))}{3 A^2 \mu^{4\eps}\Nf^2 \eps}+\Ord(1/\Nf^3) .
\ea
\es
From these, the anomalous dimensions read
\bs
\ba
& \gamma_\psi =\frac{4(S-\xib)}{A\Nf}+\frac{4}{3A^2\Nf^2}\bigg[4+(29-6\EP)S-3A\big(SC_\lambda-\xib C_\gamma\big)\bigg]+\Ord\big(1/\Nf^3\big) , \\ 
& \gamma_{m_\psi} =\frac{4(4-S)}{A\Nf}+\frac{8}{3A^2\Nf^2}\bigg[16-(46-3\EP)S+\tfrac{3}{2}A(SC_\lambda-4C_\gamma)\bigg]+\Ord\big(1/\Nf^3\big) .
\ea
\es
We will discuss these results once the anomalous dimensions of the superpartner are computed.

\subsubsection{Selectron Self-Energy at NLO}

We next consider the NLO correction to the selectron self-energy that consist of {$15$} two-loops and $2$ one-loop Feynman diagrams labeled $(a,b,\dots,{p})$. 
Taking into account the fact that mirror conjugate graphs take the same value, we are left with a total of
$14$ distinct graphs to evaluate. For each of them, we extract both the momentum and mass parts using parametrization \eqref{eq:sqed:selfenergyparamphi}. With all computations conducted, we obtain the results 
\bs
\label{eq:sqed:selectron}
\ba
\frac{1}{2}\times\GRAPHNLOphia &~:
\begin{cases}
\tilde\Sigma^{\phi(a)}_{2p}&\hspace{-0.3cm}= -\cfrac{4S(4-\xi(4-3\xi))}{3\pi^4\eps}+\Ord(\eps^0)\\
\tilde\Sigma^{\phi(a)}_{2m}&\hspace{-0.3cm}= -\cfrac{4S(4-\xi(4-3\xi))}{\pi^4\eps}+\Ord(\eps^0)
\end{cases} ,\\
\frac{1}{2}\times\GRAPHNLOphib &~:
\begin{cases}
\tilde\Sigma^{\phi(b)}_{2p}&\hspace{-0.3cm}= -\cfrac{8\EP S}{3\pi^4} + \Ord(\eps^1)\\
\tilde\Sigma^{\phi(b)}_{2m}&\hspace{-0.3cm}= -\cfrac{8\EP S}{\pi^4} + \Ord(\eps^1)
\end{cases} ,\\
2\times\GRAPHNLOphicd &~:
\begin{cases}
\tilde\Sigma^{\phi(cd)}_{2p}&\hspace{-0.3cm}= -\cfrac{16S}{9\pi^2}\left(\cfrac{2(4-3\xi)}{\eps^2}+\cfrac{(4-3\xi)(32+9\xi)}{3\eps}\right)+\Ord(\eps^0)\\
\tilde\Sigma^{\phi(cd)}_{2m}&\hspace{-0.3cm}= -\cfrac{16S}{3\pi^4}\left(\cfrac{2\xi}{\eps^2}+\cfrac{48+\xi(32-9\xi)}{3\eps}\right)+\Ord(\eps^0)
\end{cases} ,\\
\GRAPHNLOphie &~: 
\begin{cases}
\tilde\Sigma^{\phi(e)}_{2p}&\hspace{-0.3cm}= -\cfrac{128S}{9\pi^4}\left(\cfrac{1}{\eps^2}+\cfrac{14+9\xi}{3\eps}\right)+\Ord(\eps)\\
\tilde\Sigma^{\phi(e)}_{2m}&\hspace{-0.3cm}= -\cfrac{256S(1-\xi)}{3\pi^4\eps}+\Ord(\eps^0)
\end{cases} ,\\
\GRAPHNLOphif &~: 
\begin{cases}
\tilde\Sigma^{\phi(f)}_{2p}&\hspace{-0.3cm}= \cfrac{4S}{3\pi^4}\left(\cfrac{8-3\xi}{\eps^2}+\cfrac{64+9\xi(80-17\xi)}{18\eps}\right)+\Ord(\eps^0)\\
\tilde\Sigma^{\phi(f)}_{2m}&\hspace{-0.3cm}= \cfrac{4\xi S}{\pi^4}\left(\cfrac{\xi}{\eps^2}+\cfrac{32-13\xi}{2\eps}\right)+\Ord(\eps^0)
\end{cases} ,\\
2\times\GRAPHNLOphigh &~: 
\begin{cases}
\tilde\Sigma^{\phi(gh)}_{2p}&\hspace{-0.3cm}= \cfrac{8nS}{9\pi^4}\left(\cfrac{4-3\xi}{\eps^2}+\cfrac{104-75\xi}{6\eps}\right)+\Ord(\eps^0)\\
\tilde\Sigma^{\phi(gh)}_{2m}&\hspace{-0.3cm}= \cfrac{4nS}{3\pi^4}\left(\cfrac{4\xi}{\eps^2}+\cfrac{48+31\xi}{3\eps}\right)+\Ord(\eps^0)
\end{cases} ,\\
\GRAPHNLOphii &~: 
\begin{cases}
\tilde\Sigma^{\phi(i)}_{2p}&\hspace{-0.3cm}= \cfrac{4nS}{3\pi^4}\left(\cfrac{1}{\eps^2}+\cfrac{15}{2\eps}\right)+\Ord(\eps^0)\\
\tilde\Sigma^{\phi(i)}_{2m}&\hspace{-0.3cm}= -\cfrac{4nS}{\pi^4}\left(\cfrac{1}{\eps^2}+\cfrac{17}{2\eps}\right)+\Ord(\eps^0)
\end{cases} ,\\
\GRAPHNLOphij &~: 
\begin{cases}
\tilde\Sigma^{\phi(j)}_{2p}&\hspace{-0.3cm}= \cfrac{2S}{9\pi^4}\left(\cfrac{(8-3\xi)^2}{\eps^2}+\cfrac{(8-3\xi)(128+3\xi)}{3\eps}\right)+\Ord(\eps^0)\\
\tilde\Sigma^{\phi(j)}_{2m}&\hspace{-0.3cm}= \cfrac{2S}{3\pi^4}\left(\cfrac{(16-3\xi)\xi}{\eps^2}+\cfrac{768-344\xi+45\xi^2}{3\eps}\right)+\Ord(\eps^0)
\end{cases} ,\\
\GRAPHNLOphik &~: 
\begin{cases}
\tilde\Sigma^{\phi(k)}_{2p}&\hspace{-0.3cm}= -\cfrac{2nS}{9\pi^4}\left(\cfrac{8-3\xi}{\eps^2}+\cfrac{152-27\xi}{3\eps}\right)+\Ord(\eps^0)\\
\tilde\Sigma^{\phi(k)}_{2m}&\hspace{-0.3cm}= \cfrac{2nS}{3\pi^4}\left(\cfrac{\xi}{\eps^2}+\cfrac{48-17\xi}{3\eps}\right)+\Ord(\eps^0)
\end{cases} ,\\
\GRAPHNLOphil &~: 
\begin{cases}
\tilde\Sigma^{\phi(l)}_{2p}&\hspace{-0.3cm}= -\cfrac{2nS}{9\pi^4}\left(\cfrac{2-3\xi}{\eps^2}+\cfrac{44-63\xi}{3\eps}\right)+\Ord(\eps^0)\\
\tilde\Sigma^{\phi(l)}_{2m}&\hspace{-0.3cm}= \cfrac{2nS}{\pi^4}\left(\cfrac{6-\xi}{\eps^2}+\cfrac{128-15\xi}{3\eps}\right)+\Ord(\eps^0)
\end{cases} ,\\
\GRAPHNLOphim &~: 
\begin{cases}
\tilde\Sigma^{\phi(m)}_{2p}&\hspace{-0.3cm}= \cfrac{4nS\mathcal{E}}{9\pi^4\eps}+\Ord(\eps^0)\\
\tilde\Sigma^{\phi(m)}_{2m}&\hspace{-0.3cm}= \cfrac{4nS\mathcal{E}}{\pi^4\eps}+\Ord(\eps^0)\\
\end{cases} ,\\
\GRAPHNLOphin &~:
\begin{cases}
\tilde\Sigma^{\phi(n)}_{2p}&\hspace{-0.3cm}= \cfrac{4nS}{9\pi^4}\left(\cfrac{1}{\eps^2}+\cfrac{28}{3\eps}\right)+\Ord(\eps^0)\\
\tilde\Sigma^{\phi(n)}_{2m}&\hspace{-0.3cm}= -\cfrac{4nS}{\pi^4}\left(\cfrac{1}{\eps^2}+\cfrac{22}{3\eps}\right)+\Ord(\eps^0)
\end{cases} ,\\
\GRAPHNLOphio &~:
\begin{cases}
\tilde\Sigma^{\phi(o)}_{2p}&\hspace{-0.3cm}= -\cfrac{(2+n)(8-3\xi)}{4\pi^4\eps}\zeta_2C_\gamma+\Ord(\eps^0)\\
\tilde\Sigma^{\phi(o)}_{2m}&\hspace{-0.3cm}= -\cfrac{3(2+n)S\xi}{4\pi^4\eps}\zeta_2C_\gamma+\Ord(\eps^0)
\end{cases} ,\label{eq:sqed:oneloopNLOphia}\\
\GRAPHNLOphip &~:
\begin{cases}
\tilde\Sigma^{\phi(p)}_{2p}&\hspace{-0.3cm}= \cfrac{nS}{\pi^4\eps}\zeta_2C_\lambda+\Ord(\eps^0)\\
\tilde\Sigma^{\phi(p)}_{2m}&\hspace{-0.3cm}= -\cfrac{3nS}{\pi^4\eps}\zeta_2C_\lambda+\Ord(\eps^0)
\end{cases} . \label{eq:sqed:oneloopNLOphib}
\ea
\es
Again, note that the computation of the last two graph leads to the trivial results that they are simply the one-loop diagram result times the corresponding interaction correction coefficient with a sign $-C_x$ thanks to the identities \eqref{eq:sqed:NLOprop}. 
Interestingly, $C_\eps$ does not contribute at all to the selectron self-energy since there is no one-loop diagram containing an $\eps$-scalar polarization at this order due to the absence of direct coupling between the selectron and the $\eps$-scalar. 
Again, we also have explicitly checked that none of the 297 three-loop diagrams contributes to the electron self-energy at NLO.

Summing all the contributions \eqref{eq:sqed:selectron} reads
\bs
\label{eq:Sigmam:tot}
\ba
& {\Sigma^\phi_p} =\frac{(6-n+2\xib)S}{R\Nf\eps}-\frac{(6-n+2\xib)^2S}{2R^2\Nf^2\eps^2}-\frac{S}{6R^2\Nf^2\eps}\bigg[8(85+28\xib)-n(163+40\xib)-12\EP  -6R\big(nC_\lambda-2(3+\xib)C_\gamma\big)\bigg] \nonum \\[-0.2cm]
& \hspace{14cm}+\Ord(\eps^0) ,\\[-0.25cm]
& {\Sigma^\phi_m} =\frac{3(n+\xi)S}{R\Nf\eps}+\frac{9(n+\xi)^2S}{2R^2\Nf^2\eps^2}+\frac{3S}{2R^2\Nf^2\eps}\bigg[81n+12\EP-8(2+\xib)\xib -2R\big(nC_\lambda+\xi C_\gamma\big)\bigg]+\Ord(\eps^0) ,
\ea
\es
where we again use the useful notation 
\be
\bar{\xi}=(2-3\xi)/2 , \qquad 
R=A(-4p^2/\bar\mu^2)^{\eps} , \qquad 
A=3\pi^2(n+2S)/4 .
\ee
We note that $\eps$-scalars contribute to the self-energies, in part, from the polarization correction 
$C_\lambda$ (this time for the selectron only) but not from $C_\eps$ (see \eqref{eq:Ceps}).
We can now compute the renormalization functions up to NLO for the selectron field and mass using defining Equation \eqref{eq:sqed:ZSigmadef}, reading 
\bs
\ba
& Z_\phi = 1+\frac{(6-n+2\bar\xi)S}{A\mu^{2\eps}\Nf \eps} +\frac{(2(3+\bar\xi)^2-n(5+2\bar\xi))S}{A^2\mu^{4\eps}\Nf^2\eps^2} - \frac{(8-12\EP+29n-6A\mu^{2\eps}(nC_\lambda-2(3+{\bar\xi})C_\gamma))S}{6A^2\mu^{4\eps}\Nf^2\eps} \nonum \\[-0.2cm]
& \hspace{13cm}+\Ord(1/\Nf^3) , \\[-0.1cm]
& Z_{m_\phi} = 1-\frac{(4+n)S}{A \mu^{2\eps} \Nf \eps} + \frac{(8+5n)S}{A^2\mu^{4\eps}\Nf^2\eps^2} + \frac{2(28-15\EP+7n+\tfrac{3}{2}A\mu^{2\eps}(nC_\lambda+2C_\gamma))S}{3A^2\mu^{4\eps}\Nf^2\eps}+\Ord(1/\Nf^3) .
\ea
\es
The factors $\mu^{2\eps}$ are discussed under Equation \eqref{eq:sqed:Zpsi+Zmpsi}. Using the definition of the anomalous dimensions \eqref{eq:gammadef}, we derive the anomalous dimensions for the selectron field and mass, reading
\bs
\ba
& \hspace{-1cm} \gamma_\phi =\frac{2(n-6-2\xib)S}{A\Nf}+\frac{2S}{3A^2\Nf^2}\bigg[8+29n-12\EP-3A\big(nC_\lambda-(8-3\xi)C_\gamma\big)\bigg]+\Ord\big(1/\Nf^3\big) , \hspace{-1cm} \\
& \hspace{-1cm} \gamma_{m_\phi} =\frac{2(4+n)S}{A\Nf}-\frac{8S}{3A^2\Nf^2}\bigg[28-15\EP+7n+\tfrac{3}{4}A(nC_\lambda+4C_\gamma)\bigg]+\Ord\big(1/\Nf^3\big) .\hspace{-1cm} 
\ea
\es
We will discuss the results for the different cases in the next section.

\section{Critical Exponents and Observables}
\label{sec:sqed:results}

In this section, we apply the general results obtained in the previous section to the various QEDs of interest. In particular, we will present the critical exponents and discuss the observables arising from the computation of the polarization operators of the gauge multiple. We will conclude with a study of the stability of the non-trivial IR fixed point at which all physical quantities are computed.

\subsection{Results for Fermionic QED\titlemath{_3}}
As a check of our computations, we will first recover well-known results for large-$\Nf$ fQED$_3$. This can be achieved by considering our general results without supersymmetry ($S=0$, $n=2$), which yields:
\bs
\begin{empheq}[box=\widefbox]{align}
& \gamma_\psi^{\text{fQED}_3}=-\frac{4(2-3\xi)}{3\pi^2\Nf}+\frac{8(64-92\xi-(6-9 \xi)\pi^2)}{9\pi^4\Nf^2}+\Ord\left(1/\Nf^3\right) , \\
& \gamma_{m_\Psi}^{\text{fQED}_3}=\frac{32}{3\pi^2\Nf}-\frac{64(28-3\pi^2)}{9\pi^4\Nf^2}+\Ord\left(1/\Nf^3\right) , \label{eq:sqed:gammamQED3} \\
& \Pi_\gamma^{\text{fQED}_3}= -\frac{\Nf e^2}{8 \pE} \left[1+ \frac{ C_\gamma^{\text{fQED}_3}}{\Nf} +\Ord(1/\Nf^2) \right] , \\
& C_\gamma^{\text{fQED}_3}=\frac{2(92-9\pi^2)}{9\pi^2}=0.07146 .
\end{empheq}
\es
The field and mass anomalous dimensions correspond exactly to the expressions first found by Gracey with a different method. In particular, the field anomalous dimension was first derived in the Landau gauge in \cite{Gracey:1993iu} and then in an arbitrary covariant gauge in \cite{Gracey:1993sn}. The mass anomalous dimension was derived in \cite{Gracey:1993sn}. The interaction correction coefficient, $C_\gamma^{\text{fQED}_3}$, was first explicitly computed in \cite{Gusynin:2000zb,Teber:2012de,Kotikov:2013kcl}. Hence, our results are in complete agreement with those of the literature.

\subsection{Results for \titlemath{\mathcal{N}=1} SQED\titlemath{_3}}

We now consider the novel case of $\mathcal{N}=1$ SQED$_3$, i.e., taking $S=1$, $n=2$. First, it is interesting to consider the results with arbitrary $\EP$ to study the effect of DRED. In this case, our general results yield
\bs
\ba
& \gamma_\psi = \frac{2\xi}{\pi^2\Nf}{+}\frac{2(2-(12-\pi^2)\xi)}{\pi^4\Nf^2}+\Ord\left(1/\Nf^3\right) ,\\
& \gamma_\phi = {-}\frac{2(2-\xi)}{\pi^2\Nf}{+}\frac{2(26-(2-\xi)\pi^2-12\xi)}{\pi^4\Nf^2}+\Ord\left(1/\Nf^3\right) ,\\
& \gamma_{m_\psi}=\frac{4}{\pi^2\Nf}-\frac{4(14-\pi^2)}{\pi^4\Nf^2}+\Ord\left(1/\Nf^3\right) ,\\
& \gamma_{m_\phi}=\frac{4}{\pi^2\Nf}-\frac{4(46-4\mathcal{E}-3\pi^2)}{3\pi^4\Nf^2}+\Ord\left(1/\Nf^3\right) , \\
& C_\gamma=\frac{12-\pi^2}{\pi^2} , \qquad
C_\eps=\frac{(12-\pi^2)\EP}{\pi^2} , \qquad
C_\lambda=\frac{38{-}2\mathcal{E}-3\pi^2}{3\pi^2} . 
\ea
\es
Interestingly, the effect of the $\eps$-scalar is stiff but crucial. Indeed, the quantities $\gamma_\psi$ and $\gamma_\phi$, as well as $\gamma_{m_\psi}$ and $C_\gamma$, are all $\EP$-independent up to NLO. Only $C_\lambda$ and $\gamma_{m_\phi}$ depend on $\EP$. Taking $\EP=1$ so that DRED is allowed, these results simplify as
\bs
\begin{empheq}[box=\widefbox]{align}
& \gamma_\psi^{\text{SQED}_3} =\frac{2\xi}{\pi^2\Nf}{+}\frac{2(2-(12-\pi^2)\xi)}{\pi^4\Nf^2}+\Ord\left(1/\Nf^3\right) ,\\
& \gamma_\phi ^{\text{SQED}_3} = {-}\frac{2(2-\xi)}{\pi^2\Nf}{+}\frac{2(26-(2-\xi)\pi^2-12\xi)}{\pi^4\Nf^2}+\Ord\left(1/\Nf^3\right) ,\\
& \gamma_{m_\psi}^{\text{SQED}_3} =\frac{4}{\pi^2\Nf}-\frac{4(14-\pi^2)}{\pi^4\Nf^2}+\Ord\left(1/\Nf^3\right) , \label{eq:sqed:gammam}\\
& \gamma_{m_\phi}^{\text{SQED}_3} =\frac{4}{\pi^2\Nf}-\frac{4(14-\pi^2)}{\pi^4\Nf^2}+\Ord\left(1/\Nf^3\right) , \\
& \Pi_x^{\text{SQED}_3}= -\frac{\Nf e^2}{4\pE} \left[1+ \frac{C_x^{\text{SQED}_3}}{\Nf} +\Ord(1/\Nf^2) \right] , \quad x=\{\gamma,\eps,\lambda\} ,\\
& C_x^{\text{SQED}_3}=\frac{12-\pi^2}{\pi^2}=0.2159 , \quad x=\{\gamma,\eps,\lambda\} .
\end{empheq}
\es
Remarkably, the $\eps$-scalars ensure the validity of 
identities for the polarization operators
\be
\Pi_\gamma^{\text{SQED}_3}(p^2)= \Pi_\eps^{\text{SQED}_3}(p^2)= \Pi_\lambda^{\text{SQED}_3}(p^2) ,
\ee
as well as for the mass anomalous dimensions
\be
\gamma_{m_\psi}^{\text{SQED}_3}=\gamma_{m_\phi}^{\text{SQED}_3} ,
\ee
both verified up to NLO in our calculations. This is a behavior expected from SUSY that physical (gauge invariant) quantities are identical in the same multiplet \cite{Hollik:1999xh,Rupp:2000vi}. On the other hand, the field anomalous dimensions for the electron and the selectron are not equal, neither at LO nor at NLO. This is indeed due to the use of a gauge-fixing term that breaks supersymmetry (Wess--Zumino gauge). We recall here that this is expected and not an issue since the breaking of SUSY occurs only for gauge-dependent quantities that are, by definition, non-physical.

\subsection{Results for Bosonic QED\titlemath{_3}}

We now consider the second subcase of interest in this article, which is bosonic QED$_3$, i.e., taking $S=1$, $n=0$ and $\EP=0$. In this case, our general results yield
\bs
\begin{empheq}[box=\widefbox]{align}
& \gamma_\phi^{\text{bQED}_3}=-\frac{4(8-3\xi)}{3 \pi ^2 \Nf}+\frac{8(440-164\xi-3\pi^2(8-3\xi))}{9 \pi ^4 \Nf^2}+\Ord(1/\Nf^3) , \\
& \gamma_{m_\phi}^{\text{bQED}_3}=\frac{16}{3\pi^2\Nf}-\frac{32(64-3\pi^2)}{9\pi^4\Nf^2}+\Ord(1/\Nf^3) , \\
& \Pi_\gamma^{\text{bQED}_3}= -\frac{\Nf e^2}{8\pE} \left[1+ \frac{C_\gamma^{\text{bQED}_3}}{\Nf} +\Ord(1/\Nf^2) \right] , \\
& C_\gamma^{\text{bQED}_3}=\frac{2(164-9\pi^2)}{9\pi^2}=1.6926 .
\end{empheq}
\es
Note that the LO results are in accordance with those previously derived in \cite{Benvenuti:2018cwd,Khachatryan:2019veb,Khachatryan:2019ygx}, and to our knowledge, the NLO results (first published in our short paper \cite{Metayer:2022wre}) are new.

\subsection{Results for Reduced QED\titlemath{_{4,3}} (Graphene)}
\label{sec:sqed:graphene}

From the above results, we are now in a position to study QED$_{4,3}$. Let us
recall that this model is an effective description of graphene at its ultra-relativistic IR fixed point. We access its properties from those of fQED$_3$ with the
help of mapping first proposed \mbox{in \cite{Kotikov:2016yrn}}. Comparing the LO-softened photon propagator of fQED$_{3}$ \eqref{eq:sqed:LOprop:photon} with the corresponding bare photon propagator of QED$_{4,3}$ (see, e.g., \cite{Metayer:2021rco}),
\be
D^{\mu\nu\text{fQED}_3}_{1AA}(p) = \frac{8\I}{\Nf e^2\pE}\left(g^{\mu\nu}-(1-\xi)\frac{p^\mu p^\nu}{p^2}\right) ,\quad
D^{\mu\nu\text{QED}_{4,3}}_{0}(p) = \frac{\I}{2\pE}\left(g^{\mu\nu}-\frac{1-\xi}{2}\frac{p^\mu p^\nu}{p^2}\right) ,
\ee
yields the following naive map
\be
\label{eq:sqed:naivemaipQED3toQED43}
\text{fQED}_{3}\rightarrow \text{QED}_{4,3} ~~ = ~~ \left\{
\frac{1}{\pi^2\Nf}\rightarrow \bar\alpha . ~~
\xi\rightarrow\frac{1+\xi}{2} 
\right\} .
\ee
This map is enough to recover the results for the polarization at one and two-loop for QED$_{4,3}$ from the polarization of fQED$_3$ and, therefore, the corresponding correction coefficient $C_\gamma$. This map is also sufficient to recover the one-loop anomalous dimensions of the QED$_{4,3}$ model from the LO result of the fQED$_3$ model. However, it breaks at two-loop. Indeed, these models, though very similar, have two major differences that manifest at NLO. 

First, fQED$_3$ at NLO is expressed in a non-local gauge, while the QED$_{4,3}$ is not. To compensate this effect, it is enough to consider that if the two-loop polarization of fQED$_3$ is next to the gauge parameter, $\xi$, it should not be present in the QED$_{4,3}$ case. Since the two-loop polarization is proportional to $C_\gamma$, one can use the additional rule \cite{Kotikov:2016yrn}
\be
\xi\times C_\gamma^{\text{fQED}_3} \rightarrow 0 ,
\ee
to recover the proper gauge dependence at two-loop in the anomalous dimensions of QED$_{4,3}$. 

Secondly, in fQED$_3$, we have softened the photon propagator at NLO and computed additional (one loop but NLO) diagrams (see Equations \eqref{eq:sqed:NLOselecCgamma}--\eqref{eq:sqed:NLOselecClambda}, \eqref{eq:sqed:oneloopNLOphia} and \eqref{eq:sqed:oneloopNLOphib}). These diagrams are not present in QED$_{4,3}$ and are replaced by diagrams with a simple fermion loop. To take this into account, we should replace the NLO-softened propagator in fQED$_3$ with the LO one times the regular factor for a fermion loop {in QED$_{4,3}$}, i.e., $-\Nf$. Since the {factor} between the two propagators is exactly {$-C_\gamma^{\text{fQED}_3}/\Nf$} (see \eqref{eq:sqed:LONLOrelationgeugeprop:photon}), the additional needed rule reads \cite{Kotikov:2016yrn}
\be
C_\gamma^{\text{fQED}_3}\rightarrow \Nf ,
\ee
as to be applied on the anomalous dimensions only. 

Performed carefully, this mapping yields the following results for QED$_{4,3}$
\bs
\label{sqed:gamma:psi+m:improved:rqed43}
\begin{empheq}[box=\widefbox]{align}
\gamma_\Psi^{\text{QED}_{4,3}} & = -2 \bar\alpha \frac{1 - 3 \xi}{3} 
+16 \bar\alpha^2 \left(\Nf \zeta_2 + \frac{4}{27} \right) + \Ord(\bar\alpha^3) ,
\label{sqed:gamma:psi:improved:rqed43} \\
\gamma_m^{\text{QED}_{4,3}} & =\frac{32 \bar\alpha}{3} - 
64 \bar\alpha^2 \left(\Nf \zeta_2 - \frac{8}{27} \right)+ \Ord(\bar\alpha^3) ,
\label{sqed:gamma:m:improved:rqed43} \\
\Pi_\gamma^{\text{QED}_{4,3}} & = -\frac{\pi \Nf \alpha}{2\pE} \left[1+C_\gamma^{\text{QED}_{4,3}} \alpha +\Ord(\alpha^2)\right] , \\
C_\gamma^{\text{QED}_{4,3}} & = \frac{92-9\pi^2}{18\pi}=0.05612 ,
\end{empheq}
\es
which perfectly recover the results of \cite{Teber:2012de,Kotikov:2013kcl,Kotikov:2013eha,Teber:2018goo,Metayer:2021rco}. Note that the use of $\alpha$ instead of $\bar\alpha$ for the polarization is on purpose. 

Going further, following \cite{Teber:2012de,Kotikov:2013kcl}, we can use our results to compute the optical conductivity of graphene at its IR fixed point. Indeed, the polarization of the photon $\Pi^{\mu\nu}$, finite and gauge invariant for QED$_{4,3}$ (hence, physical) can be related to the optical (AC) conductivity of graphene with the Kubo formula
\be
\sigma_g(p_0) ={-} \lim_{\vec p \ra 0} \frac{\I p_0}{|\vec p |^2} \Pi^{00}(p_0,\vec p ) ,
\label{eq:rqed:optcondgra}
\ee
where $p^\mu = (p_0, \vec p )$. Since the parametrization for the photon polarization reads $\Pi^{\mu\nu}=(p^2g^{\mu\nu}-p^\mu p^\nu)\Pi$ and $\Pi^{\text{QED}_{4,3}}\sim 1/\pE$, Formula \eqref{eq:rqed:optcondgra} simplifies as
\be
\sigma_g = -\pE \Pi^{\text{QED}_{4,3}} .
\ee
After momentarily restoring the constants $\hbar$, $c$ and $\eps_0$ for clarity, it can be written as
\be
\sigma_g = \sigma_{0g} \bigg(1 + C_\gamma^{\text{QED}_{4,3}} \alpha + \Ord(\alpha^2) \bigg) , 
\qquad \sigma_{0g} = \frac{\Nf e^2}{8\hbar}=\frac{\pi \Nf e^2}{4 h} ,
\label{sigma_sg}
\ee
where $\sigma_{0g}$ is the well-known universal minimal AC conductivity of graphene. Moreover, following \cite{Peres:2010mx}, the optical conductivity of graphene is related to its transmittance ($T_g$) and its absorbance ($A_g$) 
via the relation
\be
T_g=1-A_g=\left(1+\cfrac{\sigma_g}{2\eps_0c}\right)^{-2}\approx 1-\frac{\sigma_{0g}}{\eps_0c}=1-\frac{\pi\Nf\alpha}{2} \quad \Rightarrow \quad A_g = \frac{\pi\Nf\alpha}{2} ,
\ee
where $\alpha=e^2/(4\pi\eps_0\hbar c)$ is the usual QED fine structure constant. At first order, since $\Nf=2$ (i.e., 8 elementary spinors) and $\alpha=1/137$, we obtain an absorbance of
\be
A_{0g}=\pi\alpha=0.0229 .
\ee
Moreover, $C_\gamma^{\text{QED}_{{4,3}}}$ is the interaction correction coefficient to this quantity so that we can expand the leading order absorbance to compute corrections, reading
\be
A_{g}=\pi\alpha \left[1 + \left(C_\gamma^{\text{QED}_{4,3}}- \frac{3 \pi}{4} \right)\alpha +\Ord(\alpha^2) \right] .
\ee
From perturbation theory, we expect the next corrections to be even smaller so that the first one can be taken as an error bar for the next ones, i.e., multiplying the NLO factor by $\pm1$, and since $\alpha=1/137$, we have numerically that pure standing relativistic graphene has an optical absorbance of
\begin{empheq}[box=\widefbox]{align}
A_g=(2.29 \pm 0.04)\% .
\label{eq:rqed:graptheneoppacity}
\end{empheq}
Surprisingly, this result is very close to the one found in the experiments, $A_g^{\text{exp}}=(2.3~\pm~0.2)\%$ \cite{2008PhRvL.101s6405M,2008Sci...320.1308N}, even though measurements are in the pseudo-relativistic limit ($v_F {\approx} c/300$), while our value \eqref{eq:rqed:graptheneoppacity} is computed in the ultra-relativistic limit (with electrons traveling at the speed of light, $v_F =c$) (see related discussions in \cite{Teber:2018jdh}).
\vspace{-0.1cm}

\subsection{Results for Reduced \titlemath{\mathcal{N}=1} SQED\titlemath{_{4,3}} (Super-Graphene)}
\label{sec:sqed:super-graphene}

As a non-trivial application, we will map our results for SQED$_3$ to a model of super-graphene, i.e., for SQED$_{4,3}$ (see the action \eqref{super-graphene}). We recall that this model is an effective description of an eventual pure suspended super-graphene material at its ultra-relativistic fixed point. Following the non-SUSY case, the mapping arises from comparing the LO IR-softened gauge propagators of SQED$_{3}$ \eqref{eq:sqed:LOprop} with the propagators of SQED$_{4,3}$ derived from, e.g., \cite{Herzog:2018lqz}. These propagators read 
\bs
\label{effective-propagators}
\ba
& \hat D^{\mu\nu\text{SQED}_{3}}_{1AA}(p) = \frac{4 \I}{\Nf e^2 \pE} \left(\hat g^{\mu\nu}+(1-\xi)\frac{\hat p^\mu \hat p^\nu}{p^2}\right) , 
&& \hat D^{\mu\nu\text{SQED}_{4,3}}_{0AA}(p) = \frac{\I}{2 \pE} \left(\hat g^{\mu\nu}+\frac{1-\xi}{2}\frac{\hat p^\mu \hat p^\nu}{p^2}\right) ,
\label{effective-propagators:photon} \\
& \bar D^{\mu\nu\text{SQED}_{3}}_{1AA}(p) = \frac{4 \I}{\Nf e^2 \pE} \bar g_{\mu \nu} ,
&& \bar D^{\mu\nu\text{SQED}_{4,3}}_{0AA}(p) = \frac{\I}{2 \pE} \bar g_{\mu \nu}, 
\label{effective-propagators:ep-scalar}\\
& D^{\mu\nu\text{SQED}_{3}}_{1\lambda\bar\lambda}(p) = -\frac{4 \I \fsl{p}}{\Nf e^2 \pE} ,
&& D^{\mu\nu\text{SQED}_{4,3}}_{0\lambda\bar\lambda}(p) = -\frac{\I \fsl{p}}{2 \pE}. 
\label{effective-propagators:photino}
\ea
\es
It is then straightforward to deduce the following naive mapping
\vspace{-0.1cm}
\be
\text{SQED}_3 \rightarrow \text{SQED}_{4,3} ~~ = ~~ 
\left\{ \frac{1}{\pi^22\Nf}\rightarrow \bar\alpha , ~
\xi\rightarrow\frac{1+\xi}{2}\
\right\} ,
\ee
\vspace{-0.1cm}
which is the same as the non-SUSY case up to a factor of two. This allows for {accessing} the polarization of SQED$_{4,3}$ up to two loops and also the anomalous dimensions up to one loop. In order to access the correct two-loop contribution to the anomalous dimensions for this model, as in the non-SUSY case, we first have to cancel the effect of the non-local gauge by using 
\be
\xi\times C_\gamma^{\text{SQED}_3} = 0 ,
\ee
and then cancel the effect of the NLO softening of the gauge propagators by taking
\be
C_x \rightarrow \Nf \qquad \forall x=\{\gamma,\eps,\lambda\} ,
\ee
in the anomalous dimensions. Performed carefully, this mapping yields the following results
\bs
\begin{empheq}[box=\widefbox]{align}
& \gamma_\psi^{\text{SQED}_{4,3}}=2(1+\xi)\bar\alpha+16\bar\alpha^2+\Ord(\bar\alpha^3) ,\\
& \gamma_\phi^{\text{SQED}_{4,3}}=-2(3-\xi)\bar\alpha+16(1+6\Nf\zeta_2)\bar\alpha^2+\Ord(\bar\alpha^3) ,\\
& \gamma_{m_\psi}^{\text{SQED}_{4,3}}=8\bar\alpha-32(1+3\Nf\zeta_2)\bar\alpha^2+\Ord(\bar\alpha^3) ,
\label{sqed:gamma:m:improved:sqed43}\\
& \gamma_{m_\phi}^{\text{SQED}_{4,3}}=8\bar\alpha-32(1+3\Nf\zeta_2)\bar\alpha^2+\Ord(\bar\alpha^3) , \\
& \Pi_\gamma^{\text{SQED}_{4,3}} = -\frac{\pi \Nf \alpha}{\pE} \left[1+C_\gamma^{\text{SQED}_{4,3}} \alpha +\Ord(\alpha^2)\right] , \\
& C_\gamma^{\text{SQED}_{4,3}}=\frac{12-\pi^2}{2\pi} = 0.3391 .
\end{empheq}
\es
Note that use of $\alpha$ instead of $\bar\alpha$ for polarization is on purpose. These results are in accordance with \cite{Herzog:2018lqz} at one loop. To our knowledge, the two-loop contributions are a new result. 
Note that \cite{Herzog:2018lqz} considered a super-graphene model on the boundary (on a substrate) such that the coupling $\alpha_{\text{bdry}}$ is twice as small than in our case, i.e., $\alpha_{\text{bdry}}=\alpha/2$.

Similar to the non-supersymmetric case, we can derive the optical conductivity of the hypothetical super-graphene. Using the same Kubo formula, as in \eqref{eq:rqed:optcondgra}, yields the following minimal AC conductivity of super-graphene
\be
\sigma_{sg} = \sigma_{0sg} \bigg(1 + C_\gamma^{\text{SQED}_{4,3}} \bar\alpha + \Ord(\bar\alpha^2) \bigg) , 
\qquad \sigma_{0sg} = \frac{\Nf e^2}{4\hbar} = \frac{\pi\Nf e^2}{2h} ,
\label{eq:sqed:sigma_sg}
\ee
which is twice as big than the non-SUSY one \eqref{sigma_sg}. From here, a procedure similar to the non-SUSY case yields the following optical absorbance
\begin{empheq}[box=\widefbox]{align}
A_{sg}^{\text{exp}}=(4.59\pm 0.15)\% .
\end{empheq}
Therefore, the absorbance of ultra-relativistic freestanding super-graphene is twice the value of normal graphene in the same conditions. Amusingly, this result is exactly the same as for bilayer (non-SUSY) graphene, which is experimentally twice the absorbance of graphene, i.e., $A_g\approx4.6\%$, \cite{2008arXiv0803.3718N,Peres:2010mx}.

\subsection{Stability of the IR Fixed Point}

An important question is related to the stability of the non-trivial IR fixed point
with respect to radiative corrections. As we have discussed in the Introduction, for all variants of QED$_3$ that we have studied, this fixed point arises in the large-$\Nf$ limit and, more precisely, in the limit $p_E \ll e^2 \Nf$. As $\Nf$ decreases, corrections in $1/\Nf$ increase, which calls for an examination of how the fixed point is affected. 

Following the bQED$_3$ and fQED$_3$ cases (see \cite{Appelquist:1981sf,Appelquist:1988sr}, respectively), for all the QED$_3$ models studied in this article, one can define a dimensionless effective charge
\be
g_r(\pE) = \frac{g}{\pE \left(1 - \Pi_\gamma(\pE) \right)} , \qquad g=e^2\Nf .
\label{coupling_gr}
\ee
In the case of our general (gQED$_3$) model, the photon polarization operator takes the following form
\be
\Pi_\gamma({\pE})=X \cfrac{g}{\pE} \qquad 
\text{with} \qquad 
X=-\frac{n+2S}{16}\left(1+\frac{C_\gamma}{\Nf}+\Ord(1/\Nf^2)\right) ,
\label{general_Pi_gamma}
\ee
where $C_\gamma$ encodes the effects of interactions. From (\ref{coupling_gr}) 
and (\ref{general_Pi_gamma}), one can define the beta function associated with the effective coupling, $g_r$. Its expression is given by:
\be
\beta(g_r) = \frac{\D g_r}{\D \log\pE} = - g_r \left(1 + X g_r \right) ,
\ee
and displays two fixed points. One of them is the (trivial) asymptotically free UV fixed point, $g_r^* \ra 0$. The second one is the (non-trivial) interacting IR fixed point that we are interested in, $g_r^* \ra -1/X$ (see \cite{Appelquist:1981vg,Appelquist:1981sf,Appelquist:1986fd}, as well as the more recent \cite{Gusynin:2020cra}). Summarizing
\be
g_r^*=
\begin{cases}
0 & ~~ \text{asymptotic UV fixed point} , \\
-1/X & ~~ \text{interacting IR fixed point} .
\end{cases}
\ee
By combining the above results, the non-trivial IR fixed point of the various cases of interest read
\be
g_r^*=\frac{16}{n+2S}\left(1-\frac{C_\gamma}{\Nf}+\Ord(1/\Nf^2)\right)=
\begin{cases}
\text{SQED}_3: & 4(1-0.21{6}/\Nf+\Ord(1/\Nf^2)) \\
\text{fQED}_3: & 8(1-0.071/\Nf+\Ord(1/\Nf^2)) \\
\text{bQED}_3: & 8(1-1.693/\Nf+\Ord(1/\Nf^2)) 
\end{cases}
\ee
where the results are accurate up to the NLO in the $1/\Nf$ expansion. We see that fQED$_3$ is the least affected by radiative corrections and that the latter also weakly affects SQED$_3$ (though three times more than fQED$_3$). In the case of bQED$_3$, however, the correction is of the order of $1$. A resummation yields
\be
g_r^*\big|_{\text{bQED}_3} = \frac{8}{1+1.693/\Nf+\Ord(1/\Nf^2)} ,
\ee
so that, despite being shifted by an amount in the order of $1/\Nf$, the fixed point still exists. It would be interesting to have an all-order proof of the existence of the fixed point{,} but this goes beyond the scope of the present paper.

\section{Dynamical (Matter) Mass Generation}
\label{sec:sqed:crit}

As an application of our results, we now turn to an estimate of $\Nc$, the critical number of (s)electron flavors, which is such that for $\Nf > \Nc$, the (s)electron is massless, while for $\Nf < \Nc$, a dynamical mass, with a Miransky scaling 
\cite{Appelquist:1988sr}, is generated, reading
\be
\begin{tikzpicture}[scale=1.75, baseline=(place)]
\coordinate (place) at (0,0) {};
\node [] at (1,0) {$|$};
\node [] at (0,0) {$|$};
\node [right] at (2,0) {$\Nf $};
\node [below] at (1,-0.1) {$N_c$};
\node [below] at (0,-0.1) {0};
\draw[->] (0,0)--(2,0){};
\node [above] at (0.5,0) {$m_{\text{dyn}}\neq0$};
\node [above] at (1.5,0) {$m_{\text{dyn}}=0$};
\end{tikzpicture} ,
\qquad
m_\text{dyn}\propto\exp\left(\frac{-2\pi}{\sqrt{\Nc/\Nf-1}}\right) .
\ee
As discussed in the model section, at the level of the action, the potentially generated parity-even mass terms (parity-odd masses cannot be dynamically generated \cite{Appelquist:1986qw}) are of the form \eqref{eq:sqed:massterm}. Let us remark that only the electron mass term breaks the global flavor symmetry. From SUSY, we also expect that $m_{\text{dyn}_\psi} =m_{\text{dyn}_\phi}$, which we will simply call $m_{\text{dyn}}$.

\subsection{The (Semi-Phenomenological) Gap Equation}

In principle, the critical number of fermion flavors should be derived via the self-consistent resolution of properly truncated (coupled) Schwinger--Dyson (SD) equations. Due to the complexity of the calculations, for decades, this task has been carried out only at the LO, which has resulted in several inconsistencies, such as severe gauge dependence \cite{Rembiesa:1990jd,Atkinson:1993mz} and/or broken Ward identities (see also the thesis in \cite{Bloch:1995dd}). In the case of fQED$_3$, 
following early multiloop works of Nash \cite{Nash:1989xx} and Kotikov \cite{Kotikov:1989nm,Kotikov:2011kg}, a complete
gauge-invariant prescription up to the NLO of the $1/\Nf$-expansion appeared only rather recently in \cite{Gusynin:2016som} and \cite{Kotikov:2016wrb,Kotikov:2016prf} (see also \cite{Kotikov:2020slw} 
for a recent review). In \cite{Kotikov:2016yrn}, the results were then mapped to QED$_{4,3}$, thereby 
extending the LO results of \cite{Gorbar:2001qt} to the NLO in $\al$. 

The systematic approach reviewed in \cite{Kotikov:2020slw} alleviates doubts about the validity of the SD equation approach to access the non-perturbative regime of dynamical mass generation. Nevertheless, it is very technical and difficult to apply to, e.g., SQED$_3$ where SUSY leads to a dramatic increase in the number of graphs with respect to fQED$_3$ and is also responsible for a coupled gap equation for the electron and selectron. Moreover, the complexity of intermediate steps contrasts with the very simple final form of the gap equation that reads
\be
b (d_e-2 - b) = (d_e - 2) \big(\gamma_{m1}+ \cdots \big) .
\label{gap-equation:LO:goodxi}
\ee
where $\gamma_{m}=\gamma_{m1}+\gamma_{m2}+\cdots$ expanded either in loop or large-$\Nf$. In \eqref{gap-equation:LO:goodxi}, the electron dynamical mass scaling is $m_{\text{dyn}}\sim \pE^{-b}$, and dynamical generation occurs when $b$ becomes complex. Actually, as was already noted in the early literature on four-dimensional models (see, \mbox{e.g., \cite{Bardeen:1985sm,Leung:1985sn,Miransky:1988gk,Leung:1989hw,Kondo:1992sq}}) and
reconsidered recently \cite{Gusynin:2016som,Metayer:2021rco}, the form of this gap equation can be deduced from the UV asymptotic behavior of the fermion propagator.
In \cite{Metayer:2021rco}, it was argued by the present authors that the gap equation may be quadratic in $b$ at all loop orders and, therefore, semi-phenomenologically written non-pertubatively in the form
\be
b(d_e-2-b) = \gamma_m(d_e -2 - \gamma_m) ,
\label{gap-equation}
\ee
with the gauge-invariant $\gamma_m$ as the only input. In this case, $b$ becomes complex for $(b - (d_e-2)/2)^2<0$, yielding the criterion
\be
K(\Nc)=0 \qquad \text{where} \qquad K(\Nf)=\left(\gamma_m(\Nf)-\frac{d_e-2}{2}\right)^2 , 
\label{gap-eqn:de}
\ee
from which the corresponding critical number of fermion $\Nc$ (and possibly the critical coupling $\al_c$) can be computed. 
Note that, if $\gamma_m$ would be known exactly, the gap equation would then simply yield $\gamma_m(\Nc)=(d_e-2)/2$. However, when the mass anomalous dimension is known only perturbatively up to a certain order, the gap Equation \eqref{gap-eqn:de} accordingly needs to be properly truncated, i.e., where $\gamma_m=\gamma_{1m}+\gamma_{2m}+\cdots$. Hence, we should use
\be
K(\Nf)=\frac{(d_e-2)^2}{4}{-}(d_e-2)\gamma_{1m}+\left(\gamma_{1m}^2{-}(d_e-2)\gamma_{2m}\right)+\cdots ,
\ee
and then solve $K(\Nc)=0$ (or possibly $K(\alpha_c)=0$). 
Since $\gamma_m$ is gauge-invariant by construction, the resulting $\Nc$ 
will automatically be gauge-invariant too. Moreover, as it is built from the SD formalism, it can be truncated to the accuracy at which $\gamma_m$ is known (\mbox{Equation \eqref{gap-equation}} reduces to \eqref{gap-equation:LO:goodxi} at the
LO in $1/\Nf$). From this polynomial equation, we will obtain multiple solutions for $\Nc$. The physical $\Nc$ will be taken as the largest real solution that is found, which is in accordance with perturbation theory. 
Though semi-phenomenological, such an approach is straightforward and completely gauge invariant. 

For completeness, we provide numerically the mass anomalous dimensions that we obtained in Table \ref{tab:gammanum}.
\begin{table}[h!]
\renewcommand\arraystretch{1.1}
\centering
\begin{tabular}{c|c}
{fQED$_3$} & $\gamma_{m_\psi}=1.0808/\Nf + 0.1174/\Nf^2+\Ord(1/\Nf^3)$ \\ 
{SQED$_3$} & $\gamma_{m_{{\psi}}}=0.4053/\Nf - 0.1696/\Nf^2+\Ord(1/\Nf^3)$ \\ 
{bQED$_3$} & $\gamma_{m_{{\phi}}}=0.5404/\Nf - 1.2553/\Nf^2+\Ord(1/\Nf^3)$ \\ 
\end{tabular}
\caption{Numerical mass anomalous dimensions.}
\label{tab:gammanum}
\end{table}

\phantom{a}
\vspace{-2cm}\\

\subsection{Results for {(S)}QED\titlemath{_3}}

In the following, we shall only focus on the electron mass generation and not its superpartner. Indeed, in the case of bQED$_3$ with $\Nf$ scalars, we did not find any evidence for dynamical scalar mass generation in bQED$_{3}$, suggesting that
\begin{empheq}[box=\widefbox]{equation}
\Nc^{\text{bQED}_3}=0 , \qquad (\text{all-order})
\end{empheq}
for that model, either via the SD method or via the effective gap equation method. Note that the picture seems different if one allows a non-zero quartic coupling $\lambda(|\phi|^2)^2$ in three dimensions (see, e.g., \cite{Ihrig:2019kfv}), where they obtained $\Nc^{\text{bQED}_3}(\lambda\neq0)=6.1\pm1.95$ from fixed-point collision in a four-loop expansion combined with advanced resummation techniques. The situation seems to be also different in four dimensions (see \cite{Dagotto:1989gp}).

On the other hand, for SQED$_3$ (similar to the four-dimensional case (see \cite{Shamir:1990pa,Shamir:1990pb})), we find a possibility that a selectron mass can be induced by the electron condensate, if the latter exists. As we will see in the following, our results suggest that electrons do not condense in SQED$_3$. 

Truncating the gap equation at the LO of the $1/\Nf$ expansion yields the gauge-invariant value
\begin{empheq}[box=\widefbox]{equation}
\Nc^{\text{SQED}_3}=\frac{16}{\pi^2} = 1.6211 , \qquad (\text{LO})
\end{empheq}
coinciding with the Landau gauge result of \cite{Koopmans:1989kv}, which is indeed the good gauge for SQED$_3$. The LO result suggests that an electron mass is generated for $\Nf=1$, thus seemingly breaking both flavor and SUSY symmetries. We find that higher-order corrections dramatically change this picture. Indeed, truncating the gap equation at the NLO of the $1/\Nf$ expansion, we find that 
\be
\Nc^{\text{SQED}_3}=\frac{{4}}{\pi^2}\Big(2 \pm {\I \sqrt{14-\pi^2}}\Big) = 0.8106 (1\pm1.02 \I) .
\ee
Such a complex value arises because of the negative NLO contribution (due to the selectron) to the mass anomalous dimension \eqref{eq:sqed:gammam}, as shown in Table \ref{tab:gammanum}, which prevents the gap equation from having any real valued solution. This calls for a $1/\Nf^3$ computation that is clearly outside the scope of this article. So, in order to overcome this difficulty, we shall proceed with a resummation of the seemingly alternating asymptotic series. A simple Pad\'e approximant $[1/1]$ of \eqref{eq:sqed:gammam} leads to
\be
\gamma_{m_\psi}^{\text{SQED}_3}=\gamma_{m_\phi}^{\text{SQED}_3}=\frac{{4}}{14+(\Nf-1)\pi^2} , \qquad (\text{{NLO} [1/1]}) , \qquad (\text{NLO}) .
\ee
Using this new improved value to solve the gap equation non-perturbatively, i.e., $\gamma_{m_\psi}(\Nc)=1/2$, yields
\begin{empheq}[box=\widefbox]{equation}
\Nc^{\text{SQED}_3}=\frac{\pi^2-6}{\pi^2} = 0.3921 , \qquad (\text{NLO [1/1]}).
\end{empheq}
This result is strong evidence that beyond the LO of the $1/\Nf$ expansion, no dynamical (parity-even) mass is generated for the electron in $\mathcal{N}=1$ SQED$_{3}$. 
Though a dynamical breaking of SUSY may take place in SQED$_3$ (the Witten index is not well-defined with massless matter fields; see, e.g., ref. \cite{Appelquist:1997gq} and references therein), the absence of any electron condensate suggests that SUSY is preserved, in accordance with our perturbative result $\gamma_{m_\psi}=\gamma_{m_\phi}$ up to NLO.

We then focus on the case of fQED$_3$ ($S=0$, $n=2$), for which the gap equation is known exactly up to NLO \cite{Gusynin:2016som,Kotikov:2016prf,Kotikov:2020slw}. The same procedure, this time using \eqref{eq:sqed:gammamQED3} for the mass anomalous dimension, leads at LO to 
\begin{empheq}[box=\widefbox]{equation}
\Nc^{\text{fQED}_3}=\frac{128}{3\pi^2}=4.32 , \qquad (\text{LO})
\end{empheq}
\vspace*{-0.5cm}\\
and at NLO to 
\begin{empheq}[box=\widefbox]{equation}
\Nc^{\text{fQED}_3}=\frac{16}{3\pi^2}\Big(4+\sqrt{3\pi^2-28}\Big)=2.85 , \qquad (\text{NLO})
\end{empheq}
which is in accordance with \cite{Gusynin:2016som,Kotikov:2016prf,Kotikov:2020slw}. Although the problem of a complex $\Nc$ is not encountered in this case (because the NLO term in \eqref{eq:sqed:gammamQED3} is positive, as shown in Table \ref{tab:gammanum}), we still provide for completeness the improved $\Nc$ value obtained with resummation, i.e., 
\begin{empheq}[box=\widefbox]{equation}
\Nc^{\text{fQED}_3}=\frac{2(4+3\pi^2)}{3\pi^2}=2.27 , \qquad (\text{NLO [1/1]}) .
\label{eq:ncfqed3new}
\end{empheq}
As expected from radiative correction effects, this value is smaller than the exact NLO one but still quite close, in accordance with the stability of the critical point. In striking contrast with both SQED$_3$ and bQED$_3$, this suggests that a dynamical (flavor-breaking and parity-even) mass is radiatively generated for the electron in fQED$_3$ for small values of $\Nf$, i.e., for $\Nf = 1$ and $2$. This new improved value \eqref{eq:ncfqed3new} is to compare with the extensive fQED$_3$ D$\chi$SB literature, see Table \ref{chap4:tab:Nc-values}, where seemingly all values between $0$ and $4$ (even infinite in some early studies) have been obtained over four decades. 

\newpage

\begin{table}[ht]
\renewcommand\arraystretch{1.0}
\centering 
\begin{tabular}{|c|c|c|}
\hline
$\Nc$ in fQED$_3$ & Method & Year \\
\hline \hline
$\infty$ & Schwinger--Dyson (LO) & 1984 \cite{Pisarski:1984dj} \\
\hline
$\infty$ & Schwinger--Dyson (non-pert., Landau gauge) & 1990, 1992 \cite{Pennington:1990bx,Curtis:1992gm} \\
\hline
$\infty$ & RG study & 1991 \cite{Pisarski:1991kg} \\
\hline
$\infty$ & Lattice simulations & 1993, 1996 \cite{Azcoiti:1993fb,Azcoiti:1995mi} \\
\hline
$< 4.4$ & F-theorem & 2015 \cite{Giombi:2015haa} \\ 
\hline
$(4/3)(32/\pi^2)=4.32$ & Schwinger--Dyson (LO, resum.) & 1989 \cite{Nash:1989xx} \\
\hline
$4.422$ & RG study (one loop) ($N_c^{\text{conf}} \approx 6.24$) & 2016 \cite{Janssen:2016nrm} \\
\hline
$4$ & Functional RG ($4.1<N_c^{\text{conf}}<10.0$) & 2014 \cite{Braun:2014wja} \\
\hline
$3<N_c< 4$ & RG study & 2001 \cite{Kubota:2001kk} \\
\hline
$3.5 \pm 0.5$ & Lattice simulations & 1988, 1989 \cite{Dagotto:1988id,Dagotto:1989td} \\
\hline
$3.31$ & Schwinger--Dyson (NLO, Landau gauge) & 1993 \cite{Kotikov:1989nm,Kotikov:2011kg} \\
\hline
$3.29$ & Schwinger--Dyson (NLO, Landau gauge) & 2016 \cite{Kotikov:2016wrb} \\
\hline
$32/ \pi^2 \approx 3.24$ & Schwinger--Dyson (LO, Landau gauge) & 1988 \cite{Appelquist:1988sr} \\

\hline
3.0084--3.0844 & Schwinger--Dyson (NLO, resum.) & 2016 \cite{Kotikov:2016prf} \\
\hline
$2.89$ & RG study (one loop) & 2016 \cite{Herbut:2016ide} \\
\hline
$2.85$ & Schwinger--Dyson (NLO, resummation, $\forall \xi$) & 2016 \cite{Gusynin:2016som,Kotikov:2016prf} \\
\hline
$1 + \sqrt{2} = 2.41$ & F-theorem & 2016 \cite{Giombi:2016fct} \\
\hline
\rowcolor{mygray} $2.27$ & Effective gap equation (NLO, $\forall \xi$, double resum.) & 2022 \cite{Metayer:2021rco} \\
\hline
$<9/4=2.25$ & RG study (one loop) & 2015 \cite{DiPietro:2015taa} \\
\hline
$<3/2$ & Free energy constraint & 1999 \cite{Appelquist:1999hr} \\
\hline
$1< N_c<4$ & Lattice simulations & 2004, 2008 \cite{Hands:2004bh,Strouthos:2007stc} \\
\hline
0 & Schwinger--Dyson (non-pert., Landau gauge) & 1990 \cite{Atkinson:1989fp} \\
\hline
0 & Lattice simulations & 2015, 2016 \cite{Karthik:2015sgq,Karthik:2016ppr} \\
\hline
\end{tabular}
\caption{Reproduced from \cite{Teber:2018jdh} and updated. D$\chi$SB in fQED$_3$: some values of $\Nc$ obtained over the years with different methods. The value obtained with our method is grayed. Note that recent analytical methods (including ours) converge to a value of $\Nc$ in the range $]2,3[$ such that a dynamical mass is generated for $\Nf \leq 2$. On the other hand, results from lattice simulations are inconsistent. This may partly be due to the fact that, as $\Nf=2$ is close to $\Nc$, the dynamically generated mass is so small (see estimate and discussion in \cite{Gusynin:2016som}) that it is difficult to extract from lattice {simulations}.}
\label{chap4:tab:Nc-values}
\end{table}

\subsection{Results for (S)QED\titlemath{_{4,3}} (Graphene and Super-Graphene)}

From the results, we have obtained for QED$_{4,3}$, in particular, the mass anomalous dimension at two loops \eqref{sqed:gamma:m:improved:rqed43}, we may apply our semi-phenomenological gap equation formalism and derive the critical coupling constant and critical fermion flavor number. This was conducted in \cite{Metayer:2021rco} with results that are in complete agreement with those derived from the SD equation formalism \cite{Kotikov:2016yrn}. We review them in the following and then carry on with the supersymmetric case.

The computations require the use of an RPA-like procedure, which consist of resumming the two-loop $\Nf$ dependency (see \cite{Kotikov:2016yrn,Metayer:2021rco} for more details), yielding
\begin{empheq}[box=\widefbox]{equation}
\al_{c}^{\text{QED}_{4,3}}(\Nf) =\frac{12\pi}{128-3\pi^2\Nf}, \qquad N_{c}^{\text{QED}_{4,3}}=\frac{128}{3\pi^2}=4.3230 , \qquad \text{(1-loop)} ,
\label{alc+Nc:res:one-loop:rqed43}
\end{empheq}
which, for the range of allowed non-zero values of $\Nf$, \eqref{alc+Nc:res:one-loop:rqed43} yield
\ba
& \al_{c}(\Nf=0)=0.2945, \quad
\al_{c}(\Nf=1)=0.3832, \quad 
\al_{c}(\Nf=2)=0.5481, \quad \nonum \\
& \al_{c}(\Nf=3)=0.9624, \quad 
\al_{c}(\Nf=4)=3.9415 .
\label{alc:N:res:one-loop:QED43}
\ea
Similarly, at two loops, the following can be obtained
\begin{empheq}[box=\widefbox]{equation}
\al_{c}^{\text{QED}_{4,3}}(\Nf) =\frac{36\pi}{32(6+\sqrt{6})-9\pi^2\Nf}, 
\quad N_{c}^{\text{QED}_{4,3}}=\frac{32}{9\pi^2}\left(6+\sqrt{6}\right)=3.0440 , 
\quad \text{(2-loop)} ,
\label{alc+Nc:res:two-loop:rqed43}
\end{empheq}
which, for the range of allowed non-zero values of $\Nf$, \eqref{alc+Nc:res:two-loop:rqed43} yield
\ba
& \al_{c}(\Nf=0)=0.4183, \quad
\al_{c}(\Nf=1)=0.6230, \quad
\al_{c}(\Nf=2)=1.2196, \quad
\al_{c}(\Nf=3)=28.967 .
\label{alc:N:res:two-loop:rqed43}
\ea
We recall that, in the case of graphene, we are interested in $\Nf=2$ because graphene has a total of eight spinors (two cones/sub-lattices $\times$ two valley/chirality $\times$ two spins). Moreover, in the ultra-relativistic limit we are interested in, graphene has a weak coupling constant, $\al \sim 1/137$. The result $N_{c}^{\text{QED}_{4,3}}=3$ is unreachable because it is valid in the limit $\al \ra \infty$. Moreover, the critical coupling $\al_{c}(\Nf=2)\approx1.2$ is much larger than $\al \sim 1/137$. Hence, graphene remains (semi-)metallic in the ultra-relativistic limit. This agrees with the results originally derived in \cite{Kotikov:2016yrn,Metayer:2021rco}. It is also compatible with experiments on graphene that do not find any evidence for a metal-to-insulator transition.

\begin{figure}[h!]
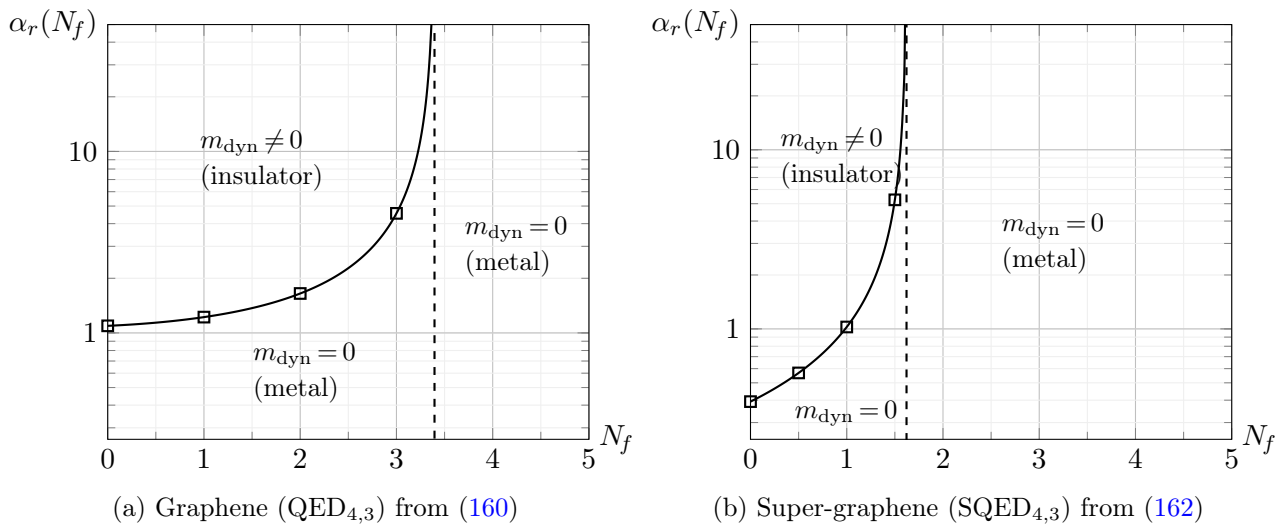

\centering
\begin{subfigure}{0.49\linewidth}
\centering
\rqedfplot
\caption{Graphene (QED$_{4,3}$) from \eqref{alc+Nc:res:two-loop:rqed43}}
\end{subfigure}
\begin{subfigure}{0.49\linewidth}
\centering
\srqedftplot
\caption{Super-graphene (SQED$_{4,3}$) from \eqref{eq:sqed43critical}}
\end{subfigure}
\caption{Phase diagrams for dynamical mass generation in (super-)graphene. Note that the relevant case for both graphene and super-graphene is $\Nf=2$. Here, insulator refers to an excitonic insulating phase, while metal refers to a semimetallic phase.}
\label{fig:sqed:phasediags}
\end{figure}

Similarly, we derive the results for the SQED$_{4,3}$ critical coupling and fermion flavor number, corresponding to a phase transition in the ultra-relativistic limit of freestanding super-graphene. Following the RPA-like procedure introduced in \cite{Kotikov:2016yrn,Metayer:2021rco}, we obtain 
\begin{empheq}[box=\widefbox]{equation}
\alpha_c^{\text{SQED}_{4,3}}(\Nf)=\frac{2\pi}{16-\pi^2\Nf} , \qquad 
\Nc^{\text{SQED}_{4,3}}=\frac{16}{\pi^2}=1.6211 , \qquad (\text{1-loop}) .
\label{eq:sqed43critical}
\end{empheq}
For the range of allowed values of $\Nf$, it leads numerically to
\vspace{-0.5cm}\\
\ba
& \alpha_c^{\text{SQED}_{4,3}}(\Nf=0)=0.3927 , \qquad
\alpha_c^{\text{SQED}_{4,3}}(\Nf=1)=1.0249 .
\ea
\vspace{-0.5cm}\\
Unfortunately, at two loops, the result is non-physical, despite trying RPA-like or Padé resummations. This is probably a parity effect (like the four-loop approach in QED$_4$, see \cite{Metayer:2021rco}), we then settle for the one-loop approach. Since the relevant value for super-graphene is also $\Nf=2$, and that $\Nc^{\text{SQED}_{4,3}}=1.6211$ at one loop, and that we expect higher-order corrections to lower $\Nc$, we expect that $\Nf=2$ will always be above $\Nc$ in SQED$_{4,3}$. Hence, super-graphene is even further away from the insulating phase than graphene. For completeness, we provide in Figure \ref{fig:sqed:phasediags} the phase diagram of (super-)graphene.
\vspace{-0.25cm}

\phantom{a}
\vspace*{-1.5cm}\\

\newpage 

\subsection{Meta-Analysis of the Results}

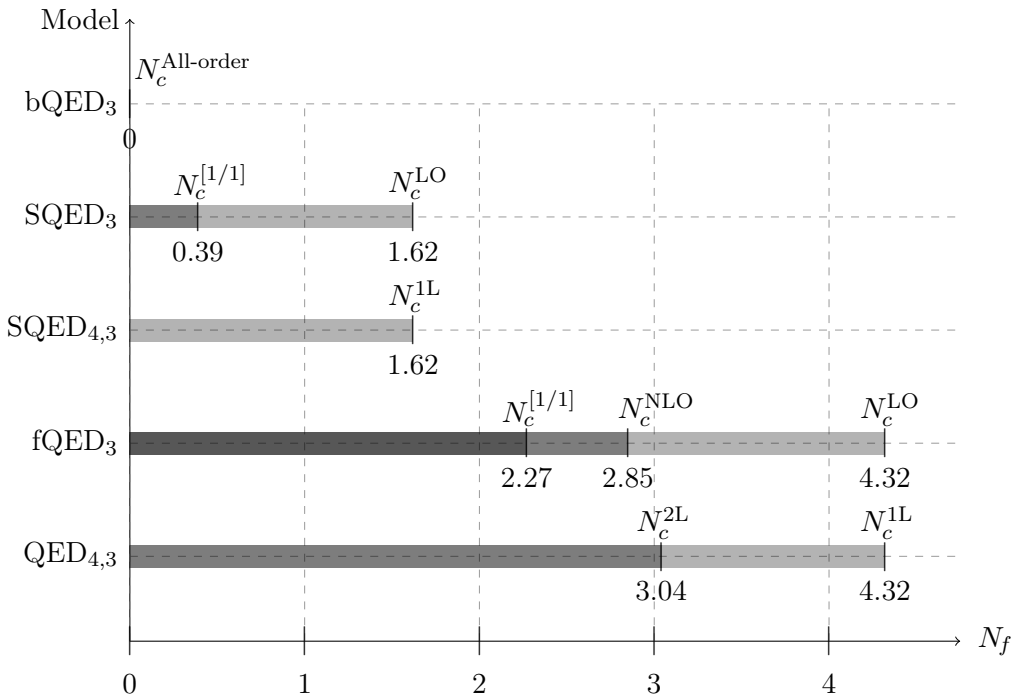
\begin{figure}[h!]
\centering
\begin{tikzpicture}[xscale=2.3,yscale=1.5]
\draw[->] (0,0.25) -- (4.75,0.25) node[right=0.1cm] {$\Nf$};
\draw[->] (0,0.25) -- (0,5.75) node[left] {Model};
\draw[step=1.0,gray,opacity=0.75,thin,dashed] (0,0.25) grid (4.75,5);
\customtick{0}{0.25}
\customtick{1}{0.25}
\customtick{2}{0.25}
\customtick{3}{0.25}
\customtick{4}{0.25}
\node[left=0.25cm] at (0.1,5) {bQED$_3$};
\custompoint{0}{5}{~~~$\Nc^{\text{All-order}}$}
\node[left=0.25cm] at (0.1,4) {SQED$_3$};
\custompoint{1.62}{4}{$\Nc^{\text{LO}}$}
\custompoint{0.39}{4}{$\Nc^{[1/1]}$}
\node[left=0.25cm] at (0.1,3) {SQED$_{4,3}$};
\custompoint{1.62}{3}{$\Nc^{\text{1L}}$}
\node[left=0.25cm] at (0.1,2) {fQED$_3$};
\custompoint{4.32}{2}{$\Nc^{\text{LO}}$}
\custompoint{2.85}{2}{\phantom{a}$\Nc^{\text{NLO}}$}
\custompoint{2.27}{2}{$\Nc^{[1/1]}$}
\node[left=0.25cm] at (0.1,1) {QED$_{4,3}$};
\custompoint{4.32}{1}{$\Nc^{\text{1L}}$}
\custompoint{3.04}{1}{$\Nc^{\text{2L}}$}
\end{tikzpicture}
\caption{All results obtained in this article for the critical number of fermion flavors below which a dynamical mass is generated in various QED models. The darker it is, the more likely the corresponding model is massive for a given $\Nf$. Note that the case of interest is generally $\Nf=1$ for all the QED$_3$ variants. For graphene (QED$_{4,3}$) and super-graphene (SQED$_{4,3}$), the case of interest is usually $\Nf=2$. 
}
\label{fig:sqed:allNC}
\end{figure}

For completeness and ease of comparison, we provide in Figure \ref{fig:sqed:allNC} a comparative plot of all the $\Nc$ values found for all the QED models studied in this article. It shows that, on the one hand, dynamical mass generation is likely to be possible in fQED$_3$, while, on the other hand, bQED$_3$ does not show any sign of boson condensation. Therefore, in $\mathcal{N}=1$ SQED$_3$, the fermionic part attempts to condense, while the bosonic part tries to prevent it from doing so. Ultimately, the selectronic part seems to overcome the electronic part such that no mass is radiatively generated in SQED$_3$, thereby staying in a conformal phase. As for the reduced QED models, the QED$_{4,3}$ theory seems to allow mass generation for small values of the fermion number, $\Nf =3$, which would be enough for a metal-to-insulating phase transition to occur, provided that the system is in a strongly coupled regime (high fine structure constant). Adding SUSY, that is, considering $\mathcal{N}=1$ SQED$_{4,3}$, further reduces the critical number of fermion flavors to $1$. In the case of graphene in the ultra-relativistic limit, the coupling is small, $\al \sim 1/137$, and the insulating phase is unreachable (both with and without supersymmetry). In all cases, our results indicate that the addition of SUSY to an Abelian gauge theory seems to suppress (rather than enhance) dynamical mass generation.

\newpage

\section{Conclusions}
\label{sec:conclusion}

In this article, we have {reviewed} the critical properties of several variants of three-dimensional QED with fermionic (fQED$_3$), bosonic (bQED$_3$) and minimally supersymmetric (SQED$_3$) charged matter. All these cases were considered in a unified way with the help of a general gQED$_3$ model. Reduced QED models and their supersymmetric extension were also studied in relation to graphene (QED$_{4,3}$) and super-graphene (SQED$_{4,3}$).

In the general framework provided by the gQED$_3$ model, we performed a complete analytical perturbative computation of matter and gauge field anomalous dimensions at the LO and the NLO in the large-$\Nf$ expansion, in the DRED scheme and for arbitrary covariant gauge fixing. All these quantities correspond to the critical exponents of the considered models at the non-trivial IR fixed point that arises in the large-$\Nf$ limit. Expanding on our previous (short) paper \cite{Metayer:2022wre}, all {calculation} details were provided. Along the way, we added perturbative proof of the Furry theorem, generalized for these models. We also studied thoroughly the effect of DRED and showed its crucial importance in ensuring that the theory remains SUSY invariant.

All of our results have a transcendental structure that is similar to that known in the case of fQED$_3$. There are, however, noticeable quantitative differences, with radiative corrections having a tendency to increase vacuum polarization in bQED$_3$ with respect to fQED$_3$ while acting oppositely on the mass anomalous dimension. 
The case of SQED$_3$ is, somehow, intermediate between fQED$_3$ and bQED$_3$ with, in particular, a tendency of the bosonic contribution from the selectron to, on the one hand, increase the overall photon polarization and, on the other hand, decrease the overall electron mass anomalous dimension.

As a first application of our results, we computed the optical conductivity of super-graphene (SQED$_{4,3}$) and showed that it leads to an optical absorbance of $\sim 4.6\%$. This result is exactly twice the absorbance of usual (non-SUSY) graphene (QED$_{4,3}$) and is a direct consequence of the enhanced effect of interactions that, if ever realized, SUSY would bring on the optical absorbance. Another application was devoted to the study of a potential dynamical (matter) mass generation. This non-perturbative phenomenon again revealed a marked difference between fQED$_3$ and bQED$_3$. In fQED$_3$, a flavor-breaking parity-invariant mass is generated for $\Nf \leq 2$ (in terms of four-component spinors), while in bQED$_3$, we did not find any evidence for a dynamically generated scalar mass. In all cases, our results indicate that the addition of SUSY to an Abelian gauge theory seems to suppress (rather than enhance) dynamical mass generation. In the case of SQED$_3$, the value found for the critical electron flavor number, $\Nc$, that is such that for $\Nf < \Nc$ a dynamical mass for the electron would be generated, is given by $\Nc=0.39$ (in terms of four-component spinors). Contrary to fQED$_3$, this strongly suggests that $\mathcal{N}=1$ SQED$_3$ remains in an interacting conformal phase for all values of $\Nf$.
In the case of SQED$_{4,3}$, we found $\Nc=1.62$, which is indeed lower than the corresponding value for QED$_{4,3}$, $\Nc=3.04$. Note that these results hold only at strong coupling. Graphene at its IR fixed point is weakly coupled ($\al \sim 1/137$) and, thus, is deep in a semimetallic phase, which is in qualitative agreement with the experiments in actual samples. 


\appendix 

\newpage

\section{Appendix on Multiloop Massless Techniques}
\label{chap:multiloop}

This Appendix provides the necessary multiloop massless techniques for the computation of two-point massless integrals encountered in this review. 

\subsection{One Loop}

We consider a Euclidean space of dimension $d$ and follow the notations of the \mbox{review in \cite{Kotikov:2018wxe}}. The one-loop massless propagator-type master topology is given by the so-called \textit{bubble} integral
\be
J(d,\vec p\, ,\al ,\beta ) 
= \generaloneloop{\al}{\beta}
= \int \frac{[\D^d k]}{k^{2\al} (\vec p- \vec k\,)^{2\beta}}
= \frac{(p^2)^{d/2-\al-\beta}}{(4\pi)^{d/2}}\,G(d,\al,\beta)\, ,
\label{def:one-loop:2-pt}
\ee
where $[\D^dk]= \D^dk /(2\pi)^d$ and $\al$, $\beta$ are the so-called indices of the propagators. For the diagrammatic representation of the integral, we used the propagator line with a generalized {index}
\be
\generalprop{\al}=\frac{1}{(p^2)^\al}\,.
\label{eq:multiloop:linedefwithindex}
\ee
In \eqref{def:one-loop:2-pt}, $G$ is the dimensionless function left after factorization of the trivial dependence in $p$ of the integral. 
At one loop, $G$ is known exactly and  has a simple expression in terms of Euler $\Gamma$-functions
\ba
G(d,\al,\beta)= \frac{a(d,\al)\,a(d,\beta)}{a(d,\al+\beta-d/2)}, 
\qquad a(d,\al)= \frac{\Gamma(d/2-\al)}{\Gamma(\al)}\, .
\label{eq:multiloop:Gres}
\ea
To fix notations, we recall the usual analytic continuation of $\Gamma(x)=(x-1)!$, reading
\ba
\Gamma(1-\eps)
= \exp\left(\gamma_E\eps+\sum_{n=2}^\infty\frac{\eps^n}{n}\zeta_n\right)
= e^{\gamma_E \eps}\left(1 + 
\frac{\zeta_2 }{2} \eps^2 + 
\frac{\zeta_3}{3} \eps^3
+\Ord(\eps^4)\right) \,,
\label{eq:gammaEuler}
\ea
where $\gamma_E=0.577$ is the Euler–Mascheroni constant and $\zeta_n$ is the Riemann zeta function, where $\zeta_2=\pi^2/6$, and $\zeta_3=1.202$ is the {Apéry} constant.
Note that, at one loop, by construction and in any dimension, $G$ vanishes exactly if one index is negative or zero, as well as being symmetrical under index exchange, i.e., 
\be
G(d,\al,\beta)=0\,, \text{ if } \al\leq0 \text{ or } \beta\leq0 \,, \qquad \text{and} \qquad
G(d,\al,\beta)=G(d,\beta,\al)\,.
\label{eq:multiloop:Grules}
\ee
In our dimension of interest, $d=3-2\eps$; there are two important examples of the evaluation of this function, reading
\bs
\ba
& G(3-2\eps,1,1)=4^\eps\pi^{3/2}e^{-\gamma_E\eps}\left(1+\frac{5}{2}\zeta_2\eps^2-\frac{1}{3}\zeta_3\eps^3+\Ord(\eps^4)\right)\,,  \\
& G(3-2\eps,1,1/2)=4^{-\eps}\pi^{-1/2}e^{-\gamma_E\eps}\left(\frac{2}{\eps}+8+(32-7\zeta_2)\eps
+\Ord(\eps^2)\right)\,.
\label{eq:multiloop:G1/2}
\ea
\es

\newpage

\subsection{Two Loop}

At two loops, one can encounter various massless propagator-type topologies. Nevertheless, all two-loop topologies can be encompassed by a general topology, the {\it {diamond diagram}}. This is the two-loop massless propagator-type master-topology integral and is given by
\ba
J(d,\vec p\, ,\al_i) =
\generaltwoloop{\al_1}{\al_2}{\al_3}{\al_4}{\al_5}
& = \int \frac{[\D^d k_1 ]\,[\D^d k_2]}{
k_1^{2\al_1}\,
k_2^{2\al_2}\,
(\vec p - \vec k_2)^{2\al_3}\,
(\vec p- \vec k_1)^{2\al_4}\,
(\vec k_{12})^{2\al_5}} =\frac{(p^2)^{d-\sum\al_i}}{(4\pi)^d}\,G(d,\al_i)\,, 
\label{eq:multiloopdiamond}
\ea
where $\vec k_{12}=\vec k_1-\vec k_2$ and $G(d,\al_1,\dots,\al_5)$ is dimensionless and unknown for arbitrary indices $\{\al_i\}_{i=1-5}$. This integral has two kinds of symmetries: first under the exchange $k_1 \leftrightarrow k_2$ and second under the shifts $k_1\rightarrow k_1-p$ and $k_2\rightarrow k_2-p$. This generates the following relations
\bs
\ba
& J(d,\vec p,\al_1,\al_2,\al_3,\al_4,\al_5)=J(d,\vec p,\al_2,\al_1,\al_4,\al_3,\al_5) \qquad (k_1 \leftrightarrow k_2)\,, \\
& J(d,\vec p,\al_1,\al_2,\al_3,\al_4,\al_5)=J(d,\vec p,\al_4,\al_3,\al_2,\al_1,\al_5) \qquad (k_1\rightarrow k_1-p \text{ and } k_2\rightarrow k_2-p)\,,
\ea
\es
The first relation is then a mirror reflection along the vertical axis, i.e., it exchanges $\alpha_1 \leftrightarrow \alpha_2 $ and $\alpha_3\leftrightarrow \alpha_4$, while the second one is the mirror reflection along the horizontal axis, i.e., it exchanges $\alpha_1\leftrightarrow\alpha_4$ and $\alpha_2\leftrightarrow\alpha_3$. 

This diamond integral is the only one we have to consider at two loops for two-point massless diagrams as it encompasses all the other two-loop topologies. Indeed, since lines with zero index contract {like}
\be
\phantom{\bigg|}
\generalprop{\al}\vdot\generalprop{0}\vdot\generalprop{\beta} ~~ = ~~ \generalprop{\al}\vdot\generalprop{\beta} ~~ = ~~ \generalprop{\al+\beta} ~~ \,,
\ee
we have that all possible two-loop sub-topologies can be written using the diamond diagram with well-chosen zero indices, reading
%
\bs
\ba
& \generaltwoloop{\al_1}{\al_2}{\al_3}{\al_4}{0} = J(d,\vec p ,\al_1,\al_2,\al_3,\al_4,0) = \doublebubble{\al_1}{\al_4}{\al_2}{\al_3} = \frac{(p^2)^{d-\sum \alpha_i}}{(4\pi)^d} G(d,\al_1,\al_4) G(d,\al_2,\al_3)\,, \nonum \\[-0.8cm]
& 
\\[0.1cm]
&\generaltwoloop{0}{\al_1}{\al_2}{\al_3}{\al_4} = J(d,\vec p,0,\al_1,\al_2,\al_3,\al_4) = \eye{\al_1}{\al_3}{\al_2}{\al_4} = \frac{(p^2)^{d-\sum \alpha_i}}{(4\pi)^d} G(d,\al_1,\al_2+\al_3+\al_4-d/2) \nonum \\[-0.8cm]
& \hspace{10cm} \times G(d,\al_3,\al_4)\,,
\\[0.1cm]
& \generaltwoloop{0}{\al_1}{0}{\al_2}{\al_3} = J(d,\vec p ,0, \alpha_1, 0,\alpha_2, \al_3) = \sunset{\al_1}{\al_3}{\al_2}  = \frac{(p^2)^{d-\sum \alpha_i}}{(4\pi)^d} G(d,\al_1,\al_2+\al_3-d/2) \nonum \\[-0.8cm]
& \hspace{10cm} \times G(d,\al_2,\al_3)\,,
\ea
\es
%
where these three sub-topologies have been reduced to the exactly known one-loop master topology, $G(d,\alpha,\beta)$. Indeed, the first one (double bubble) is obviously the multiplications of two one-loop bubbles, and upon closer inspection, the two others (eye and sunset) are convolutions of one-loop bubbles.

\newpage

We are then left with the case where all $\alpha_i$ are non-zero, like, for example, $J(d,\vec p, 1,1,1,1,1)$. In this case, IBP techniques can be used \cite{Vasiliev:1981,Chetyrkin:1981,Tkachov:1981} and allow the derivation of identities of the form 
\ba
\generaltwoloop{\al_1}{\al_2}{\al_3}{\al_4}{\al_5} ~ = ~ \frac{1}{2\al_1+\al_4+\al_5-d} \Vast[
& p^2 \al_4 \times \generaltwoloop{\al_1}{\al_2}{\al_3}{\al_4^+}{\al_5}
~ - ~ \al_4 \times \generaltwoloop{\al_1^-}{\al_2}{\al_3}{\al_4^+}{\al_5} \nonum \\
& ~ + ~ \al_5 \times \generaltwoloop{\al_1}{\al_2^-}{\al_3}{\al_4}{\al_5^+}
~ - ~ \al_5 \times \generaltwoloop{\al_1^-}{\al_2}{\al_3}{\al_4}{\al_5^+} \Vast]\,
\ea
where $\alpha_i^\pm=\alpha_i\pm1$. Several similar IBP identities can be derived, and altogether, they form a powerful reduction algorithm. One can show that, ultimately, every two-loop integral $J(d,\vec p, \al_1,\al_2,\al_3,\al_4,\al_5)$, with integer indices $\alpha_i$, can be reduced as a linear combination of a small set of master integrals. The implementation of the IBP identities and the reduction process can be conveniently automated with the \textsc{Mathematica} versatile package \textsc{LiteRed} by Roman Lee \cite{Lee:2012cn,Lee:2013mka}.

Using IBP reduction techniques allows us to reduce any two-loop integral into masters that can be expressed with the trivial one-loop function, $G(d,\alpha,\beta)$, together with two non-trivial two-loop masters
\vspace*{-0.75cm}\\
\bs
\ba
& \generaltwoloop{}{}{}{}{\frac{1}{2}}~=~\frac{(p^2)^{d-9/2}}{(4\pi)^d} G(d,1,1,1,1,1/2)\,,\label{app:master1} \\[-0.2cm]
& \generaltwoloop{}{\frac{1}{2}}{}{\frac{1}{2}}{} ~ = ~ \frac{(p^2)^{d-4}}{(4\pi)^d} G(d,1,1/2,1,1/2,1)\,.\label{app:master2} 
\ea
\es
The first one \eqref{app:master1} is computed from the more general integral, $G(d,1,1,1,1,\alpha)$, which was evaluated exactly in \cite{Kotikov:1996} and reads
\ba
\label{eq:multiloop:G(1111a)}
&G({d},1,1,1,1,\al) =
-2\, \Gamma(\lambda)\Gamma(\lambda-\al) \Gamma(1-2\lambda+\al) \\
&\qquad \times \left [ \frac{\Gamma(\lambda)}{\Gamma(2\lambda)\Gamma(3\lambda-\al-1)}\,
\sum_{n=0}^{\infty}\,\frac{\Gamma(n+2\lambda)\Gamma(n+1)}{n!\,\Gamma(n+1+\al)}\,\frac{1}{n+1-\lambda+\al}
+\frac{\pi \cot \pi (2\lambda-\al)}{\Gamma(2\lambda)} \right ]\, ,
\nonum
\ea
where {$\lambda=(d-2)/2=(1-2\eps)/2$}. Note that \eqref{eq:multiloop:G(1111a)} may be written with a generalized hypergeometric function, ${}_3F_2$, of argument $1$, as
\be
\setlength\arraycolsep{1pt}
\hspace{-1cm}
{}_3 F_2\left(\begin{matrix} 1 ~~~ \alpha -\lambda +1 ~~~ 2 \lambda \\ ~~ \alpha +1 ~~~ \alpha -\lambda +2 ~~ \end{matrix}\bigg|1\right) 
= \frac{(\alpha -\lambda +1) \Gamma(\alpha+1)}{\Gamma (2 \lambda )} \sum_{n=0}^{\infty } \frac{\Gamma(n+2\lambda)\Gamma (n+1)}{n! \Gamma (n+\alpha +1)}\frac{1}{n+1-\lambda+\alpha} \,. \hspace{-1cm}
\label{res:I(al):Kotikov}
\ee
{There is also an equivalent representation with a ${}_3F_2$ of argument $-1$ in an earlier \mbox{work \cite{Kazakov:1983}}}, see \cite{Kotikov:2018wxe} for a review. Therefore, in our case of interest, $\alpha=1/2$ and $d=3-2\eps$, it can be expanded in $\eps$-series. This step is non-trivial since expanding generalized hypergeometric functions in a series is, in general, very hard. For our case, it can be achieved in an automated way using the \textsc{Mathematica} package \textsc{HypExp} \cite{Huber:2006,Huber:2008}, and we obtain
\be
G(3-2\eps,1,1,1,1,1/2)= 3\pi4^{-\eps}e^{-{2}\gamma_E\eps}\Big(2\zeta_2+21\zeta_3 \eps+\Ord(\eps^2)\Big)\,.
\ee
The second master integral \eqref{app:master2} is highly non-trivial and can, in principle, be computed from the results of Appendix B in \cite{Kotikov:2013eha} (based on the Gegenbauer polynomial \mbox{technique \cite{Kotikov:1996}}), where the general integral $G(d,1,\alpha,1,\beta,1)$ was derived exactly as a combination of two generalized hypergeometric functions, $_3F_2$, of argument $1$, reading
%
\ba
& G(d,1,\alpha,1,\beta,1)= \nonum\\
& \frac{\Gamma (\lambda ) \Gamma (-\alpha +\lambda +1) \Gamma (-\beta +\lambda +1)}{\alpha  (\alpha -2 \lambda ) (\alpha -\lambda ) \Gamma (\alpha ) (\beta -\lambda ) \Gamma (\beta +1) \Gamma (2 \lambda ) \Gamma (2 \lambda -\alpha ) \Gamma (\alpha +\beta -\lambda +1) \Gamma (-\alpha -\beta +3 \lambda )}\bigg[
\nonum \\
& \hspace{0cm}  \alpha  \Gamma (2 \lambda ) \Gamma (2 \lambda -\alpha ) \Gamma (\alpha +\beta -2 \lambda +1) \Gamma (\alpha +\beta -\lambda +1) \, _3F_2\big(1,2 \lambda ,2 \lambda -\alpha ;\beta +1,-\alpha +2 \lambda +1;1\big)
\nonum \\
& \hspace{0cm} -(\alpha -2 \lambda ) \Gamma (\beta +1) \sin (\pi  (\beta -2 \lambda )) \csc (\pi  (\alpha +\beta -3 \lambda -1)) \Big(
\nonum \\
& \hspace{0cm} \Gamma (\lambda ) \Gamma (2 \lambda ) \Gamma (\alpha -\lambda +1) \Gamma (\beta -\lambda +1) \, _3F_2\big(\alpha ,\alpha -\lambda +1,2 \lambda ;\alpha +1,\alpha +\beta -\lambda +1;1\big) 
\nonum \\
& \hspace{0cm} +\pi  \alpha  \Gamma (\alpha ) \sin (\pi  (-\alpha +\lambda +1)) \Gamma (2 \lambda -\alpha ) \csc ^2(\pi  (\beta -2 \lambda )) \Gamma (\alpha +\beta -\lambda +1)
\Big)
\bigg] \,,
\ea
%
where $\lambda=(d-2)/2$. However, even for our special case of interest, i.e., $d=3-2\eps$ and $\alpha=\beta=1/2$, this result is extremely hard to expand in a series of $\eps$ due to the presence of half integer indices, implying branch-cuts. The \textsc{Mathematica} package \textsc{HypExp} \cite{Huber:2006,Huber:2008} is not able to expand it in a series. Upon studying the result with very high numerical precision and cross-checking our results with the numerical (Monte Carlo) sector decomposition \textsc{Mathematica} package \textsc{FIESTA} \cite{Smirnov:2009,Smirnov:2011,Smirnov:2014} and the study in \cite{Pikelner:2020mga} (which used PSLQ \mbox{techniques \cite{Ferguson:1999}}), we were able to (re-)derive the needed result, reading 
\be
G(3-2\eps,1,1/2,1,1/2,1)= \frac{8}{3\pi}\Big(\pi^2C+24\text{Cl}_4(\pi/2)+\Ord(\eps)\Big)\,,
\label{eq:multiloop:Ghard}
\ee
where $\text{Cl}_z(\theta)$ is the Clausen function, defined for an even {$z$} index as
\be
{\text{Cl}_z(\theta)=\sum_{n=1}^\infty \frac{\sin(n\theta)}{n^z}} \,, \quad
\text{Cl}_{2}(\pi/2)= C = 0.9160 \,,  \quad 
\text{Cl}_{4}(\pi/2)= 0.9889 \,,
\ee
where $C$ is the Catalan number. From our computation, the result \eqref{eq:multiloop:Ghard} is correct numerically at least up to 100-digit precision, and we, therefore, consider it exact.


\newpage

\begin{multicols}{2}
\footnotesize
\bibliographystyle{texcode/custom.bst}
\bibliography{main.bib}
\end{multicols}

\end{fmffile}

\end{document}